%% file: main.tex
\documentclass{aa}  
\pdfoutput=1
\usepackage[varg]{txfonts} 
\usepackage{natbib}
\usepackage{graphicx}
\usepackage{amsmath}	
\usepackage{mathtools}
\usepackage{amssymb}	
\usepackage{hyperref}
\hypersetup{
    colorlinks   = true, 
    urlcolor     = blue, 
    linkcolor    = blue, 
    citecolor   = blue 
}
\usepackage{multirow}
\usepackage{tabularx}
\usepackage{siunitx}
\usepackage{bm}
\usepackage{xcolor}
\usepackage[export]{adjustbox}
\usepackage{url}
\usepackage{float}

\usepackage[rightcaption]{sidecap}

\usepackage{placeins}

\defcitealias{Conti2017MNRAS}{FC17}
\newcommand{\FC}{\citetalias{Conti2017MNRAS}}

\defcitealias{Kannawadi2019AA}{K19}
\newcommand{\K}{\citetalias{Kannawadi2019AA}}

\begin{document} 

   \title{KiDS-Legacy calibration: Unifying shear and redshift calibration with the SKiLLS multi-band image simulations}
   \titlerunning{KiDS-Legacy calibration: multi-band image simulations}
   \authorrunning{S.-S. Li et al.}

\input{authors}

   \date{Received XXX, XXXX; accepted YYY, YYYY}

 
  \abstract
  {
  We present SKiLLS, a suite of multi-band image simulations for the weak lensing analysis of the complete Kilo-Degree Survey (KiDS), dubbed KiDS-Legacy analysis. The resulting catalogues enable joint shear and redshift calibration, enhancing the realism and hence accuracy over previous efforts. To create a large volume of simulated galaxies with faithful properties and to a sufficient depth, we integrated cosmological simulations with high-quality imaging observations. We also improved the realism of simulated images by allowing the point spread function (PSF) to differ between CCD images, including stellar density variations and varying noise levels between pointings. Using realistic variable shear fields, we accounted for the impact of blended systems at different redshifts. Although the overall correction is minor, we found a clear redshift-bias correlation in the blending-only variable shear simulations, indicating the non-trivial impact of this higher-order blending effect. We also explored the impact of the PSF modelling errors and found a small yet noticeable effect on the shear bias. Finally, we conducted a series of sensitivity tests, including changing the input galaxy properties. We conclude that our fiducial shape measurement algorithm, \textit{lens}fit, is robust within the requirements of lensing analyses with KiDS. As for future weak lensing surveys with tighter requirements, we suggest further investments in understanding the impact of blends at different redshifts, improving the PSF modelling algorithm and developing the shape measurement method to be less sensitive to the galaxy properties.
  }
   \keywords{gravitational lensing: weak -- methods: data analysis -- methods: statistical -- techniques: image processing}

   \maketitle
%

\section{Introduction}
\label{Sec:intro}
\input{Sec1_intro}

\section{Input mock catalogues}
\label{Sec:input}
\input{Sec2_input}
\section{KiDS+VIKING 9-band image simulations}
\label{Sec:sim}
\input{Sec3_ImSim}

\section{Shape measurements with the updated \textit{lens}fit}
\label{Sec:shape}
\input{Sec4_shape}

\section{Shear biases for the updated \textit{lens}fit}
\label{Sec:shear}
\input{Sec5_results}

\section{Sensitivity analysis}\label{Sec:sensi}
\input{Sec6_sensitivity}

\section{Discussion and conclusions}\label{Sec:discussion}
\input{Sec7_conclusion}


\begin{acknowledgements}
    We thank Fedor Getman for providing the deep VST-COSMOS catalogue and Arun Kannawadi for reading the manuscript and providing useful comments. We also wish to thank other members of the KiDS-Legacy Calibration Working Group (especially Benjamin Joachimi, Benjamin St\"{o}lzner and Anna Wittje) for informative discussions and suggestions through numerous teleconferences. This work used the compute resources from the Academic Leiden Interdisciplinary Cluster Environment (ALICE) provided by Leiden University. We acknowledge support from: the Netherlands Research School for Astronomy (SSL); the Royal Society and Imperial College (KK); the Netherlands Organisation for Scientific Research (NWO) under Vici grant 639.043.512 (HHo); and the UK Science and Technology Facilities Council (STFC) under grant ST/N000919/1 (LM) and ST/V000594/1 (CH). We also acknowledge support from the European Research Council (ERC) under grant agreement No. 647112 (CH) and No. 770935 (HHi, JLvdB, AHW); the Max Planck Society and the Alexander von Humboldt Foundation in the framework of the Max Planck-Humboldt Research Award endowed by the Federal Ministry of Education and Research (CH, MY); the Deutsche Forschungsgemeinschaft Heisenberg grant Hi 1495/5-1 (HHi); the Polish National Science Center through grants no. 2020/38/E/ST9/00395, 2018/30/E/ST9/00698, 2018/31/G/ST9/03388 and 2020/39/B/ST9/03494 (MBi); the Polish Ministry of Science and Higher Education through grant DIR/WK/2018/12 (MBi); the University of Western Australia through a Scholarship for International Research Fees and Ad Hoc Postgraduate Scholarship (MBr); and the ARC Centre of Excellence for All Sky Astrophysics in 3 Dimensions (ASTRO 3D) through project number CE170100013 (CL). The results in this paper are based on observations made with ESO Telescopes at the La Silla Paranal Observatory under programme IDs: 088.D-4013, 092.A-0176, 092.D-0370, 094.D-0417, 177.A-3016, 177.A-3017, 177.A-3018 and 179.A-2004, and on data products produced by the KiDS consortium. The KiDS production team acknowledges support from: Deutsche Forschungsgemeinschaft, ERC, NOVA and NWO-M grants; Target; the University of Padova, and the University Federico II (Naples). Contributions to the data processing for VIKING were made by the VISTA Data Flow System at CASU, Cambridge and WFAU, Edinburgh. The \textsc{SURFS}-\textsc{Shark} simulations were produced at the Pawsey Supercomputing Centre with funding from the Australian Government and the Government of Western Australia. Author contributions: All authors contributed to the development and writing of this paper. The authorship list is given in three groups: the lead authors (SSL, KK, HHo, LM) followed by two alphabetical groups. The first alphabetical group includes those who are key contributors to both the scientific analysis and the data products. The second group covers those who have either made a significant contribution to the data products, or to the scientific analysis.

\end{acknowledgements}

  \bibliographystyle{aa} 
  \bibliography{reference} 

\begin{appendix} 

\input{Sec999_appendix}

\end{appendix}

\end{document}

%% file: authors.tex
\author{Shun-Sheng Li\inst{1}
        \and
        Konrad Kuijken\inst{1}
        \and
        Henk Hoekstra\inst{1}
        \and
        Lance Miller\inst{2}
        \and
        Catherine Heymans\inst{3,4}
        \and
        Hendrik Hildebrandt\inst{4}
        \and \\
        Jan Luca van den Busch\inst{4}
        \and
        Angus H. Wright\inst{4}
        \and 
        Mijin Yoon\inst{4}
        \and
        Maciej Bilicki\inst{5}
        \and
        Mat{\'\i}as Bravo\inst{6}
        \and
        Claudia del P. Lagos\inst{6,7}
}

\institute{Leiden Observatory, Leiden University, Niels Bohrweg 2, 2333 CA Leiden, the Netherlands\\
\email{ssli@strw.leidenuniv.nl}
\and
Department of Physics, University of Oxford, Denys Wilkinson Building, Keble Road, Oxford OX1 3RH, UK
\and
Institute for Astronomy, University of Edinburgh, Royal Observatory, Blackford Hill, Edinburgh, EH9 3HJ, UK
\and
Ruhr University Bochum, Faculty of Physics and Astronomy, Astronomical Institute (AIRUB), German Centre for Cosmological Lensing, 44780 Bochum, Germany
\and
Center for Theoretical Physics, Polish Academy of Sciences, al. Lotnik\'{o}w 32/46, 02-668 Warsaw, Poland
\and
International Centre for Radio Astronomy Research (ICRAR), M468, University of Western Australia, 35 Stirling Hwy, Crawley, WA 6009, Australia
\and
ARC Centre of Excellence for All Sky Astrophysics in 3 Dimensions (ASTRO 3D), Australia
            }

%% file: Sec1_intro.tex
Weak gravitational lensing, the small deflection of light rays caused by inhomogeneous matter distributions, is a powerful tool for observational cosmology as an unbiased tracer of gravity~(see \citealt{Bartelmann2001PhR340291B}, for a review). It allows us to study the underlying distribution of both baryonic and dark matter~(see \citealt{Refregier2003ARAA41645R,Hoekstra2008ARNPS5899H,Kilbinger2015RPPh78h6901K}, for some reviews). Together with redshift estimates for the sources, the cosmological lensing signal can even quantify the growth of the cosmic structure and infer the properties of dark energy (e.g. \citealt{Hu1999ApJ522L21H,Huterer2002PhRvD65f3001H}). Recent weak lensing surveys, including the Kilo-Degree Survey + VISTA Kilo-degree INfrared Galaxy (KiDS+VIKING) survey (\citealt{Jong2013ExA3525D,Edge2013Msngr15432E})\footnote{\url{https://kids.strw.leidenuniv.nl}}, the Dark Energy Survey (DES, \citealt{Dark2016MNRAS4601270D})\footnote{\url{https://darkenergysurvey.org}}, and the Hyper Suprime-Cam (HSC) survey (\citealt{Aihara2018PASJ70S4A})\footnote{\url{https://hsc.mtk.nao.ac.jp/ssp/}}, have provided some of the tightest cosmological constraints on the clumpiness of matter in the local Universe~\citep{Heymans2021AA646A140H,Abbott2022PhRvD105b3520A,Hamana2020PASJ7216H}. The upcoming so-called Stage IV surveys, such as the ESA \textit{Euclid} space mission~\citep{Laureijs2011arXiv11103193L}\footnote{\url{https://sci.esa.int/web/euclid/}}, the Rubin Observatory Legacy Survey of Space and Time~(LSST, \citealt{Ivezic2019ApJ873111I})\footnote{\url{https://www.lsst.org/}}, and the NASA \textit{Nancy Grace Roman} space telescope~\citep{Spergel2015arXiv150303757S}\footnote{\url{https://roman.gsfc.nasa.gov/}}, will advance the field significantly by increasing the statistical power of weak lensing measurements by more than an order of magnitude. 

While promising, measuring the weak lensing signals to the desired accuracy in practice is demanding~(see \citealt{Mandelbaum2018ARAA56393M}, for a recent review). In particular, the observed images of distant galaxies are smeared by the point spread function (PSF) and contain pixel noise, biasing the measurements of galaxy shapes~(e.g. \citealt{Paulin2008AA48467P,Massey2013MNRAS429661M,Melchior2012MNRAS4242757M,Refregier2012MNRAS4251951R}). These issues drove the early development of many shape measurement methods and triggered a series of community-wide blind challenges based on image simulations, including the Shear TEsting Programme~(STEP, \citealt{Heymans2006MNRAS3681323H,Massey2007MNRAS37613M}) and the Gravitational LEnsing Accuracy Testing~(GREAT, \citealt{Bridle2010MNRAS4052044B,Kitching2012MNRAS4233163K,Mandelbaum2015MNRAS4502963M}). These early efforts illuminated some crucial issues and paved the way to calibrate the systematic biases for an actual survey using image simulations. 

Early applications of simulation-based calibration have already demonstrated that the calibration accuracy depends on how well the simulation matches the survey under consideration, especially the observational conditions and the galaxy properties~(e.g. \citealt{Miller2013MNRAS4292858M,Hoekstra2015MNRAS449685H,Hoekstra2017MNRAS4683295H,Samuroff2018MNRAS4754524S}). Therefore, recent implementations carefully mimic the data processing procedures and use morphological measurements from deep imaging surveys to reproduce the measured galaxy properties for a specific survey~(e.g. \citealt{Mandelbaum2018MNRAS4813170M}; \citealt{Kannawadi2019AA} hereafter K19; \citealt{MacCrann2022MNRAS5093371M}). Alternately, newer methods, such as the Bayesian Fourier Domain~\citep{Bernstein2014MNRAS4381880B} and \textsc{Metacalibration}~\citep{Huff2017arXiv170202600H,Sheldon2017ApJ84124S}, seek an unbiased estimate of the shear either using deeper data as a prior or directly calibrating the measurements using the observed data.

Recent studies have highlighted the effect of blending. The blending effect occurs when two or more objects are close together in the image plane, so their light distributions overlap. It introduces biases during both the selection and measurement processes. For example, \citet{Hartlap2011AA528A51H} found that the rejection of recognised blends alters the selection function of the final sample (see also \citealt{Chang2013MNRAS4342121C}). In some circumstances, blended systems are so close that they appear as single objects. These unrecognised blends increase the shape noise by decreasing the number density and widening the measured ellipticity dispersion (e.g. \citealt{Dawson2016ApJ81611D,Mandelbaum2018MNRAS4813170M}). Even if the blended objects are below the detection limit, they still introduce correlated noise that affects the detection and measurement of the adjacent bright galaxies (e.g. \citealt{Hoekstra2015MNRAS449685H,Hoekstra2017MNRAS4683295H,Samuroff2018MNRAS4754524S}), an effect that becomes even more dramatic when the clustering of galaxies is considered~\citep{Martinet2019AA627A59E}. Given all of these concerns, it is essential for image simulations to contain faint objects and physical clustering features. 

More concerns arise when considering a tomographic analysis, which is at the core of current and future weak lensing surveys. From the shear estimate side, the tomographic binning approach introduces further selections that link the shear bias to redshift estimates~(\K, \citealt{MacCrann2022MNRAS5093371M}). From the redshift estimate side, redshift calibration methods need mock photometric catalogues to verify their performance. These mock catalogues must resemble the target data in object selections and photometric measurements, which are challenging to address at the catalogue level~\citep{Hoyle2018MNRAS.478..592H,Wright2020AA637A100W,Busch2020AA642A200V,DeRose2022PhRvD.105l3520D}.

All these issues become even more challenging for the KiDS-Legacy analysis, the weak lensing analysis of the complete KiDS. It covers the entire $1350~{\rm deg}^2$ survey area, a ${\sim}35\%$ increase over the latest KiDS release (KiDS-DR4, \citealt{Kuijken2019AA625A2K}). More importantly, thanks to the deeper $i$-band observations and dedicated observations in spectroscopic survey fields, the KiDS-Legacy analysis aims to unleash the power of high-redshift samples (up to a redshift of $z{\sim}2$). The improved statistical power, however, makes a higher demand on the shear and redshift calibrations, including an assessment of the cross-talk between the systematic errors in the shear and redshift estimates. 

In this paper, we present SKiLLS (\textsc{SURFS}-based \@KiDS-Legacy-Like \@Simulations), the third generation of image simulations for KiDS following SCHOo\textit{l}~(Simulations Code for Heuristic Optimization of \textit{lens}fit, \citealt{Conti2017MNRAS}, hereafter FC17) and COllege~(COSMOS-like lensing emulation of ground experiments, \K). By simulating multi-band imaging that includes realistic galaxy evolution and clustering in terms of colour, morphology and number density, SKiLLS allows for the simultaneous measurement of shear and photometric redshifts from the same simulation. This study, therefore, provides the first joint calibration of these two key observables for cosmic shear analyses. With our approach, we provide a natural solution to address the expected cross-talk between shear and redshift bias, accounting for the impact of blends that carry different shears~\citep{Dawson2016ApJ81611D,Mandelbaum2018MNRAS4813170M,MacCrann2022MNRAS5093371M}. We also release our simulation pipeline, which contains customisable features for general use by other surveys\footnote{\url{https://github.com/KiDS-WL/MultiBand_ImSim.git}}.

The remainder of this paper is structured as follows. In Sect.~\ref{Sec:input}, we build input catalogues for image simulations. Then in Sect.~\ref{Sec:sim}, we detail the creation and processing of the KiDS-like multi-band images, starting from instrumental setups and ending with photometric catalogues. Section~\ref{Sec:shape} reviews our fiducial shape measurement algorithm, \textit{lens}fit~\citep{Miller2007MNRAS382315M,Miller2013MNRAS4292858M,Kitching2008MNRAS390149K}, with some improvements introduced for the KiDS-Legacy analysis. The shear calibration results for the updated \textit{lens}fit measurements are presented in Sect.~\ref{Sec:shear}, and the sensitivity test is conducted in Sect.~\ref{Sec:sensi}. Finally, we conclude in Sect.~\ref{Sec:discussion}. 

Throughout the paper, we define the complex ellipticity of an object as
\begin{equation}
    \label{eq:shape}
    {\bm \epsilon} \equiv \epsilon_1 + {\rm i}\epsilon_2 = \left(\frac{1-q}{1+q}\right)~\exp(2{\rm i}\phi)~,
\end{equation}
where $q$ and $\phi$ denote the axis ratio and the position angle of the major axis, respectively. In terms of the quadrupole moments of the measured surface brightness $Q_{ij}$, this definition equals
\begin{equation}
\label{eq:shapeQ}
    {\bm \epsilon} = \frac{Q_{11} - Q_{22} + 2{\rm i}Q_{12}}{Q_{11} + Q_{22} + 2~(Q_{11}Q_{22}-Q_{12}^2)^{1/2}}~.
\end{equation}
As stated by \citet{Bartelmann2001PhR340291B}, this ellipticity definition is convenient because it directly links to the weak lensing shear signal ${\bm \gamma}$ via the estimator
\begin{equation}
    \label{eq:shear}
    {\bm \gamma} = \frac{\sum_iw_i~\epsilon_i}{\sum_iw_i}~,
\end{equation}
where $w_i$ is a weight assigned per object to account for individual measurement uncertainties\footnote{Strictly speaking, the expectation value of the ellipticity is ${\bm \gamma}/(1-\kappa)$, where $\kappa$ is the convergence. But as $\kappa\ll1$ in the weak lensing regime, we can safely neglect this term.}. Although the cosmic shear analysis uses higher-order statistical measures, such as the two-point correlation functions~(e.g. \citealt{Kaiser1992ApJ...388..272K}), the simple estimator presented in Eq.~(\ref{eq:shear}) is commonly used for constraining the shear bias from image simulations~(e.g. \citealt{Heymans2006MNRAS3681323H}).

%% file: Sec2_input.tex

To generate mock images, we need input catalogues of galaxies and stars with realistic morphology, photometry and clustering. We detail our procedure for building these catalogues in this section. Section~\ref{Sec:inputGalaxies} describes how we create the mock galaxy catalogue by combining deep observations with up-to-date cosmological and galactic simulations. Section~\ref{Sec:inputStars} shows how we generate stellar multi-band magnitude distributions from a population synthesis code.

\subsection{Galaxies: SURFS-Shark simulations with COSMOS morphology}
\label{Sec:inputGalaxies}

Our input galaxy catalogue is a compilation of simulations and observations to balance the sample volume and the realism of galaxy morphology. We review the simulation part, including the clustering and multi-band photometry in Sect.~\ref{Sec:inputGalPhoto}. As for the galaxy morphology, which is crucial for the shear calibration, we learn it from observations with the learning algorithm detailed in Sect.~\ref{Sec:inputGalShape}.

\subsubsection{Generating synthetic galaxies from simulations}
\label{Sec:inputGalPhoto}

To jointly calibrate the shear and redshift estimates, we must base the image simulations on wide and deep ($z>2$) cosmological simulations, where the true redshift is known. In the previous KiDS redshift calibration, \citet{Busch2020AA642A200V} used the MICE Grand Challenge
(MICE-GC) simulation, an $N$-body light-cone simulation that covers an octant of the sky~\citep{Fosalba2015MNRAS4482987F}. However, the MICE simulation has a redshift limit of $z{\sim}1.4$, preventing its use for calibrating the high-redshift samples in the KiDS-Legacy analysis (up to $z{\sim}2$). Therefore, we switched to another public $N$-body simulation from the Synthetic UniveRses For Surveys (\textsc{SURFS}, \citealt{Elahi2018MNRAS}).

The \textsc{SURFS} simulation we adopted has a box size of $210h^{-1}~{\rm cMpc}$ (cMpc stands for comoving megaparsec), containing $1536^3$ particles with a mass of $2.21\times 10^8h^{-1}{\rm M}_{\odot}$, and a softening length of $4.5h^{-1}~{\rm ckpc}$ (ckpc stands for comoving kiloparsec). It assumes a $\Lambda$CDM cosmology with parameters from \citet{Planck2016AA}. The final halo catalogues and merger trees are constructed from $200$ snapshots starting at redshift $z=24$, using the phase-space halo-finder code \textsc{VELOCIraptor}~\citep{Canas2019MNRAS4822039C,Elahi2019PASA3621E} and the halo tree-builder code \textsc{treefrog}~\citep{Elahi2019PASA3628E}. We refer to \citet{Lagos2018MNRAS} for details on the building and \citet{Poulton2018PASA3542P} for validating the halo catalogues and merger trees.


The galaxy properties, including the star formation history and the metallicity history, are from an open-source semi-analytic model named \textsc{Shark}\footnote{\url{https://github.com/ICRAR/shark}}~\citep{Lagos2018MNRAS}. The model parameters are tuned to reproduce the $z=0, 1$ and $2$ stellar-mass functions~\citep{Wright2018MNRAS4803491W}, the $z=0$ black hole-bulge mass relation~\citep{McConnell2013ApJ764184M} and the mass-size relations at $z=0$~\citep{Lange2016MNRAS4621470L}. Any other observables are predictions of the model, which also match well with observations~(see \citealt{Lagos2018MNRAS} for more details). As for weak lensing calibration, the most crucial property is the redshift evolution of the galaxy number density (e.g. \citealt{Hoekstra2017MNRAS4683295H}), which we checked in detail in Appendix~\ref{Sec:modifPhoto} and found it to be sufficient for KiDS.

The light cones from the \textsc{Shark} outputs are created using the code \textsc{stingray}~\citep{Chauhan2019MNRAS.488.5898C}, an improved version of the code used by \citet{Obreschkow2009ApJ...703.1890O}. It first tiles the simulation boxes together to build a complex 3D field along the line of sight, then draws galaxy properties from the closest available time-step, resulting in spherical shells of identical redshifts. A possible issue would be the same galaxy appearing once in every box but with different intrinsic properties due to cosmic evolution. To avoid this problem, \textsc{stingray} randomises galaxy positions by applying a series of operations consisting of $90\deg$ rotations, inversions, and continuous translations. We refer to \citet{Chauhan2019MNRAS.488.5898C} for more details about the light-cone construction. 

The final mock-observable sky covers ${\sim}108~{\rm deg}^2$ with minimum repetition of the large-scale structure. The sample variance bias caused by the replicating structure is negligible for our direct shear and photometric redshift calibration. Since we learn galaxy morphology from deep observations, our input galaxy sample is still limited mainly by the observational data we have, which only covers ${\sim}1~{\rm deg}^2$ (see Sect.~\ref{Sec:inputGalShape} for details). We test the robustness of our calibration results against this sample variance bias using the sensitivity analysis detailed in Sect.~\ref{Sec:sensi}.

The multi-band photometry is drawn from a stellar population synthesis technique implemented in the \textsc{ProSpect}\footnote{\url{https://github.com/asgr/ProSpect}} and \textsc{Viperfish}\footnote{\url{https://github.com/asgr/Viperfish}} packages. \textsc{ProSpect}~\citep{Robotham2020MNRAS495905R} is a high-level package combining the commonly used stellar synthesis libraries with physically motivated dust attenuation and re-emission models; while \textsc{Viperfish} is a light wrapper to aid the interface with the \textsc{Shark} outputs. We refer to \citet{Lagos2019MNRAS} for detailed predictions, validations and a demonstration that the predicted results agree with observations in a broad range of bands from the far-ultraviolet to far-infrared, without any fine-tuning with observations.

\begin{figure}
   \centering
   \includegraphics[width=\hsize]{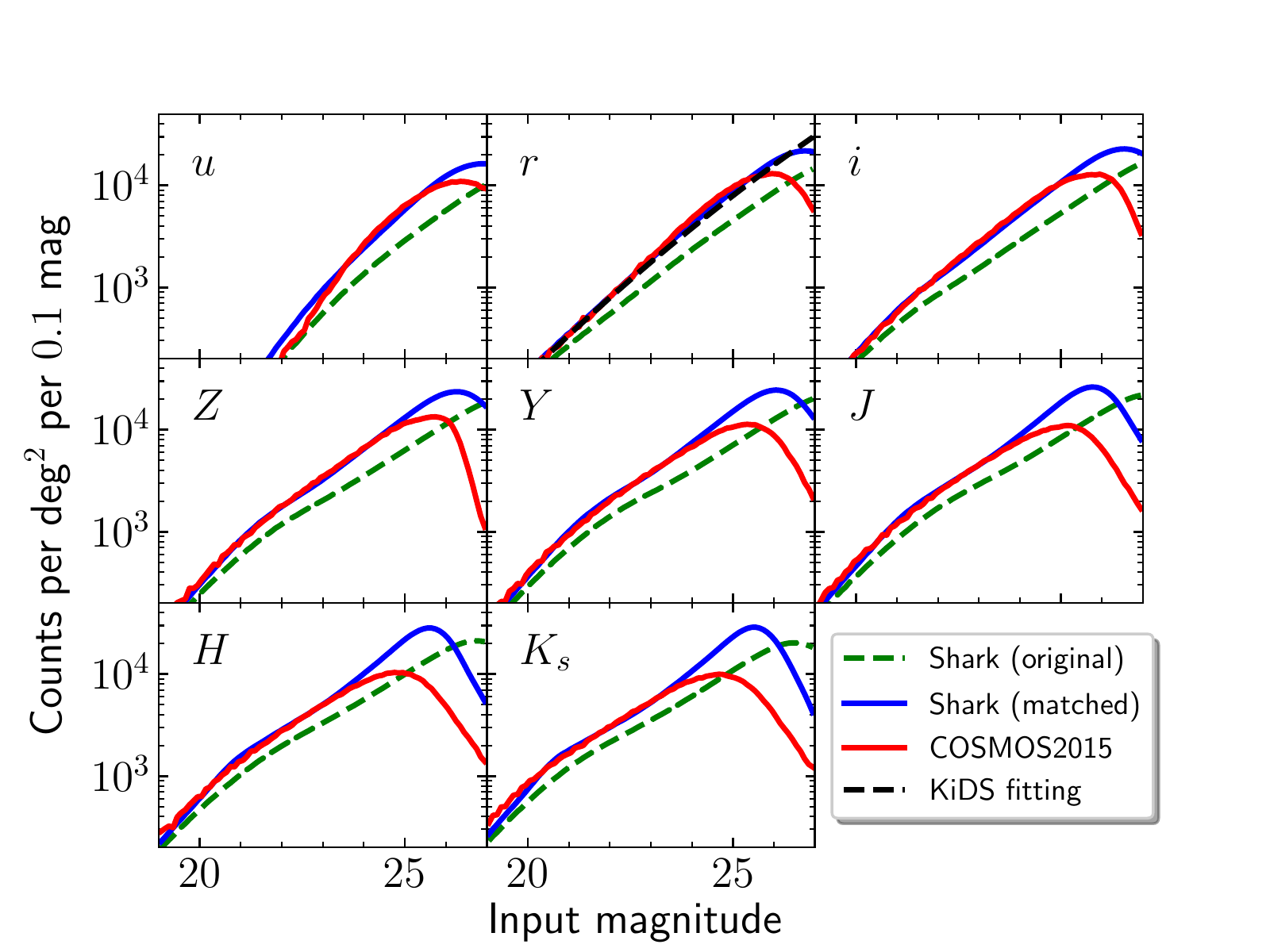}
   \caption{Number of galaxies per square degree per $0.1$ magnitude in the input apparent magnitudes. The green dashed lines are from the original \textsc{SURFS}-\textsc{Shark} mock catalogue, whilst the blue solid lines denote the modified results. The red solid lines correspond to the COSMOS2015 observations with flags applied for the UltraVISTA area inside the COSMOS field after removing saturated objects and bad areas ($1.38~{\rm deg}^2$ effective area, Table 7 of \citealt{Laigle2016ApJS}). The analytical fitting result in the $r$-band (black dashed line) is from \FC. The $g$-band photometry is not in the COSMOS2015 catalogue and, thus, not shown in the plot. We note that the COSMOS2015 catalogue is incomplete at $K_s\gtrsim 24.5$~\citep{Laigle2016ApJS}.}
    \label{fig:photoIn}%
    \end{figure}

For our purpose, we care most about the nine-band photometry covered by the KiDS+VIKING data, so we compared the synthetic near-infrared and optical magnitude distributions to observations from the COSMOS2015 catalogue~\citep{Laigle2016ApJS}. Figure~\ref{fig:photoIn} shows the magnitude distributions of eight filters available in both \textsc{Shark} and COSMOS2015 catalogues, together with an analytical fitting result from Eq.~(4) of {\FC}. The counts in the original simulations are ${\sim}35\%$ lower than the observations with some variation between filters. As this affects the blending level and then the shear bias~\citep{Hoekstra2015MNRAS449685H,Hoekstra2017MNRAS4683295H}, we calibrated the original synthetic photometry for a better agreement. The technical details are presented in Appendix~\ref{Sec:modifPhoto}. In short, we found that the differences in the magnitude distributions stem from the difference in stellar mass-to-light ratio between the simulations and observations. Therefore, we scaled the original \textsc{Shark} magnitudes using a modification factor derived from the stellar mass-to-light ratio difference. The modification is the same for all bands, preserving the intrinsic colours of individual galaxies. The modified magnitudes now agree with the observations within ${\sim}3\%$. 
%

   \begin{figure*}
   \centering
   \includegraphics[width=\textwidth]{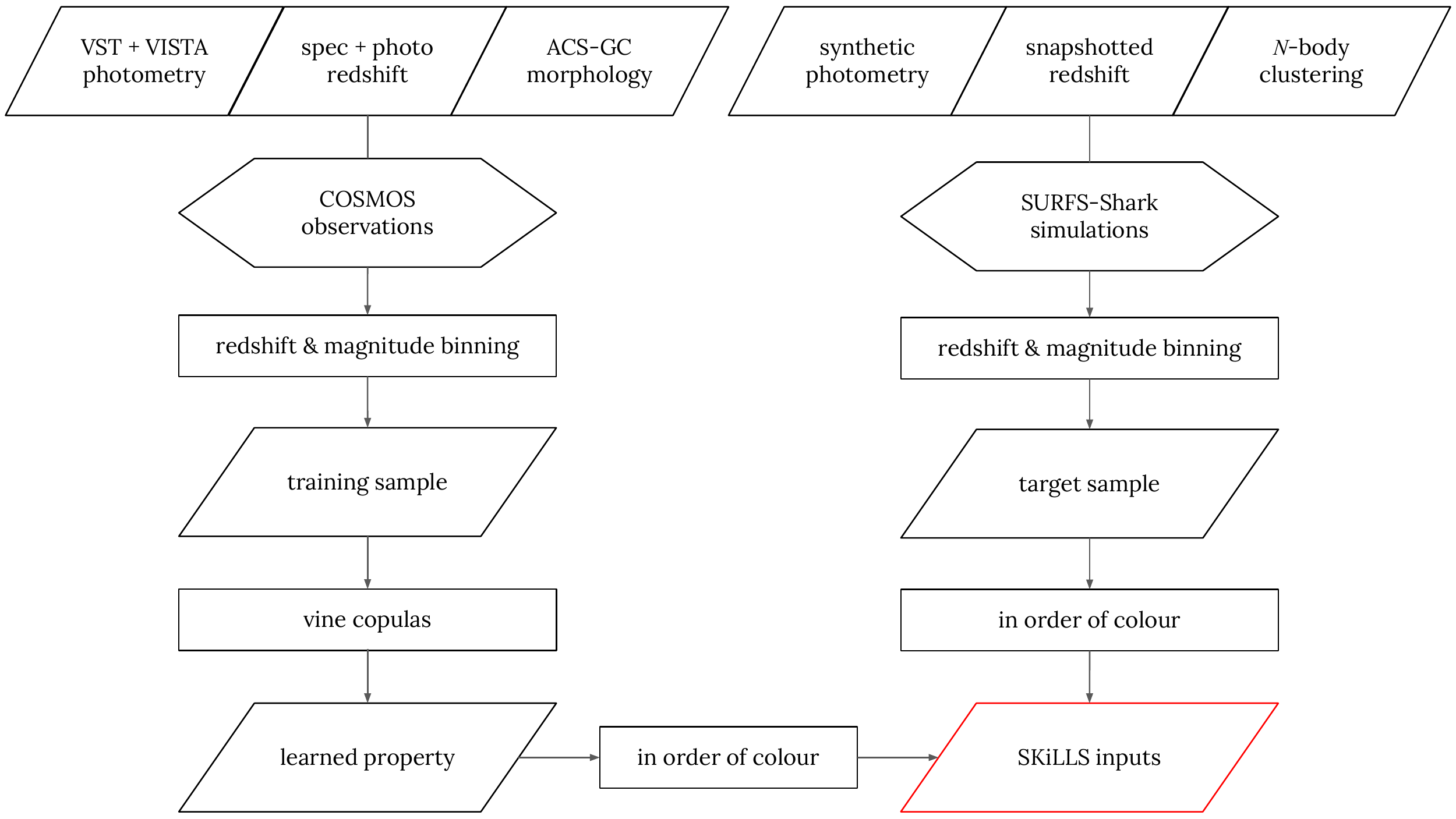}
   \caption{Flowchart summarising the algorithm to construct the SKiLLS input mock catalogue. The SKiLLS galaxies inherit the synthetic multi-band photometry and $N$-body 3D positions from the \textsc{SURFS}-\textsc{Shark} simulations, whilst the morphology is learned from the observations in the COSMOS field using an algorithm based on the vine-copula modelling (see Sect.~\ref{Sec:inputGalShape} for details).}
              \label{fig:algorithm}
    \end{figure*}

We later noticed that \citet{Bravo2020MNRAS4973026B} proposed a similar fine-tuning method when working with the panchromatic Galaxy And Mass Assembly (GAMA) survey. They used an abundance matching method by comparing the number counts between \textsc{Shark} and GAMA after fine binning in redshift and $r$-band apparent magnitude. They tuned magnitudes for all \textsc{Shark} galaxies with $r{<}21.3$ to match the number counts in GAMA. Their modifications are consistent with our results, albeit targeting different magnitude ranges.

\subsubsection{Learning galaxy morphology from observations}
\label{Sec:inputGalShape}

Simulating galaxies with realistic morphology is essential for accurate shear calibration. Following {\K}, we represent the galaxy morphology using the S\'ersic profile~\citep{sersic1963BAAA641S} with three parameters: the effective radius determining the galaxy size (also known as the half-light radius), the S\'ersic index describing the concentration of the brightness distribution, and the axis ratio determining the galaxy ellipticity. We learned these structural parameters from deep observations accounting for their mutual correlations and their correlations to galaxy photometry and redshift. Figure~\ref{fig:algorithm} shows the workflow for the learning algorithm. 

We start with a `reference' sample comprising morphology, photometry and redshifts from several deep observations. The structural parameters are adopted from the catalogue produced by \citet{Griffith2012ApJS}, who fitted S\'ersic models to the galaxy images taken by the Advanced Camera for Surveys (ACS) instrument on the \textit{Hubble} Space Telescope (HST). We used their results derived from the COSMOS survey and cleaned the sample by only preserving objects with a good fit $\left({\rm FLAG\_GALFIT\_HI}=0\right)$ and reasonable size (half-light radius between $0\farcs 01$ and $10\arcsec$) to avoid contamination. We note that this catalogue was also used by {\K} and proved to be sufficient for KiDS-like simulations. 

The $r$-band photometry is derived from a deep VST-COSMOS catalogue using $24$ separate VST observations of the COSMOS field taken from KiDS and the SUpernova Diversity And Rate Evolution (SUDARE) survey (\citealt{Cappellaro2015AA...584A..62C,Cicco2019AA...627A..33D}). These observations have a maximum seeing of $0\farcs 82$, close to the KiDS $r$-band image qualities. To ensure consistent measurements, we conducted the stacking and detection processes using the same pipeline as the standard KiDS data processing. The stacked image has an average seeing of $0\farcs 75$ and a total exposure time of \num[]{42120} seconds, which is a factor of ${\sim}23$ over a standard KiDS observation. The limiting magnitude of the final deep catalogue is more than one magnitude deeper than usual KiDS catalogues. To include colour information, we also used the $K_s$-band photometry from the COSMOS2015 catalogue~\citep{Laigle2016ApJS}, as it originates from the UltraVISTA project~\citep{McCracken2012AA544A156M} that shares the same instruments with the VIKING near-infrared observations. 

The redshifts are taken from the catalogue compiled by \citet{Busch2022AA...664A.170V}. It contains observations from several spectroscopic and high-quality photometric surveys in the COSMOS field. The spectroscopic redshifts were collected from G10-COSMOS~\citep{Davies2015MNRAS}, DEIMOS~\citep{Hasinger2018ApJ}, hCOSMOS~\citep{Damjanov2018ApJS}, VVDS~\citep{LeFevre2013AA}, LEGA-C~\citep{vanderwel2016ApJS}, FMOS-COSMOS~\citep{Silverman2015ApJS}, VUDS~\citep{lefevre2015AA}, C3R2~\citep{Masters2017ApJ,Masters2019ApJ,Euclid2020AA,Stanford2021ApJS}, DEVILS~\citep{Davies2018MNRAS} and zCOSMOS~(\textit{private comm. from M. Salvato}), while the photometric redshifts were from the PAU survey~\citep{Alarcon2021MNRAS} and COSMOS2015~\citep{Laigle2016ApJS}. For sources with multiple measurements, a specific `hierarchy' was defined with orders based on the quality of measured redshifts to choose the most reliable redshift estimates (see Appendix A in \citealt{Busch2022AA...664A.170V}, for details). Given the high quality of the redshift estimates, we treated them as true redshifts.

All catalogues mentioned above overlap in the COSMOS field, so we can combine them by cross-matching objects based on their sky positions. The final reference catalogue has \num{75403} galaxies with all the necessary information. It has a limiting magnitude of $27$ in the $r$-band but suffers incompleteness after $m_{\rm r}\gtrsim 24.5$. We verified that the incompleteness at the faint end does not bias the overall morphological distribution by comparing it to measurements from the \textit{Hubble} Ultra Deep Field observations~\citep{Coe2006AJ132926C}.

We aim to inherit not only the individual distributions of structural parameters but also their mutual dependence and possible correlations with redshifts and magnitudes. To achieve this goal, we developed a learning algorithm based on a novel statistical inference technique, dubbed vine copulas~(e.g. \citealt{joe2014dependence,CzadoClaudia2019ADDw}). A brief introduction to the technique is presented in Appendix~\ref{Sec:vineCopulas}. In short, a copula-based method models joint multi-dimensional distributions by separating the dependence between variables from the marginal distributions. It is popular in studies concerning dependence modelling, given its flexibility and reliability. In practice, we first divided galaxies into $30\times40$ bins based on their redshifts and $r$-band magnitudes. Each bin contains a similar number of reference galaxies. Then in each bin, we built a data-driven vine-copula model from the measured $r-K_s$ colour and morphological parameters using the public \texttt{pyvinecopulib} package\footnote{\url{https://github.com/vinecopulib/pyvinecopulib}}. The learned vine-copula model can be sampled to produce an arbitrary number of vectors of parameters from the constrained multi-dimensional distributions. We decided to generate the same number of vectors as the available \textsc{Shark} galaxies and assign them to the \textsc{Shark} galaxies in the order of $r-K_s$ colour. This approach allows us to mimic observations from the underlying distributions rather than repeatedly sampling from the measured values.

\begin{figure*}
   \includegraphics[width=0.5\hsize]{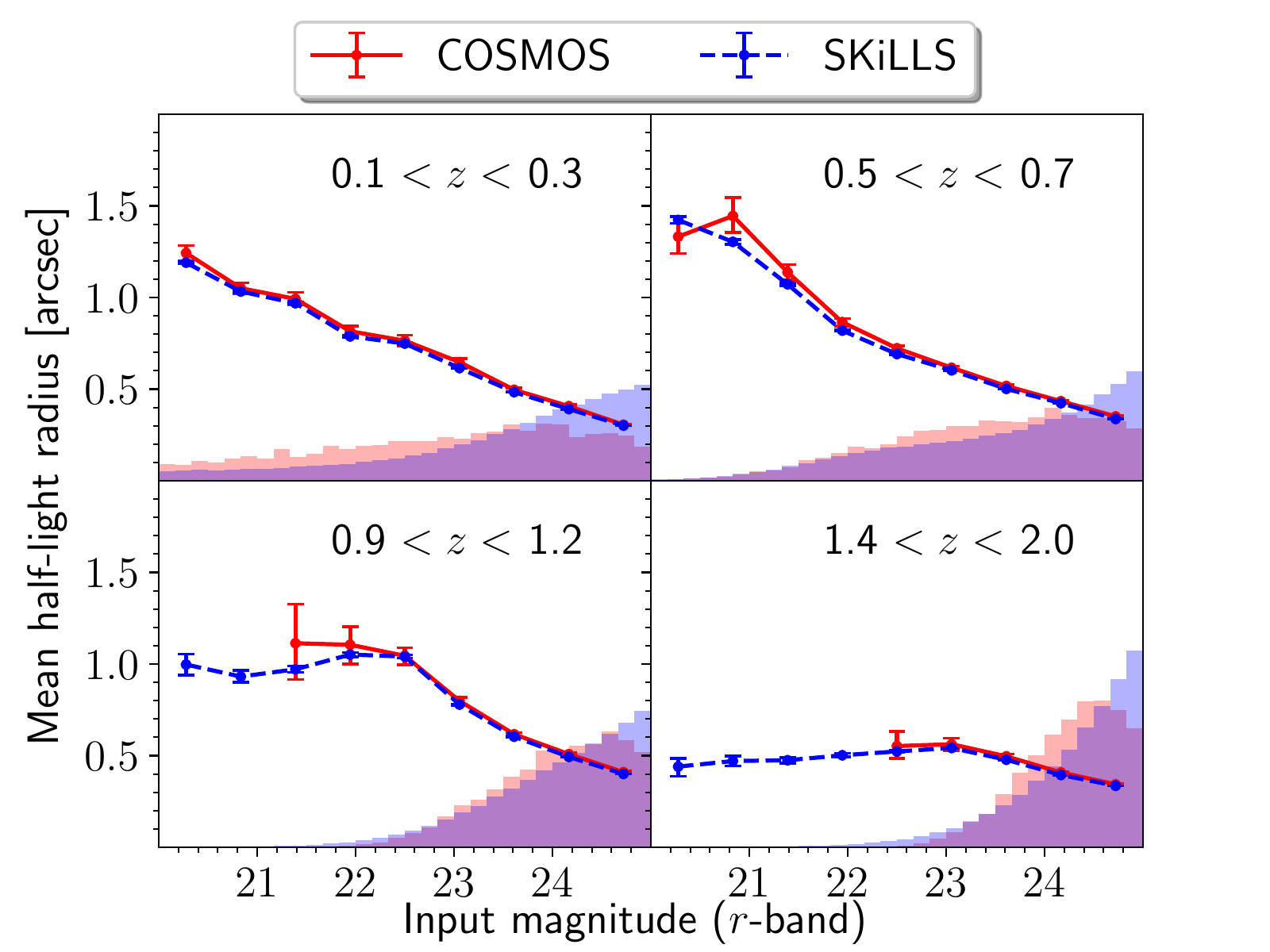}
   \includegraphics[width=0.5\hsize]{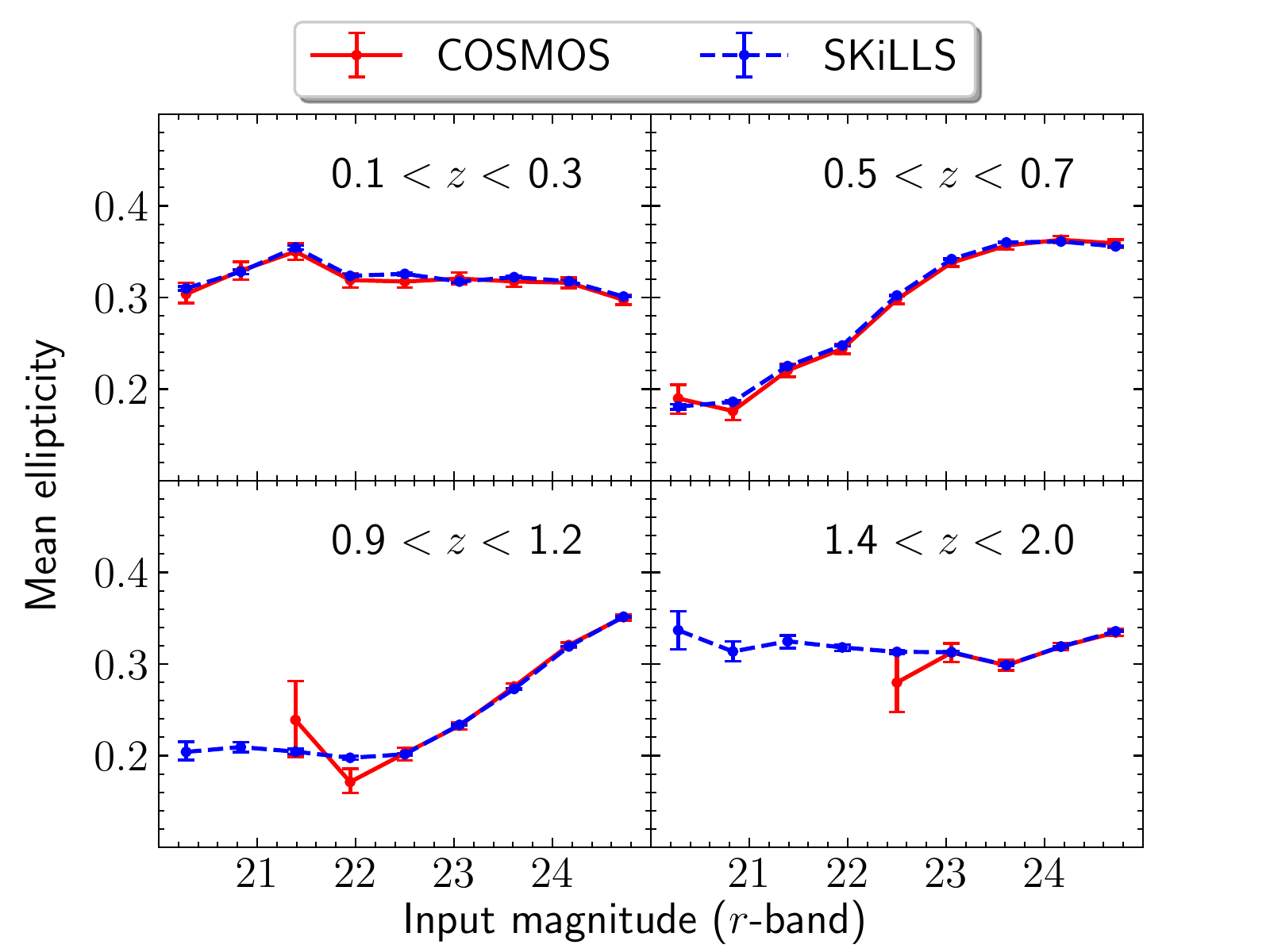}
      \caption{Comparison of the overall magnitude-morphology relations in several redshift bins.
      The red solid and blue dashed lines denote the training and target samples, respectively. The left panel shows the mean half-light radius as a function of $r$-band magnitude, whilst the right panel presents the mean ellipticity as a function of $r$-band magnitude. The statistical uncertainties shown are calculated from \num{500} bootstraps. The left panel also shows the histograms of the normalised magnitude distributions, demonstrating that the extra high-redshift bright galaxies in the simulation contribute little to the overall population.}
         \label{fig:MagMorZbin}
   \end{figure*}

\begin{figure*}
   \includegraphics[width=0.5\hsize]{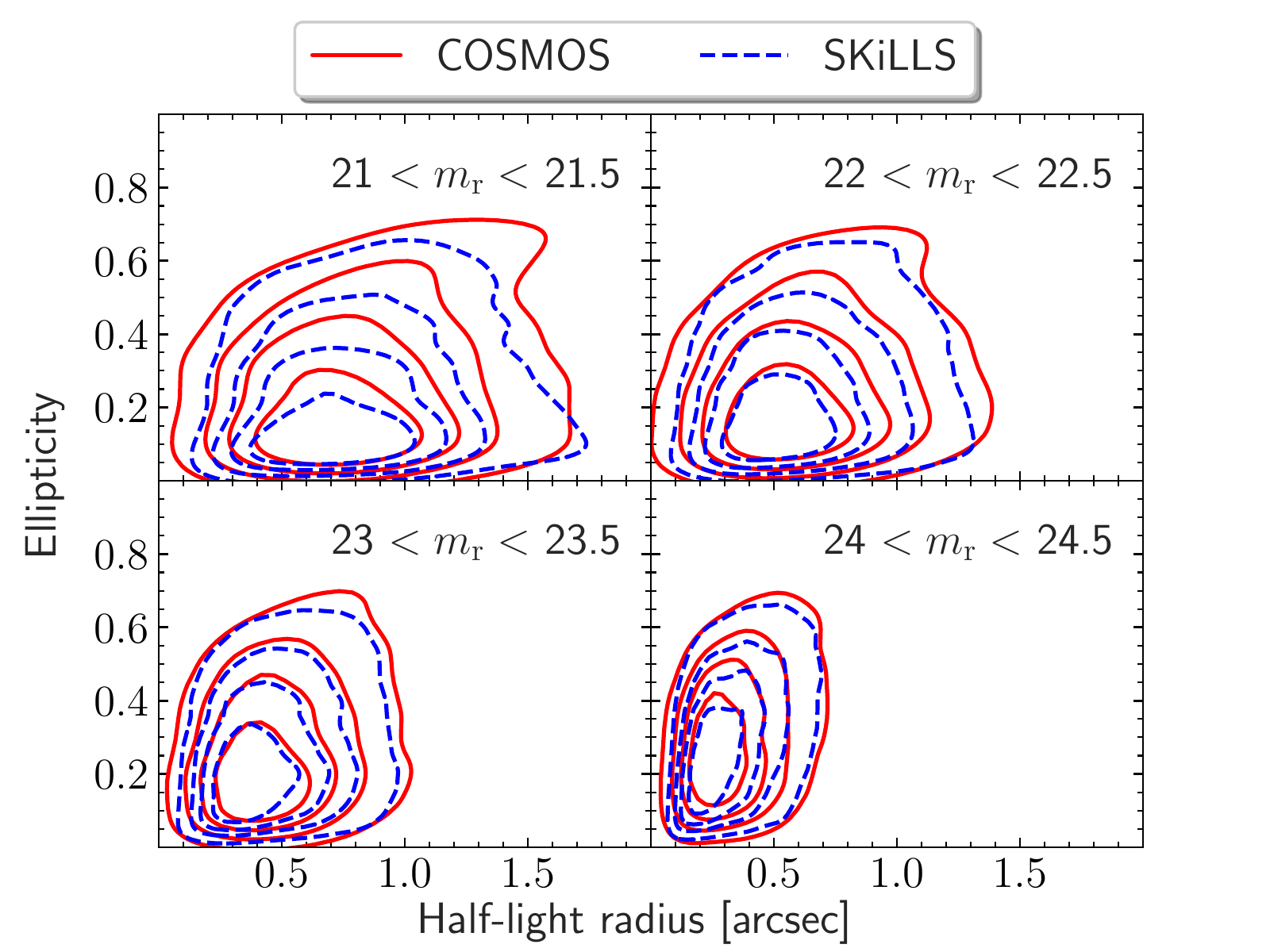}
   \includegraphics[width=0.5\hsize]{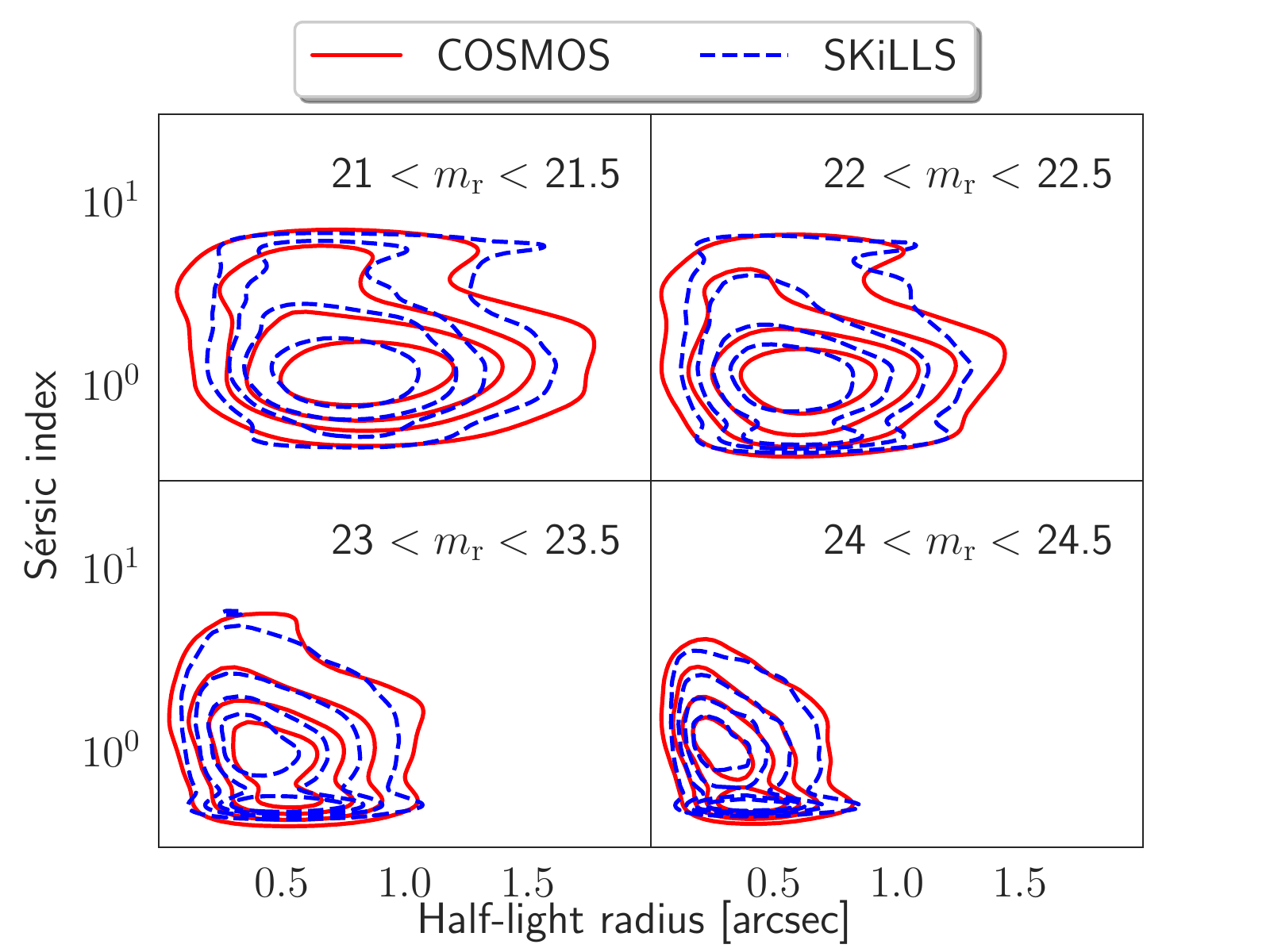}
      \caption{Two-dimensional kernel density plots of morphological parameters in several magnitude bins. The red solid and blue dashed lines denote the training and target samples, respectively. The left panel shows the correlation between the size and ellipticity, whilst the right panel presents the correlation between the size and S\'ersic index. The plotted contour levels are $20\%$, $40\%$, $60\%$, $80\%$.}
         \label{fig:Mor2dMagbin}
   \end{figure*}

Figure~\ref{fig:MagMorZbin} shows the correlations between the magnitude and the two critical structural parameters: half-light radius and ellipticity, in several redshift bins. We see that the learned sample follows the average trends of the reference sample. Figure~\ref{fig:Mor2dMagbin} presents two-dimensional contour plots in several magnitude bins to better inspect the underlying distributions of morphological parameters. We again see agreements in correlations between the size and ellipticity and between the size and concentration, proving that our copula-based algorithm captures the multi-dimensional dependence from the reference sample.

\subsection{Stars: Point objects with synthetic photometry}
\label{Sec:inputStars}

\begin{figure}
   \centering
   \includegraphics[width=\hsize]{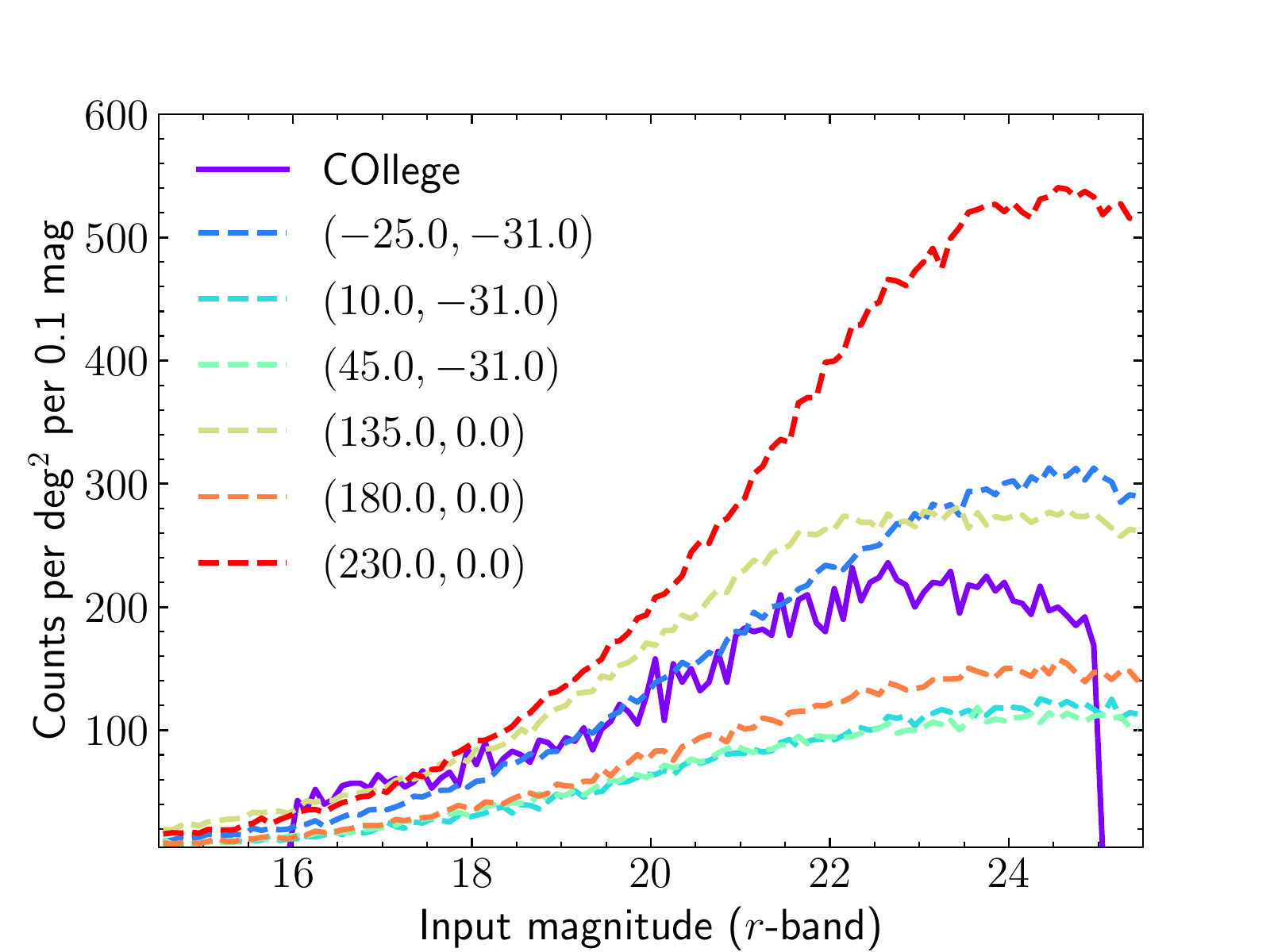}
      \caption{Input magnitude distributions in the $r$-band for the six stellar catalogues used by SKiLLS. Labels indicate the pointing centres (RA, Dec), except for `COllege', which denotes the stellar catalogue used by {\K}.}
         \label{fig:stars}
   \end{figure}

We treated stars as perfect point objects. Their multi-band photometry was obtained from the population synthesis code, \textsc{Trilegal}~(\citealt{Girardi2005AA}, with version 1.6 and the default model from its website\footnote{\url{http://stev.oapd.inaf.it/cgi-bin/trilegal}}). We generated six stellar catalogues at galactic coordinates evenly spaced across the KiDS footprint to capture the variation of stellar densities between KiDS tiles. Each catalogue spans $10~{\rm deg}^2$. When simulating a specific tile image covering $1~{\rm deg}^2$, we selected the stellar catalogue whose central pointing is closest to the target tile, then randomly drew ten per cent of stars from that catalogue as the input. Figure~\ref{fig:stars} shows the $r$-band magnitude distributions of the six stellar catalogues compared to the catalogue used by the COllege simulations. The broader coverage of stellar densities is noticeable, marking one of the improvements in SKiLLS. Also, stars in SKiLLS have nine-band magnitudes consistently predicted from a library of stellar spectra (see \citealt{Girardi2005AA}, for details), while in COllege, stars only have $r$-band magnitudes.

%% file: Sec3_ImSim.tex
This section details the creation and processing of the multi-band mock images. We start with the creation of KiDS-like optical images (Sect.~\ref{Sec:simKids}) and VIKING-like infrared images (Sect.~\ref{Sec:simViking}), then summarise the SKiLLS fiducial setups in Sect.~\ref{Sec:simFiducial}. We end the section with the measurement of colours and photometric redshifts (Sect.~\ref{Sec:simPhoto}).

\subsection{KiDS-like optical images}
\label{Sec:simKids}

Each KiDS pointing consists of four-band optical images taken with the OmegaCAM camera at the VLT Survey Telescope~\citep{Kuijken2011Msngr1468K}: $u$, $g$, $r$ and $i$. The $r$-band images are the primary products used for the shear measurement, while the remaining bands are only for photometric measurements. The science array of the OmegaCAM camera has a ${\sim}1^{\circ}\times 1^{\circ}$ field of view covered by $8\times4$ CCD images, each of size $2048\times 4100$ pixels with an average resolution of $0\farcs214$. Although the CCDs are mounted as closely as possible, a narrow gap between the neighbouring CCDs is technically inevitable. The average gap sizes between the pixels of neighbouring CCDs are:
\begin{itemize}
    \item between the long sides of the CCDs: $1.5~{\rm mm}~(100~{\rm pixels})$
    
    \item central gap along the short sides: $0.82~{\rm mm}~(55~{\rm pixels})$
    
    \item wide gap along short sides: $5.64~{\rm mm}~(376~{\rm pixels})$
    
\end{itemize}
To avoid `dead zones' caused by these gaps, each tile image incorporates multiple dithered exposures (five in the $g$, $r$ and $i$ bands, four in the $u$ band). The dithers form a staircase pattern with steps of $25\arcsec$ in RA and $85\arcsec$ in declination to match the gaps between CCDs~\citep{Jong2013ExA3525D}. 

KiDS raw observations are processed with two independent pipelines: the \textsc{Astro}-WISE pipeline designed for the photometric measurements~\citep{McFarland2013ExA3545M,Jong2015AA582A62D}\footnote{\url{http://www.astro-wise.org/}}, and the \textsc{theli} pipeline optimised for the shape measurements~\citep{Erben2005AN326432E,Schirmer2013ApJS20921S,Kuijken2015MNRAS4543500K}\footnote{\url{https://www.astro.uni-bonn.de/theli/}}. While the former is applied to all four-band observations, the latter is only used for the $r$-band observations, as KiDS only measures galaxy shapes for lensing in the $r$-band images. The main difference between the \textsc{Astro}-WISE and \textsc{theli} pipelines is in the co-addition process, where the former resamples all exposures to the same pixel grid with a uniform $0\farcs20$ pixel size, while the latter preserves the original pixels to maintain image fidelity as much as possible. 

  \begin{figure}
  \centering
  \includegraphics[width=\hsize]{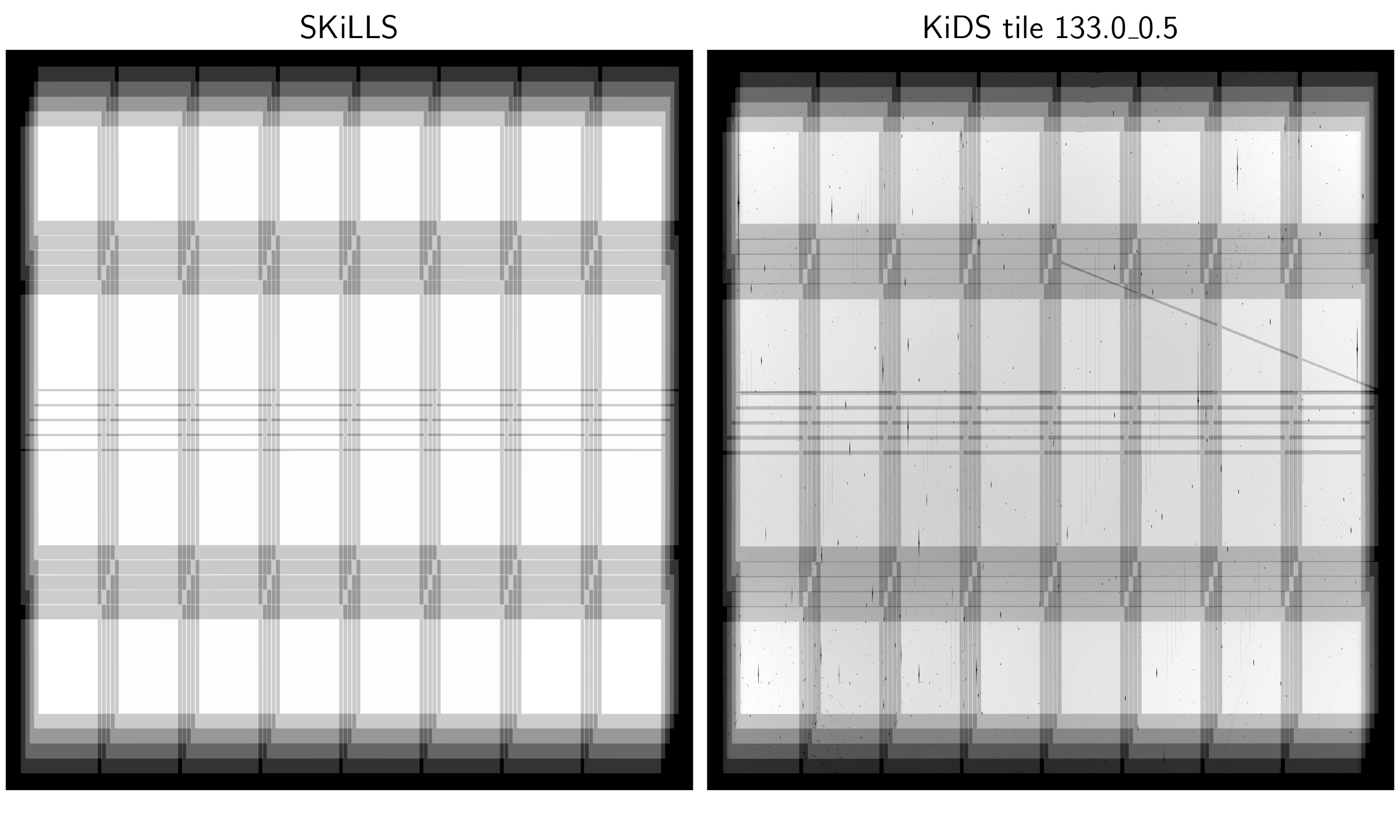}
      \caption{Comparison of the \textsc{theli} weight image produced by SKiLLS (left panel) to a randomly selected example from KiDS (right panel). The $8\times4$ CCDs cover a ${\sim}1$ square-degree sky area. The shallow regions are caused by the gaps in individual exposures. The same level of agreement is also achieved for the \textsc{Astro}-WISE co-added images.}
         \label{fig:weightMap}
  \end{figure}
We kept all these features in mind when generating SKiLLS optical images. We created raw exposures using the \textsc{GalSim} pipeline\footnote{\url{https://github.com/GalSim-developers/GalSim}}~\citep{galsim_2015AC10121R}, with galaxies and stars from the mock catalogues described in Sect.~\ref{Sec:input}. The underlying canvas mimicked the science array of the OmegaCAM camera, including pixels and gaps. Galaxies and stars were mapped to the canvas using the gnomonic (TAN) projection of their original sky coordinates. Following the KiDS image processing, we stacked exposures using the \textsc{SWarp} software~\citep{Bertin2010asclsoft10068B}, with the identical setups as in the KiDS pipelines, including \textsc{Astro}-WISE-like images re-gridded to a uniform $0\farcs20$ pixel size and \textsc{theli}-like images preserving the original $0\farcs 214$ pixel size. Figure~\ref{fig:weightMap} compares a co-added \textsc{theli} weight image from SKiLLS to a randomly selected tile from KiDS. It shows that the SKiLLS images contain the main features of KiDS images, including the gaps and dither patterns, albeit lacking subtle features, such as the inhomogeneous backgrounds between CCDs and masks of satellites. 
  
Besides the image layout, we need information on the pixel noise and point spread function (PSF) to mimic observational conditions. We extracted this information from the fourth public data release of KiDS (KiDS-DR4, or DR4 for short, \citealt{Kuijken2019AA625A2K}). It has a total of $1006$ square-degree survey tiles with stacked $ugri$ images along with their weight maps, masks and source catalogues. We selected a representative sample of $108$ tiles and replicated their properties in our image simulations (see Sec.~\ref{Sec:simFiducial} for details). For the raw pixel noise, we adopted Gaussian distributions with variances estimated from the \textsc{Astro}-WISE weight maps corrected with a boost factor of ${\sim}1.145~(=(0.214/0.2)^2)$ to account for the re-gridding effect. For the PSF, we used two approaches, depending on the different usages of the images.

For the $r$-band images from which galaxy shapes are measured, we used the position-dependent PSF models for individual exposures. These PSF models, constructed from well-identified stars, are in the form of two-dimensional polynomial functions and can recover a PSF image in the pixel grid for any given image position~(see \citealt{Miller2013MNRAS4292858M,Kuijken2015MNRAS4543500K,Giblin2021AA645A105G} for details). In practice, we recovered $32$ PSF images for each exposure using the centre positions of the CCD images. The recovered PSF images contain modelling uncertainties, which can introduce artificial spikes when being used to simulate bright stars. Therefore, we applied a cosine-tapered window to the original PSF image to suppress the modelling noise at its outskirts. The two edges of the window function are defined at $5$ and $10$ times the full-width half-maximum (FWHM) of the target PSF to preserve features in the central region as much as possible. With these recovered PSF images, we can treat the $32$ CCD images separately using their own PSFs, a significant improvement from the constant PSF used in previous work. The recovered PSF image is also superior to a Moffat profile as it captures more delicate features of complex PSFs, such as ellipticity gradients. 
    
For other optical bands where only photometry is measured, we still adopted the Moffat profile, given that the photometric measurement is insensitive to the detailed profile of PSF. We estimated the Moffat parameters by modelling bright stars identified in the \textsc{Astro}-WISE images. Since the photometry is measured from the stacked images and is less sensitive to the gentle PSF variation within a given tile, we kept the PSF model invariant for all exposures for simplicity. To alleviate the Moffat fitting bias introduced by the pixelisation of CCD images, we applied the first-order correction to the measured Moffat parameters using image simulations. Specifically, we simulated the pixelated PSF image using measured Moffat parameters and then remeasured them with the same fitting code. The difference between the remeasured and input values is the correction factor and is subtracted from the initially measured value. Our test shows that this correction can suppress the original percent-level bias down to a sub-percent level, which is sufficient for our photometry-related purpose.

\subsection{VIKING infrared images}
\label{Sec:simViking}

To improve the accuracy of photometric redshifts, KiDS includes near-infrared (NIR) measurements from the VISTA Kilo-degree Infrared Galaxy (VIKING) survey~\citep{Edge2013Msngr15432E}. The two surveys share an almost identical footprint. We refer to \citet{Wright2019AA632A34W} for details of the VIKING imaging and its usage in KiDS. Briefly, the VIKING data have three levels of products: exposures, paw-prints, and tiles. Given the complex NIR backgrounds, the VIKING survey first takes multiple exposures in quick succession with small jitter steps for reliable estimation of the noisy background. These exposures are then stacked together to create the second level of product: the `paw-print'. A paw-print still contains gaps between individual detectors, so six paw-prints with a dither pattern are used to produce a contiguous tile image. However, these co-added tiles have non-contiguous PSF patterns caused by the large dithers between successive paw-prints. Therefore, in the KiDS+VIKING analyses, photometry is done on individual paw-prints instead of the co-added tiles. The dither pattern of paw-prints causes multiple flux measurements per source (typically four in the case of the $J$-band and two in the other bands). The final flux estimate for each source is a weighted average of the individual measurements with the weights derived from individual flux errors. 

Given the complexity of the VIKING observing strategy, we simplified the NIR-band observations in SKiLLS with single images per square degree of KiDS tile. To compensate for the simplified images, we considered the overlap between individual paw-prints when estimating the observational conditions. As we show in Sect.~\ref{Sec:simPhoto}, this simplified approach can still achieve realistic photometry, which is the only important quality we seek from the NIR-band images.

Specifically, we created a `flat-field image' for each paw-print with the same size and pixel scale. Its pixel value equals the absolute standard deviation of the background pixel values on the corresponding paw-print. For each KiDS pointing, we selected all VIKING paw-prints that overlap in the given one square-degree sky area and stacked their flat-field images with shifts accounting for the different sky pointings of the paw-prints. We took the median pixel value of the co-added flat-field image as the final pixel noise of the corresponding KiDS pointing. In doing so, we captured various overlapping VIKING paw-prints in individual KiDS pointings. Following the typical situations of the KiDS+VIKING data~\citep{Wright2019AA632A34W}, we only preserved KiDS pointings with at least two paw-prints in the $ZYHK_s$-bands and at least four paw-prints in the $J$-band. This requirement reduced the number of pointings from $1006$ to $979$, which is still plentiful for our purpose. As for the PSF, we employed a constant Moffat profile for each KiDS pointing. The PSF FWHM is a weighted average from overlapping VIKING paw-prints with the weights determined by their noise levels. In order to determine the Moffat concentration index for a given FWHM value, we fitted Moffat profiles to bright stars in some representative paw-prints. The Moffat fitting bias introduced by the pixelisation is corrected using the same method introduced in Sect.~\ref{Sec:simKids}. We found the relationship between the Moffat index $n$ and FWHM (arcsec) in VIKING images to be: $\ln(n) = 66.56~\exp\left(-6.36~{\rm FWHM}\right) + 0.90$. This empirical formula is used to pair each FWHM with a unique Moffat index.

\subsection{SKiLLS fiducial setup}
\label{Sec:simFiducial}

\begin{figure}
  \centering
  \includegraphics[width=\hsize]{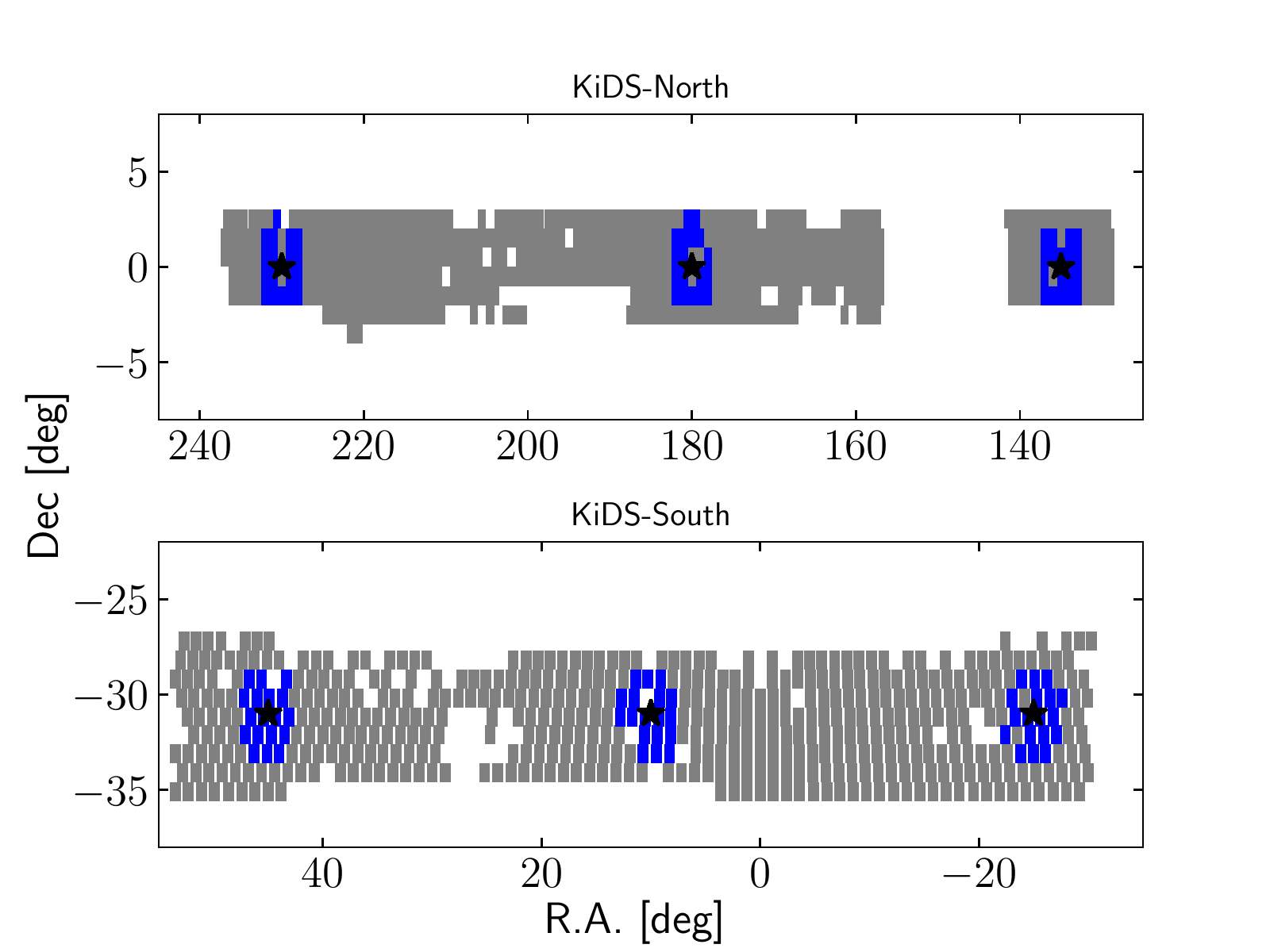}
      \caption{Sky distribution of the KiDS-DR4 tiles. Tiles shown in blue are included in the SKILLS fiducial run ($108$ tiles); The grey blocks show all KiDS-DR4 tiles that have nine-band noise and PSF information ($979$ tiles). The black stars indicate the centres of the stellar catalogues generated from \textsc{Trilegal}~\citep{Girardi2005AA}.}
         \label{fig:footprint}
  \end{figure}

Since we have $108~{\rm deg}^2$ of \textsc{Shark} galaxies as described in Sect.~\ref{Sec:inputGalaxies}, we selected $108$ KiDS pointings for the SKiLLS fiducial run. Figure~\ref{fig:footprint} shows the sky locations of the selected $108$ tiles along with the $979$ KiDS-DR4 tiles that have the nine-band noise and PSF information. Clusters of the selected blocks pair with the six stellar catalogues generated from \textsc{Trilegal} so that SKiLLS captures the stellar density variation across the whole KiDS survey (see Sect.~\ref{Sec:inputStars}). 

\begin{figure}
  \centering
  \includegraphics[width=\hsize]{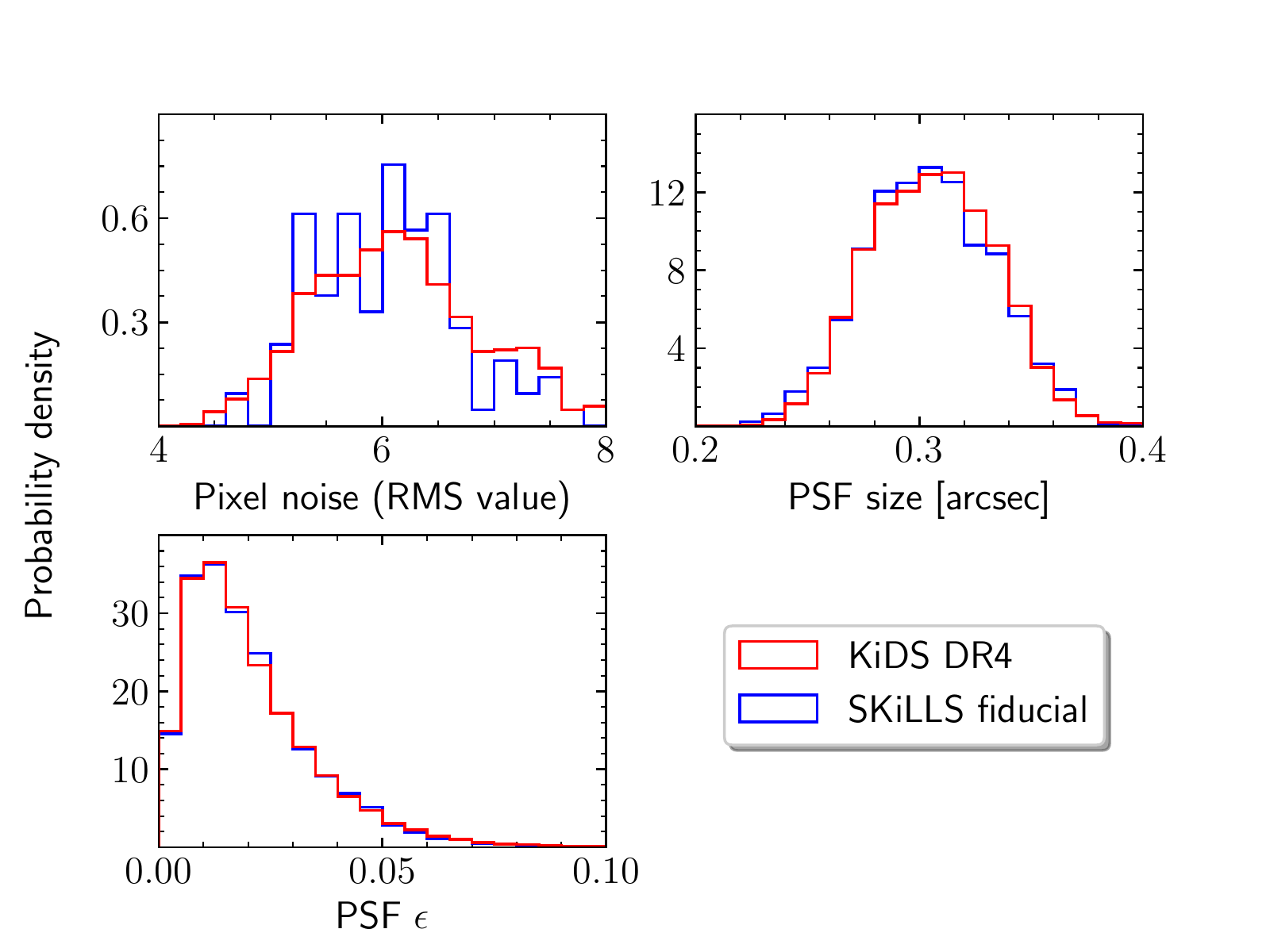}
      \caption{Comparing normalised histograms of the pixel noise (top left), PSF size (top right) and PSF ellipticity (bottom left) between KiDS-DR4 (red) and SKiLLS (blue) for the $r$-band images. The PSF size and ellipticity are measured from the recovered PSF image using a circular Gaussian window of sigma $2.5$ pixels.}
         \label{fig:noiseRband}
  \end{figure}

\begin{figure*}
   \includegraphics[width=0.5\hsize]{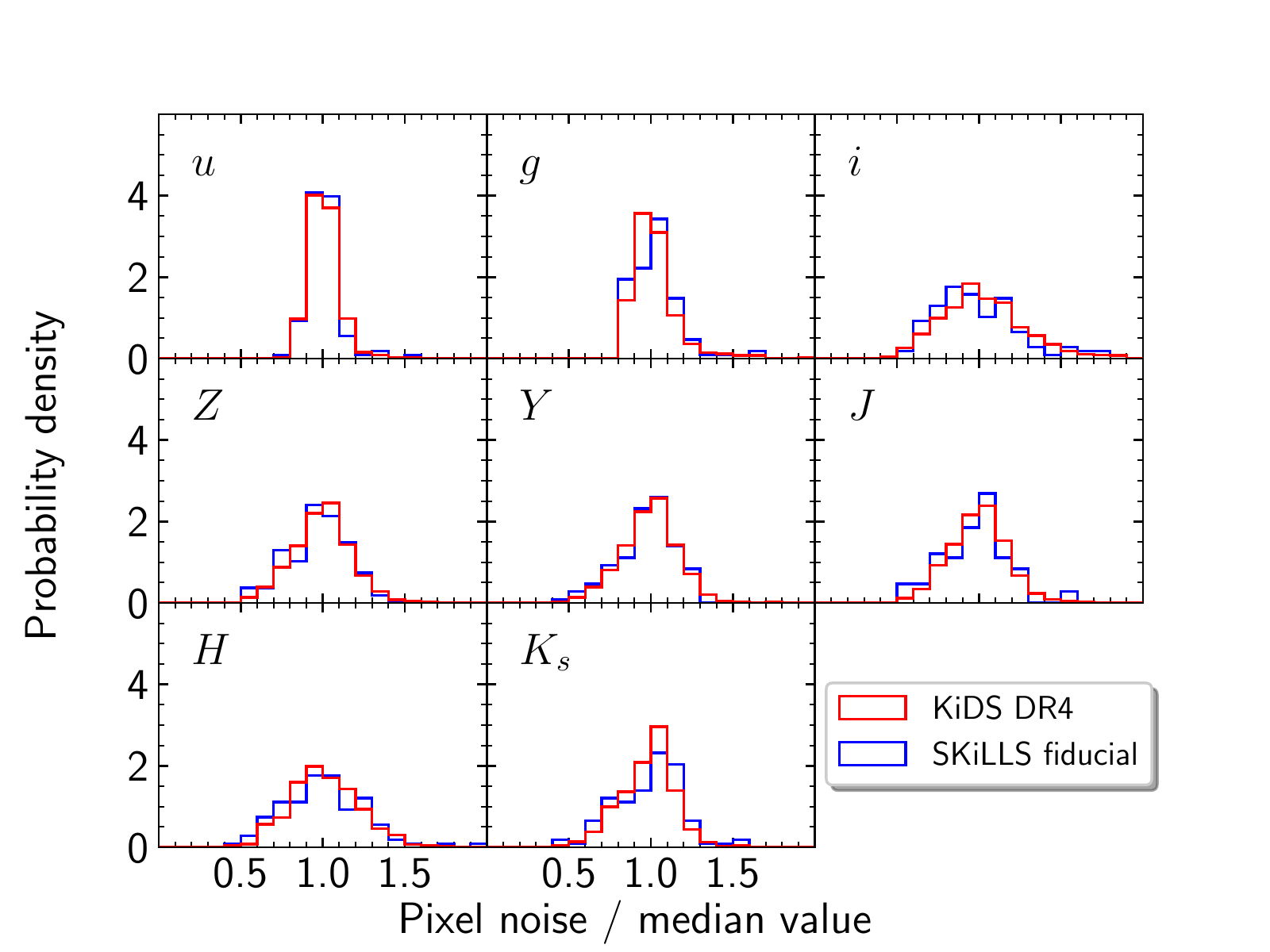}
   \includegraphics[width=0.5\hsize]{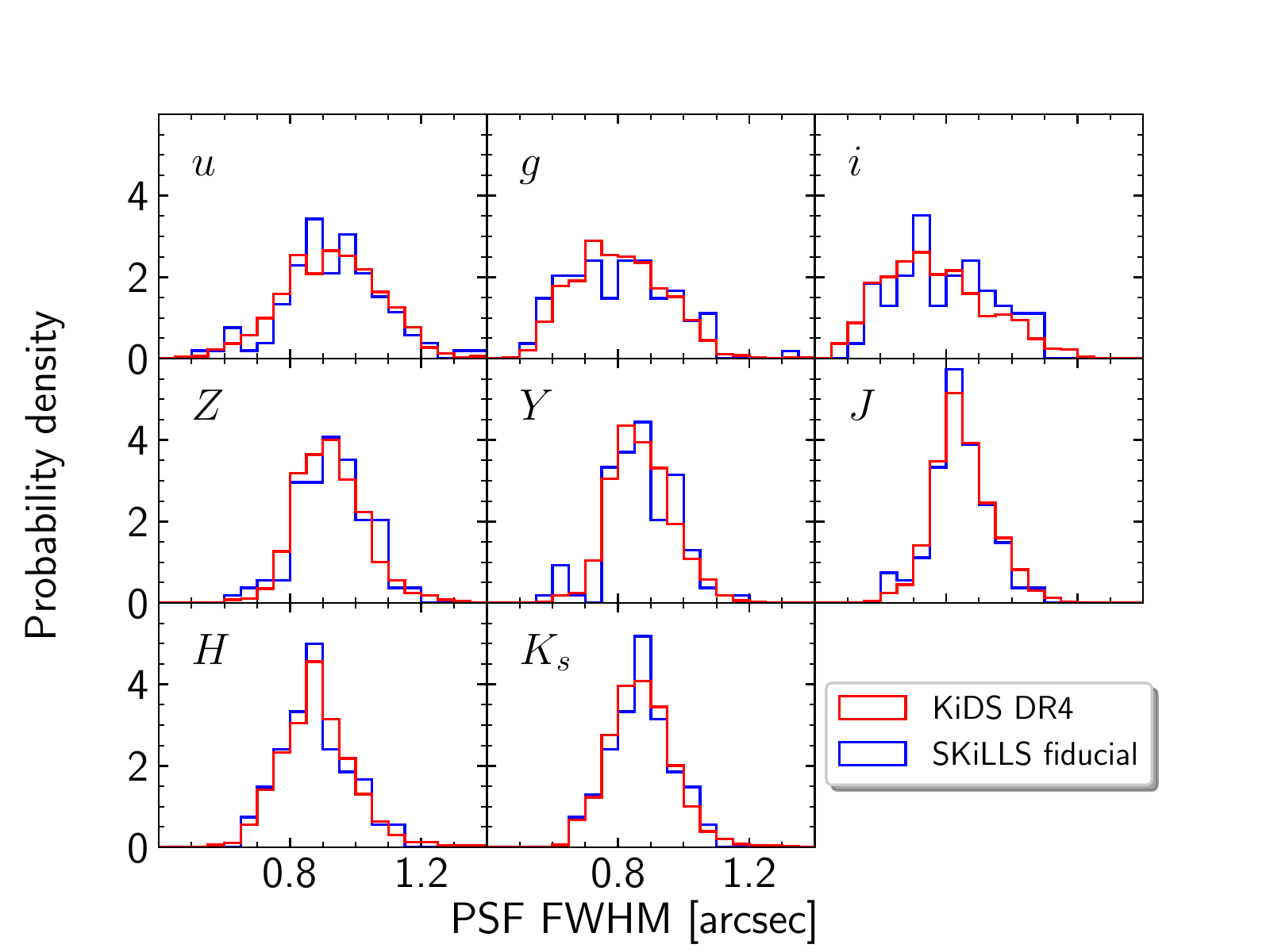}
      \caption{Comparing normalised histograms of the pixel noise (left) and PSF FWHM (right) between KiDS-DR4 (red) and SKiLLS (blue) for the bands only used for photometry. Equivalent comparisons for the lensing $r$-band images are presented in Fig.~\ref{fig:noiseRband}. The pixel noise values are divided by the median values in the whole sample for individual bands, so they can be shown in the same range.}
         \label{fig:noiseOthers}
   \end{figure*}

Figure~\ref{fig:noiseRband} compares the $r$-band noise and PSF properties between the SKiLLS selected tiles and all usable KiDS-DR4 tiles. We measured the PSF size and ellipticity using the weighted quadrupole moments with a circular Gaussian window of dispersion $2.5$ pixels, the typical galaxy size in the KiDS sample. The PSF size is defined as 
\begin{equation}
\label{eq:PSFsize}
    r_{\rm PSF}\equiv (Q_{11}Q_{22}-Q_{12}^2)^{1/4}~, 
\end{equation}
where $Q_{ij}$ are the weighted quadrupole moments, and the PSF ellipticity is defined by Eq.~(\ref{eq:shapeQ}). Figure~\ref{fig:noiseRband} shows that the selected tiles represent the KiDS-DR4 data well. Because we vary PSF for individual CCD images and exposures, the $108$ SKiLLS images cover \num{17280} different PSF models, a significant extension of the $65$ PSF models used by {\FC} and {\K}. That also explains the smooth distributions of the PSF parameters. Figure~\ref{fig:noiseOthers} shows similar comparisons for other bands. Again we see fair agreements across all bands. As KiDS-DR4 already covers ${\sim}75\%$ of the whole survey, we expect a similar agreement to the KiDS-Legacy data. The wide coverage of the noise and PSF properties also makes the SKiLLS results more robust than previous simulations and simplifies sensitivity tests (see Sect.~\ref{Sec:sensi} for details).


\subsection{Photometry and photometric redshifts}
\label{Sec:simPhoto}

With the simulated multi-band images, we can measure colours and estimate photometric redshifts (photo-$z$s) for simulated galaxies using the same tools developed in KiDS with minor adjustments.

For galaxy colours, we used the \textsc{GAaP} (Gaussian Aperture and PSF) pipeline~\citep{Kuijken2015MNRAS4543500K,Kuijken2019AA625A2K}. It provides accurate multi-band colours by accounting for PSF differences between filters and optimises signal-to-noise ratio (S/N) by down-weighting the noise-dominated outskirts. The latter is possible because the photo-$z$ estimation only needs the ratio of the fluxes from the same part of a galaxy in the given bands rather than the total light. A prerequisite for the \textsc{GAaP} pipeline is a detection catalogue with source positions and aperture parameters, which we measured from the \textsc{theli}-like $r$-band images using the \textsc{SExtractor} code~\citep{Bertin1996AAS117393B}. Once the detection catalogue is ready, we can obtain the list-driven photometry by running the \textsc{GAaP} algorithm on the $u$, $g$, $r$ and $i$ \textsc{Astro}-WISE-like images and the $Z$, $Y$, $J$, $H$ and $K_s$ simple images. In short, the \textsc{GAaP} method includes three major steps: 
\begin{enumerate}
    \item Homogenising PSFs by convolving the whole image with a spatially variable kernel map modelled from high S/N stars. The resulting image has a simple Gaussian PSF, for which estimating the PSF-independent Gaussian aperture flux is possible. The main side effect is that the convolution process introduces correlated noise between neighbouring pixels, complicating the estimation of measurement uncertainties. \textsc{GAaP} handles this by tracking the noise covariance matrix through the whole process.

    \item Defining an elliptical Gaussian aperture function for each source using the size and shape parameters measured by \textsc{SExtractor} on the $r$-band detection images. In practice, users must customise the minimum and maximum \textsc{GAaP} aperture sizes to balance the S/N and the effect of blending. Following the KiDS fiducial setup, we set the maximum aperture to $2\arcsec$ to avoid contamination from neighbouring sources. We conducted two separate runs by setting the minimum aperture to $0\farcs7$ and $1\farcs0$. When used as the input for the photo-$z$ estimation, a source-by-source decision was made to optimise the flux errors across the nine bands (see \citealt{Kuijken2019AA625A2K} for details).     

    \item Performing the aperture photometry on the PSF-Gaussianised images for each band using the defined aperture functions. It is worth stressing that \textsc{GAaP} aims to provide robust colours for the high S/N parts of galaxies; it underestimates the total fluxes for extended sources by design.
\end{enumerate}

\begin{figure}
  \centering
  \includegraphics[width=\hsize]{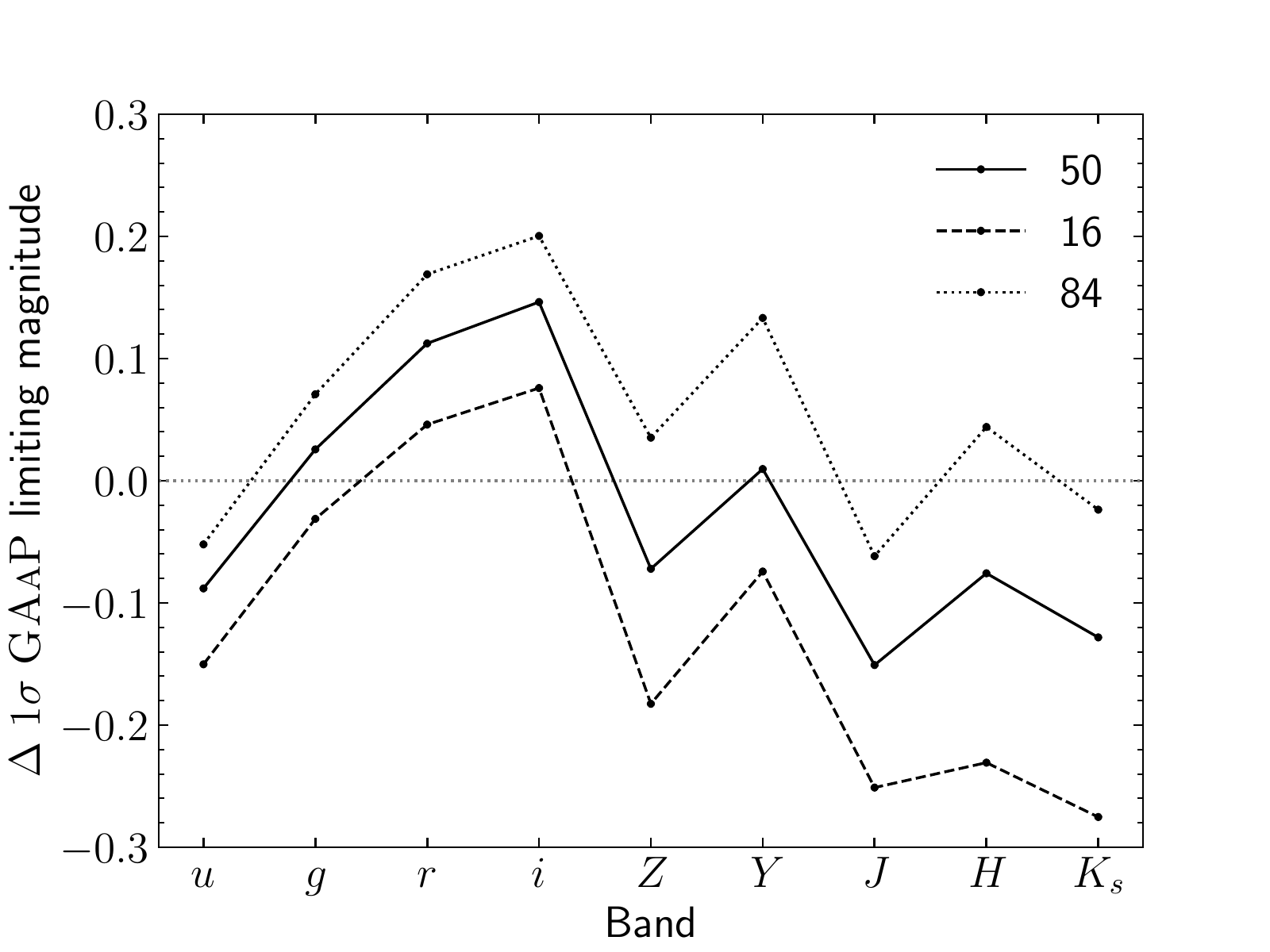}
      \caption{Differences of the image's median $1\sigma$ \textsc{GAaP} limiting magnitudes for the nine bands (simulation - data). The three lines indicate the $16$, $50$ and $84$ percentiles from the $108$ tiles included in the SKiLLS fiducial run. The larger scatters in the NIR bands are partially caused by the simplified simulating strategy.} 
         \label{fig:gaapLimiting}
  \end{figure}

\begin{figure*}
   \includegraphics[width=0.5\hsize]{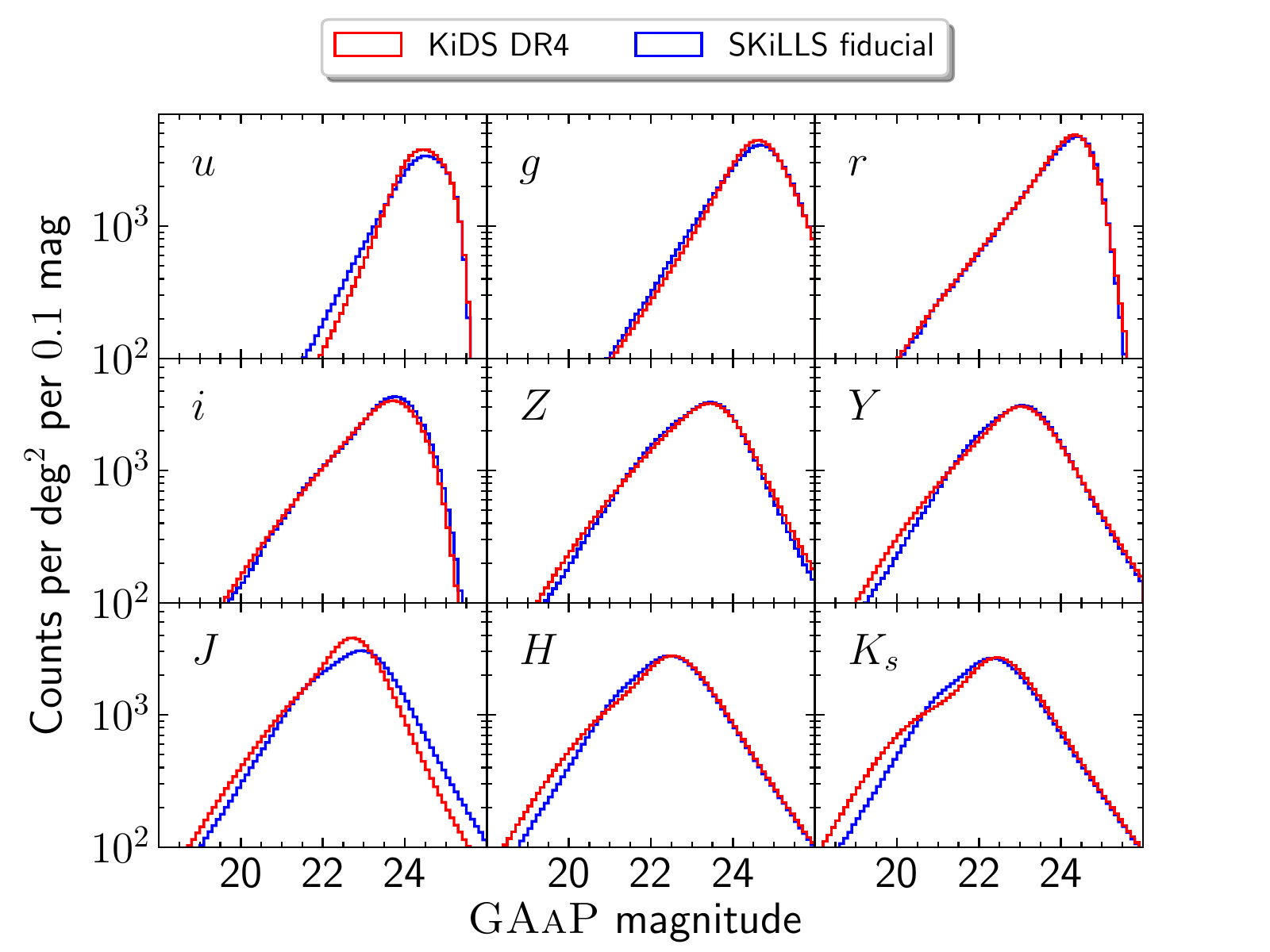}
   \includegraphics[width=0.5\hsize]{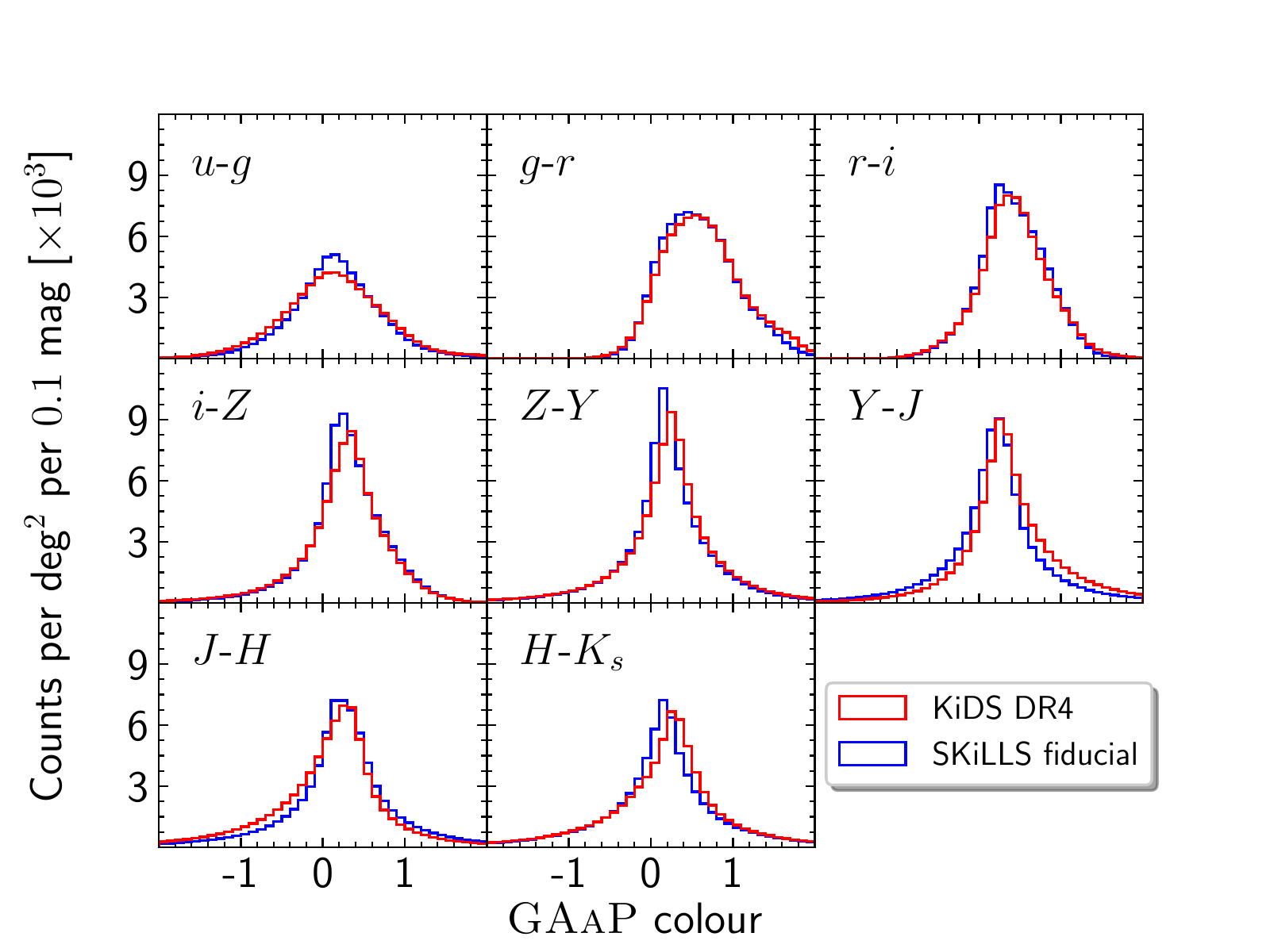}
      \caption{Comparison of the \textsc{GAaP} magnitudes (left panel) and colours (right panel) for KiDS-DR4 (red) and SKiLLS (blue). The results include all galaxies with valid photometric measurements (the \textsc{GAaP} flags in nine bands equal to $0$). Shape-measurement-related selections are not yet applied.}
         \label{fig:gaapPhotometry}
   \end{figure*}

Figure~\ref{fig:gaapLimiting} compares the nine-band $1\sigma$ \textsc{GAaP} limiting magnitudes between the KiDS-DR4 data and SKiLLS fiducial results. We calculated the median limiting magnitudes for tiles in both KiDS and SKiLLS and then compared their differences. We see a general agreement for all the bands, verifying our noise and PSF modelling. Noticeably, even for the NIR bands where we simplified the VIKING observations with single images, the differences are still tolerable, albeit with larger uncertainties. Figure~\ref{fig:gaapPhotometry} compares the \textsc{GAaP} photometric distributions between the simulation and data. Once again, we see a decent agreement in both magnitude and colour distributions.

For the photo-$z$ estimation, we implemented the public Bayesian Photometric Redshift~(\textsc{bpz}; \citealt{Benitez2000ApJ}) code with the re-calibrated template set from \citet{Capak2004PhDT} and the Bayesian redshift prior from \citet{Raichoor2014ApJ}. We closely followed the settings in the KiDS-DR4 analysis~\citep{Kuijken2019AA625A2K} unless it conflicts with the simulation input. For example, we set ZMAX to $2.5$, the limiting redshift of SKiLLS galaxies, instead of $7.0$ as in the data. We tested the choice of ZMAX in the simulations and found that only $0.1\%$ of the test sample resulted in estimates differing more than $0.1$, which means most of the objects have similar photo-$z$ estimates and end up in the same tomographic bins for these two choices. Moreover, the \textsc{Shark} photometry in the $u$, $g$, $r$, $i$ and $Z$ bands is based on the Sloan Digital Sky Survey (SDSS) photometric system, which is slightly different from the KiDS/VIKING system~\citep{Kuijken2019AA625A2K}. We corrected these slight differences in the measured \textsc{GAaP} magnitudes in order to use the KiDS/VIKING filters to run the \textsc{bpz} code. The detailed procedures and comparisons are described in Appendix~\ref{Sec:FilterTransform}. Overall, the modification is minor and has a negligible impact on the magnitude, colour distributions, and final shear biases. Still, it improves the agreement between the simulation and the data in the photo-$z$ distributions. Unless specified otherwise, we base our fiducial results on the transformed photometry.

\begin{figure*}
  \centering
   \includegraphics[width=\hsize]{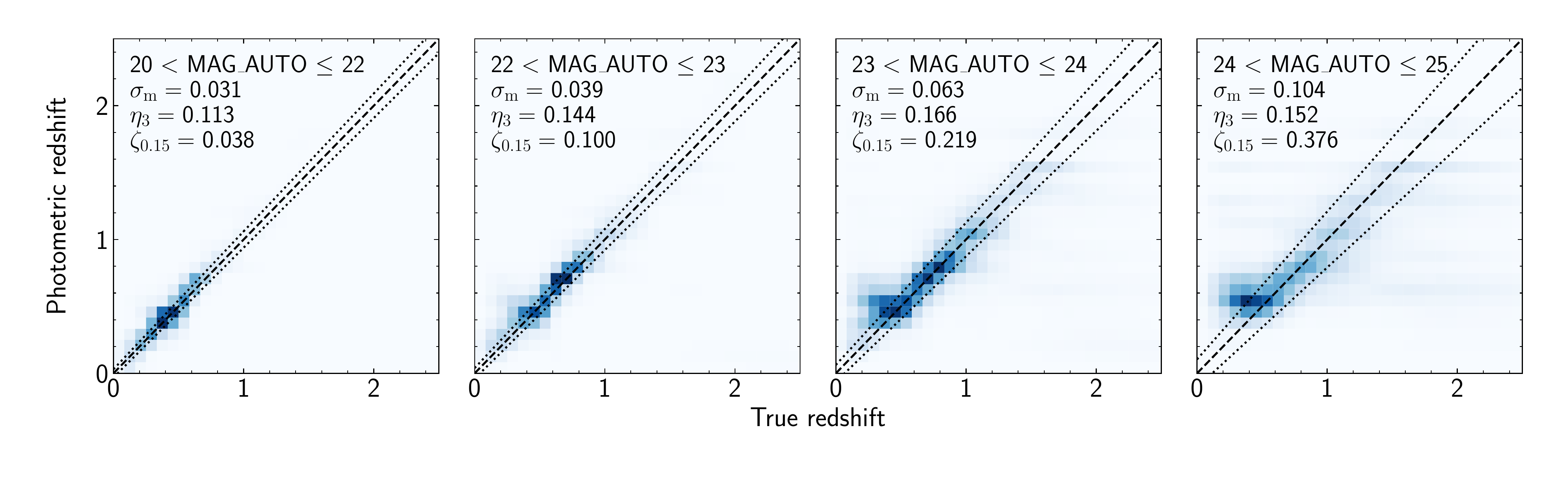}
     \caption{Photometric redshifts vs.\ true redshifts in several apparent $r$-band magnitude bins. The annotated statistics are: the normalised median-absolute-deviation ($\sigma_{m}$) of the quantity $\Delta z/(1+z)$, the fraction of sources with $|\Delta z/(1+z)|>3\sigma_{m}$ ($\eta_{3}$) and the fraction of sources with $|\Delta z/(1+z)|>0.15$ ($\zeta_{0.15}$). The dashed lines correspond to the one-to-one relation, and the dotted lines show $|\Delta z/(1+z)|=\sigma_{m}$.}
         \label{fig:bpzPer}
   \end{figure*}

Figure~\ref{fig:bpzPer} compares the estimated photo-$z$ to the true redshift from the input \textsc{SURFS}-\textsc{Shark} simulations in several measured magnitude bins. It shows the photo-$z$ vs.\ true redshift distributions, along with annotated statistics based on the distributions of $(z_{\rm B}-z_{\rm true})/(1+z_{\rm true})\equiv\Delta z/(1+z)$ values. We see the \textsc{bpz} code works well in SKiLLS and is at the same level as in KiDS~\citep{Wright2019AA632A34W}. More detailed verification of the SKiLLS photo-$z$ performance is presented in the companion redshift calibration paper~(van den Busch et al., in prep). 

As for the redshift calibration, our end-to-end approach, which starts with image simulation followed by object detection, PSF homogenisation, forced multi-band photometry, and photo-$z$ estimation, is a significant improvement compared to previous catalogue-level simulations (e.g. \citealt{Hoyle2018MNRAS.478..592H,Busch2020AA642A200V,DeRose2022PhRvD.105l3520D}). The image-simulation-based approach not only yields more realistic observational uncertainties but also naturally accounts for the blending effect, which is hard to address at the catalogue level. As for the shear calibration, these photo-$z$ estimates are essential for performing tomographic selections~(\K). Our approach that directly measures the photo-$z$s from simulated images accounts for various measurement uncertainties of photo-$z$s, hence a tomographic selection consistent with how it is done in the data. Moreover, using the same mock catalogue in both shear and redshift calibration unites these two long-separated processes in the KiDS-Legacy analysis.

%% file: Sec4_shape.tex
The primary task of any weak lensing survey is to measure the shapes of galaxy images. Previous KiDS analyses tackled this task using a likelihood-based code, dubbed \textit{lens}fit~\citep{Miller2007MNRAS382315M,Miller2013MNRAS4292858M,Kitching2008MNRAS390149K}. It is the default shape measurement algorithm for the KiDS-Legacy analysis, with some updates described in this section. We test SKiLLS using this updated \textit{lens}fit code\footnote{Nevertheless, we note that SKiLLS can also calibrate other algorithms, such as the KiDS \textsc{Metacalibration} catalogue (Yoon et al., in prep.).}. 

\subsection{The self-calibration version of \textit{lens}fit}
\label{Sec:lensfit}

The \textit{lens}fit code, first developed for CFHTLenS~\citep{Heymans2012MNRAS427146H}, follows a Bayesian model-fitting approach. We refer to \citet{Miller2013MNRAS4292858M} for its detailed formalism. In brief, it first performs a joint fit to individual exposures using a PSF-convolved galaxy model, which yields a likelihood distribution of seven parameters: 2D position, flux, scalelength, bulge-to-total flux ratio and complex ellipticity. Then it deduces the ellipticity parameters from the likelihood-weighted mean values by marginalising other parameters with priors as described by \citet{Miller2013MNRAS4292858M}. For each ellipticity estimate, an inverse-variance weight is also determined from~\citep{Miller2013MNRAS4292858M}
\begin{equation}
    \label{eq:LFwei}
    w_i \equiv \left[ \frac{\sigma_{\epsilon,~i}^2~\epsilon_{\rm max}^2}{\epsilon_{\rm max}^2-2\sigma_{\epsilon,~i}^2} + \sigma_{\epsilon,~ {\rm pop}}^2 \right]^{-1}~, 
\end{equation}
where $\sigma_{\epsilon,~i}$ is the uncertainty of the measured ellipticity, $\sigma_{\epsilon,~ {\rm pop}}$ is the ellipticity dispersion of the galaxy population (intrinsic shape noise), and $\epsilon_{\rm max}$ is the maximum allowed ellipticity in the \textit{lens}fit model-fitting. As for KiDS data, we adopted $\sigma_{\epsilon,~ {\rm pop}}=0.253$ and $\epsilon_{\rm max}=0.804$.

The code has evolved as KiDS progressed. The most significant is a self-calibration scheme for noise bias, as detailed in \FC. The pixel noise in a given image skews the likelihood, which biases the estimate of individual galaxy ellipticities. It is a complex function of the signal-to-noise ratio, galaxy properties and PSF morphology, making it difficult to predict accurately. Thus, \textit{lens}fit conducts an approximate correction using the measurements themselves, that is a self-calibration. The basic idea is to simulate a test galaxy with parameters measured from the first run, then remeasure the test galaxy using the same pipeline. The difference between the remeasured and input values serves as a correction factor for the corresponding parameter. Since its introduction, self-calibration has been a standard part of \textit{lens}fit, given its promising overall performance~(\citealt{Mandelbaum2015MNRAS4502963M}; \FC; \K). We keep this feature for the KiDS-Legacy analysis.

\subsection{Updates for KiDS-Legacy analysis}
\label{Sec:lensfitUpdate}

A long-standing mystery of all previous \textit{lens}fit analyses has been the presence of a small but significant residual bias in $\epsilon_2$ that is uncorrelated with the PSF and the underlying shear~\citep{Miller2013MNRAS4292858M,Hildebrandt2016MNRAS.463..635H,Giblin2021AA645A105G}. We now understand that this feature arises from an anisotropic error in the original likelihood sampler, which has been corrected in our algorithm. However, we found that this correction inadvertently increases the fraction of residual PSF contamination in the weighted average signal (see the discussion in \citealt{Giblin2021AA645A105G}). Besides, object selection and galaxy weights are also known to introduce bias~(e.g. \citealt{Kaiser2000ApJ537555K}, \citealt{Bernstein2002AJ123583B}, \citealt{Hirata2003MNRAS343459H}, \citealt{Jarvis2016MNRAS4602245J} and \FC). These selection biases can be more severe than the raw measurement bias and hence cannot be ignored even for a perfect self-calibration measurement algorithm. 

{\FC} presented a method to isotropise weights using an empirical correction scheme, which has been adopted in previous KiDS studies to mitigate these biases. Unfortunately, we found this approach to be insufficient for the improved \textit{lens}fit algorithm. Furthermore, we found the approach to be sensitive to the sample volume, and therefore hard to apply consistently to the data and simulations. So, we introduce a new empirical correction scheme that mitigates the PSF contamination to the weighted shear signal.

\subsubsection{Weight correction}
\label{Sec:weightCorr}

We start with the PSF leakages in the reported weight. For galaxies with comparable surface brightness, those aligned with the PSF tend to have a higher integrated signal-to-noise ratio than those cross-aligned with the PSF. This orientation preference causes the asymmetry of the measurement variance (the $\sigma_{\epsilon,~i}^2$ term in Eq.~\ref{eq:LFwei}), which can be measured using a linear function to the first order
\begin{equation}
    \label{eq:alphaVar}
    S_i = \alpha_{S}\epsilon_{{\rm PSF},~i,~{\rm proj}} + \mathcal{N}\left[\langle S\rangle,~\sigma_S\right]~,
\end{equation}
where $S_i\equiv \sigma_{\epsilon,~i}^2$ refers to the measurement variance, and $\epsilon_{{\rm PSF},~i,~{\rm proj}} \equiv {\rm Real}\left({\bm \epsilon}_{{\rm PSF},~i}~{\bm \epsilon}^*_{{\rm obs},~i}\right)$ is the scalar projection of the PSF ellipticity in the direction of the galaxy ellipticity. The $\alpha_{S}$ term quantifies the PSF contamination in the measurement variance, while $\mathcal{N}\left[\langle S\rangle,~\sigma_S\right]$ denotes the noise, which we assume follows a Gaussian distribution with a mean of $\langle S\rangle$ and standard deviation of $\sigma_S$.

Following {\FC}, we estimate the PSF contamination as a function of the integrated signal-to-noise ratio ($\nu_{\rm SN}$) reported by \textit{lens}fit and the resolution, which is defined as 
\begin{equation}
    \label{eq:resolution}
    \mathcal{R}\equiv \frac{r_{\rm PSF}^2}{r_{\rm PSF}^2+r_{ab}^2}~,
\end{equation}
where $r_{ab}\equiv r_{\rm e}\sqrt{q}$ is the circularised galaxy size with $r_{\rm e}$ and $q$ denoting the scalelength along the major axis and the axis ratio, respectively. The PSF size $r_{\rm PSF}$ is defined by Eq.~(\ref{eq:PSFsize}). By construction, the resolution $\mathcal{R}$ has a value between $0$ and $1$, with a larger value corresponding to a more poorly resolved object.

   \begin{figure}
   \centering
   \includegraphics[width=\hsize]{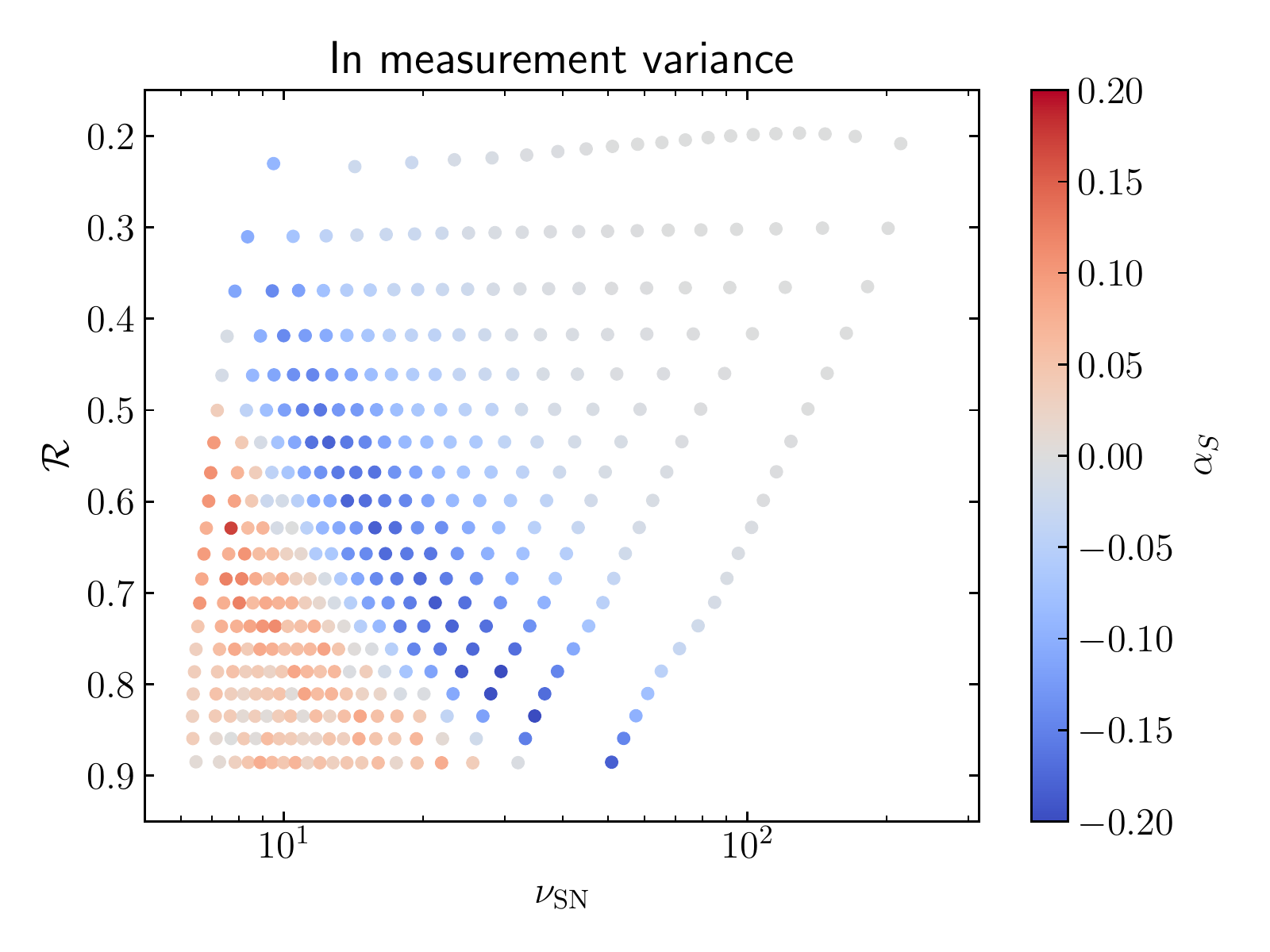}
      \caption{PSF leakage in the measurement variance as a function of S/N and $\mathcal{R}$. We note that the larger $\mathcal{R}$ corresponds to a poorer resolution by definition (Eq.~\ref{eq:resolution}).}
         \label{fig:alphaS}
   \end{figure}
%
   \begin{figure}
   \centering
   \includegraphics[width=\hsize]{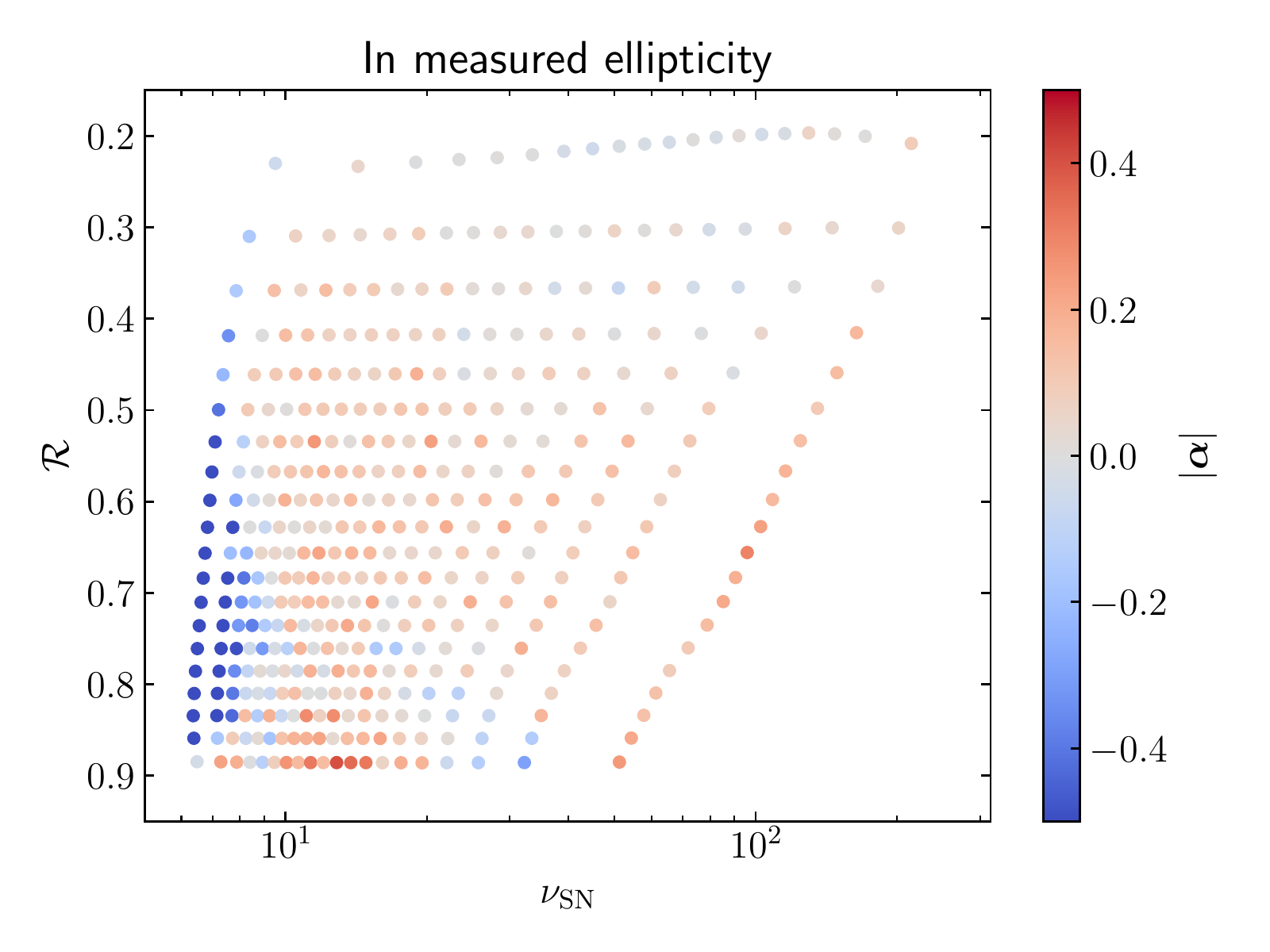}
   \caption{PSF leakage in the measured ellipticity after the weight calibration as a function of S/N and $\mathcal{R}$. We note that the larger $\mathcal{R}$ corresponds to a poorer resolution by definition (Eq.~\ref{eq:resolution}).}
         \label{fig:alpha}
   \end{figure}
When estimating $\alpha_{S}$, we first divide galaxies into an irregular $20\times 20$ grid of $\nu_{\rm SN}$ and $\mathcal{R}$, each containing the same number of objects. Then in each bin, we perform a linear regression using Eq.~(\ref{eq:alphaVar}) to measure $\alpha_{S}$. Figure~\ref{fig:alphaS} shows the measurements for the KiDS-DR4 re-run with the updated \textit{lens}fit. It demonstrates a clear correlation between the estimated $\alpha_{S}$ and the $\nu_{\rm SN}$ and $\mathcal{R}$. We derive the corrected measurement variance for individual galaxies through $\sigma^2_{\epsilon,~i,~{\rm corr}}=\sigma^2_{\epsilon,~i}-\alpha_{S}\epsilon_{{\rm PSF},~i,~{\rm proj}}$, where the value of $\alpha_{S}$ is determined based on which $\nu_{\rm SN}$-$\mathcal{R}$ bin the target galaxy is assigned to. The corrected \textit{lens}fit weight is then calculated with
\begin{equation}
    \label{eq:LFweiNew}
    w_{{\rm corr},~i} \equiv \left[ \frac{\sigma^2_{\epsilon,~i,~{\rm corr}}~\epsilon_{\rm max}^2}{\epsilon_{\rm max}^2-2\sigma^2_{\epsilon,~i,~{\rm corr}}} + \sigma_{\epsilon,~ {\rm pop}}^2 \right]^{-1}~, 
\end{equation}
following Eq.~(\ref{eq:LFwei}). We verified that this approach is sufficient to remove the overall weight bias and is robust against the binning scheme.

\subsubsection{Ellipticity correction}
\label{Sec:eCorr}

In addition to the weight bias, there is still some residual PSF leakage in the measured ellipticity because of the residual noise bias and selection effects. To first order, this residual PSF bias can be formulated as
\begin{equation}
    \label{eq:alpha}
    {\bm \epsilon}_{{\rm obs},~i} = {\bm \epsilon}_{{\rm true},~i} + {\bm \alpha}~{\bm \epsilon}_{{\rm PSF},~i} + {\bm c} + \mathcal{N}\left[0,~{\bm \sigma}_{\epsilon}\right]~,
\end{equation}
where ${\bm \epsilon}_{{\rm obs},~i}$ is the measured ellipticity, ${\bm \epsilon}_{{\rm true},~i}$ is the underlying true ellipticity, ${\bm \alpha}$ is the fraction of the PSF ellipticity ${\bm \epsilon}_{{\rm PSF},~i}$ that leaks into the measured ellipticity, and ${\bm c}$ is an additive term uncorrelated with the PSF. $\mathcal{N}\left[0,~{\bm \sigma}_{\epsilon}\right]$ denotes the noise in individual shape measurements, which are assumed to follow a Gaussian distribution of mean $0$ and standard variation ${\bm \sigma}_{\epsilon}$. We note that all parameters in Eq.~(\ref{eq:alpha}) are complex numbers (${\bm \alpha} = \alpha_1+{\rm i}\alpha_2$). We focus on the ${\bm \alpha}$ term, as the ${\bm c}$ term with the improved likelihood sampler is now small in practice, and the $\mathcal{N}\left[0,~{\bm \sigma}_{\epsilon}\right]$ vanishes for an ensemble of galaxies.

Like the weight bias correction, we first estimate ${\bm \alpha}$ in the $20\times~20$ grid of $\nu_{\rm SN}$ and $\mathcal{R}$ using a linear regression of Eq.~(\ref{eq:alpha}). Figure~\ref{fig:alpha} shows the amplitude of ${\bm \alpha}$ in the 2D $\nu_{\rm SN}$ and $\mathcal{R}$ plane. We see modest values in most situations, except for the low $\nu_{\rm SN}$ cases, where it drops abruptly to negative values. We confirmed that the negative tail is mainly from the selection effects by measuring the PSF leakage using the input ellipticity in simulations. This non-trivial negative tail prevents us from using the direct correction approach introduced in the weight bias correction section. Therefore, we propose a hybrid approach, with a fitting procedure for the overall trend and a direct correction for residuals. Specifically, we first fit the measured ${\bm \alpha}$ as a function of $\nu_{\rm SN}$ and $\mathcal{R}$, using a function of the form
\begin{equation}
    \label{eq:alphaPoly}
    {\bm \alpha}_{\rm p}(\nu_{{\rm SN}},~ \mathcal{R}) = {\bm a}_0 + {\bm a}_1 \nu_{\rm SN}^{-2} + {\bm a}_2 \nu_{\rm SN}^{-3} + {\bm b}_1 \mathcal{R} + {\bm c}_1 \mathcal{R}~\nu_{\rm SN}^{-2}~,
\end{equation}
whose coefficients are constrained using the weighted mean results from the $20\times~20$ grid. Then, we correct the raw measurements of individual galaxies using ${\bm \epsilon}_{{\rm obs},~i,~{\rm tmp}}={\bm \epsilon}_{{\rm obs},~i} - {\bm \alpha}_{\rm p}(\nu_{{\rm SN}, ~i},~ \mathcal{R}_i)~{\bm \epsilon}_{{\rm PSF},~i}$, where the polynomial ${\bm \alpha}_{\rm p}(\nu_{{\rm SN}, ~i},~ \mathcal{R}_i)$ is determined from the target galaxy's $\nu_{{\rm SN}, ~i}$ and $\mathcal{R}_i$. After removing the overall trend, we use the corrected ${\bm \epsilon}_{{\rm obs},~i,~{\rm tmp}}$ to measure the residual ${\bm \alpha}_{\rm r}$, which changes mildly across the 2D $\nu_{\rm SN}$ and $\mathcal{R}$ plane. Therefore, we can conduct the direct correction through ${\bm \epsilon}_{{\rm obs},~i, ~{\rm corr}}={\bm \epsilon}_{{\rm obs},~i,~{\rm tmp}}-{\bm \alpha}_{\rm r}~{\bm \epsilon}_{{\rm PSF},~i}$, where the values of ${\bm \alpha}_{\rm r}$ for individual galaxies are determined based on which $\nu_{\rm SN}$-$\mathcal{R}$ bin they are assigned. This two-step approach balances performance and robustness. We verified that the corrected measurements have negligible PSF leakages and the results are robust against the binning scheme. 

   \begin{figure*}
   \includegraphics[width=0.33\hsize]{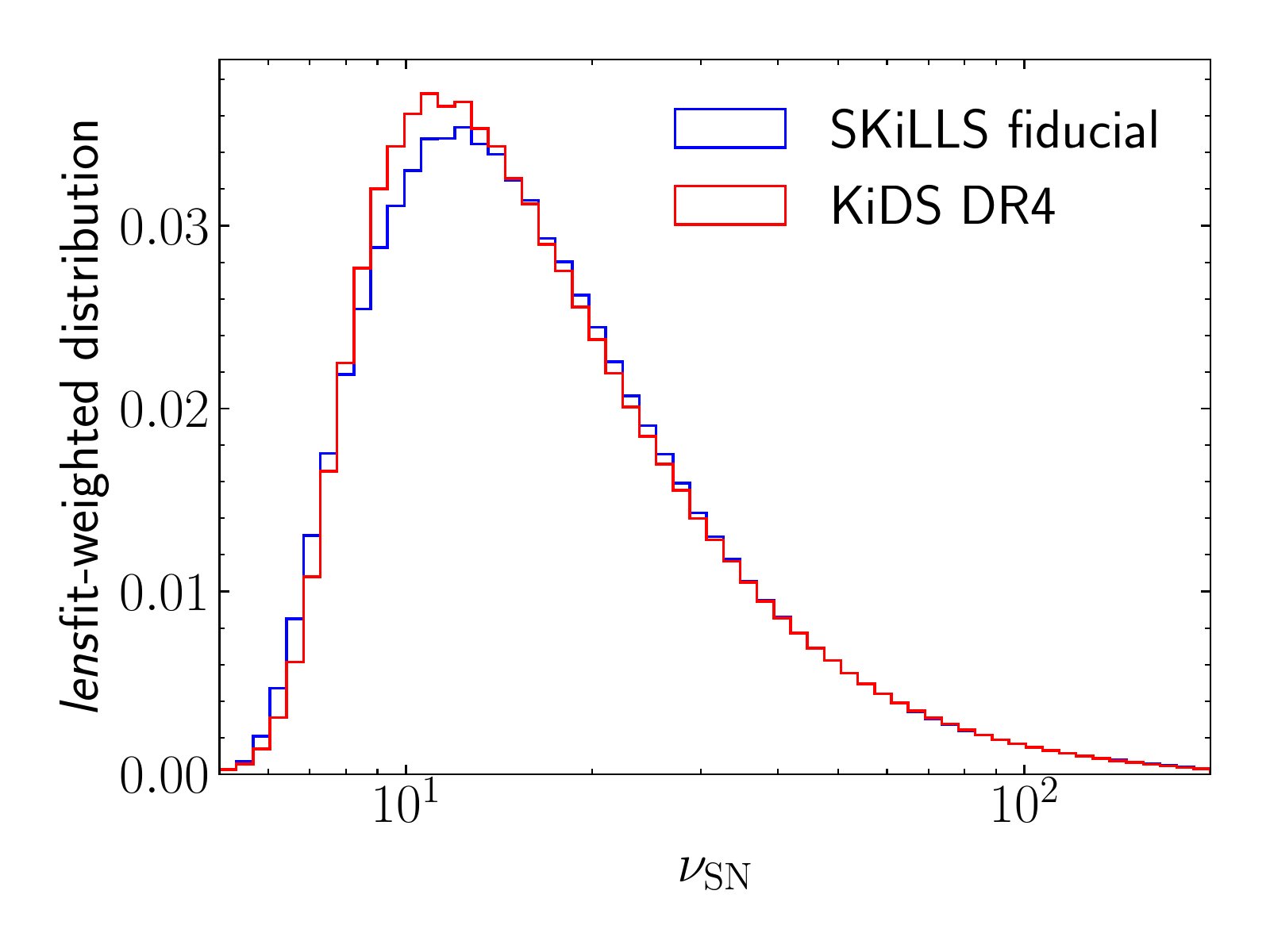}
   \includegraphics[width=0.33\hsize]{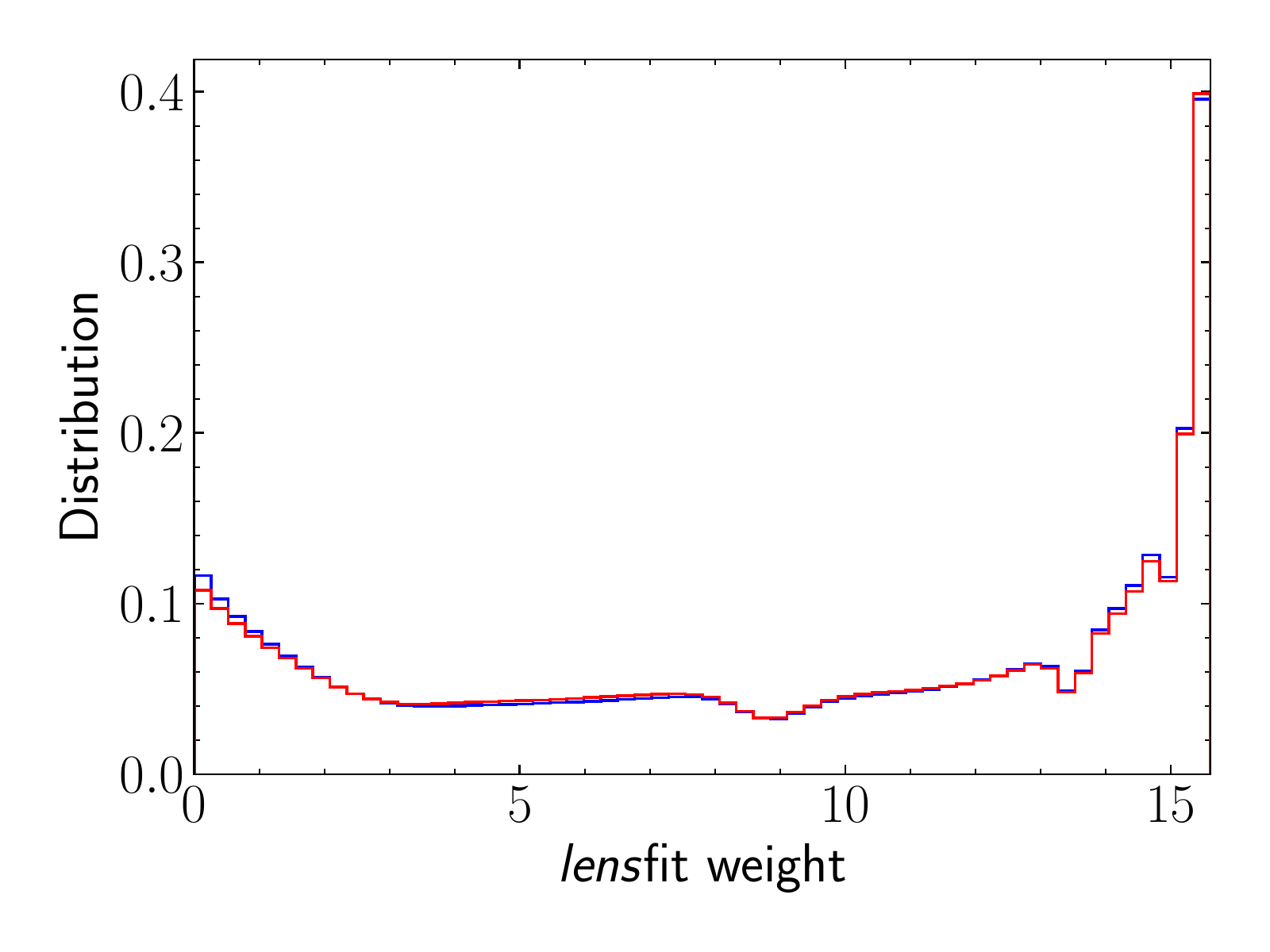}
   \includegraphics[width=0.33\hsize]{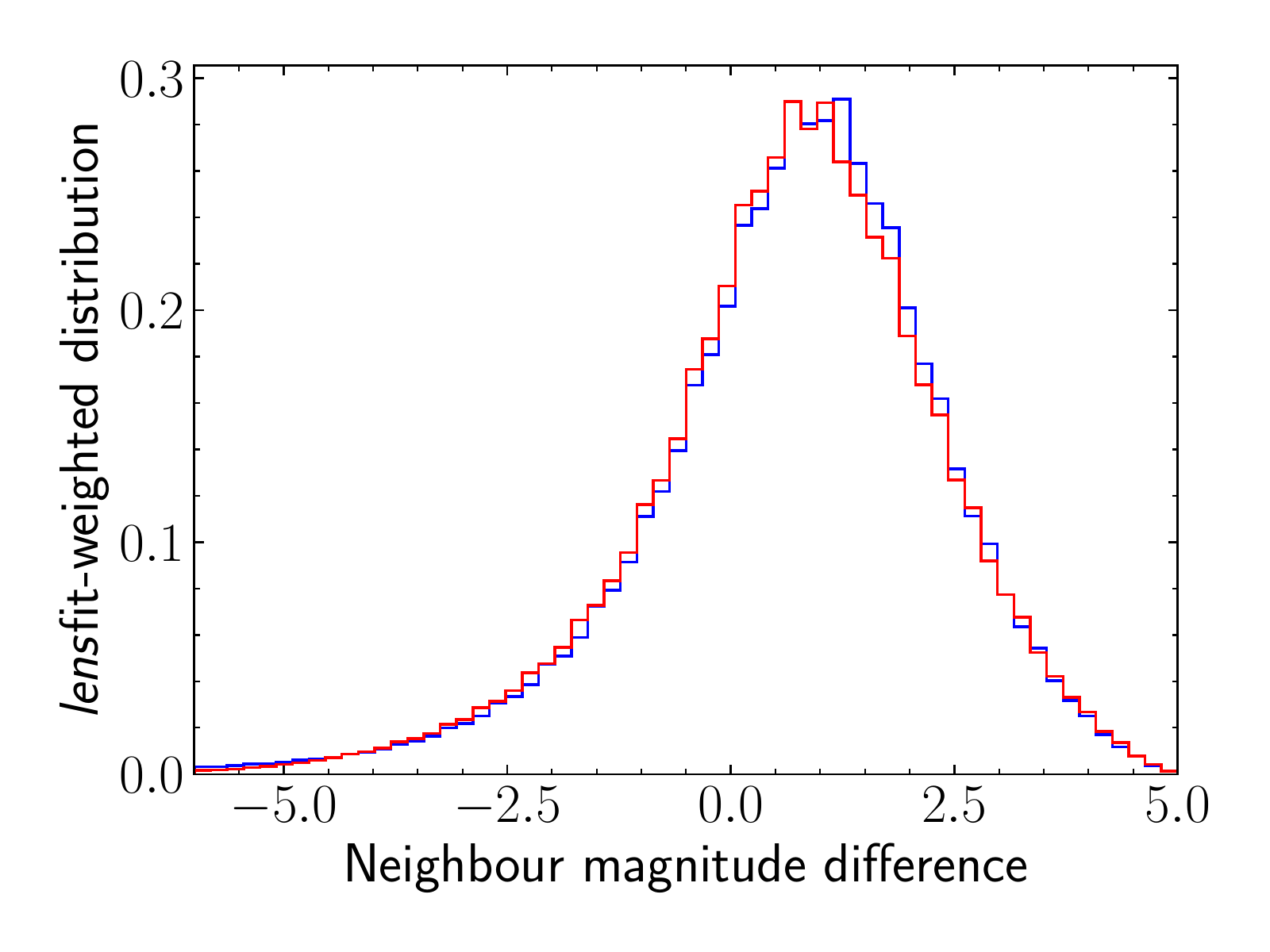}
   \includegraphics[width=0.33\hsize]{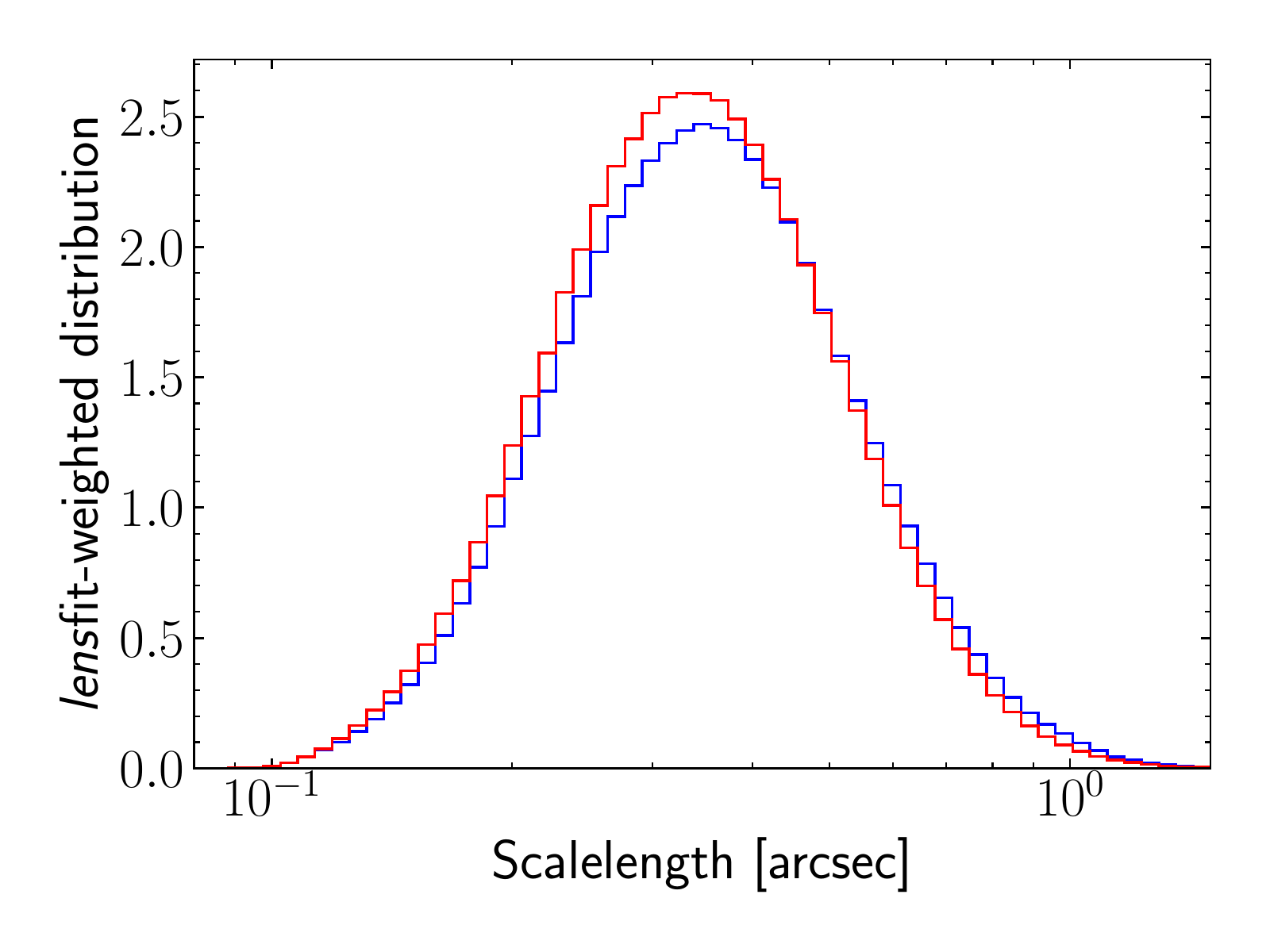}
   \includegraphics[width=0.33\hsize]{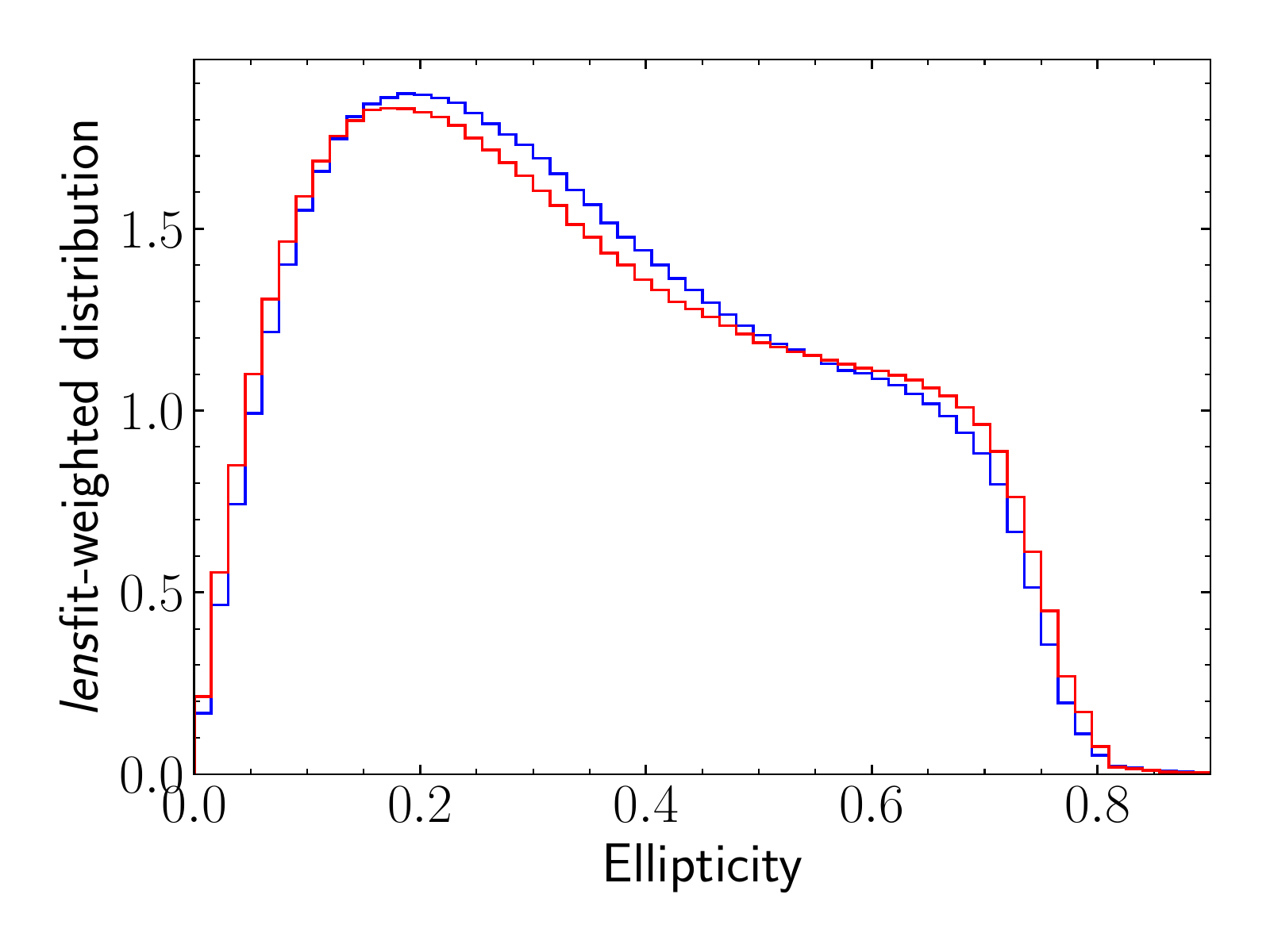}
   \includegraphics[width=0.33\hsize]{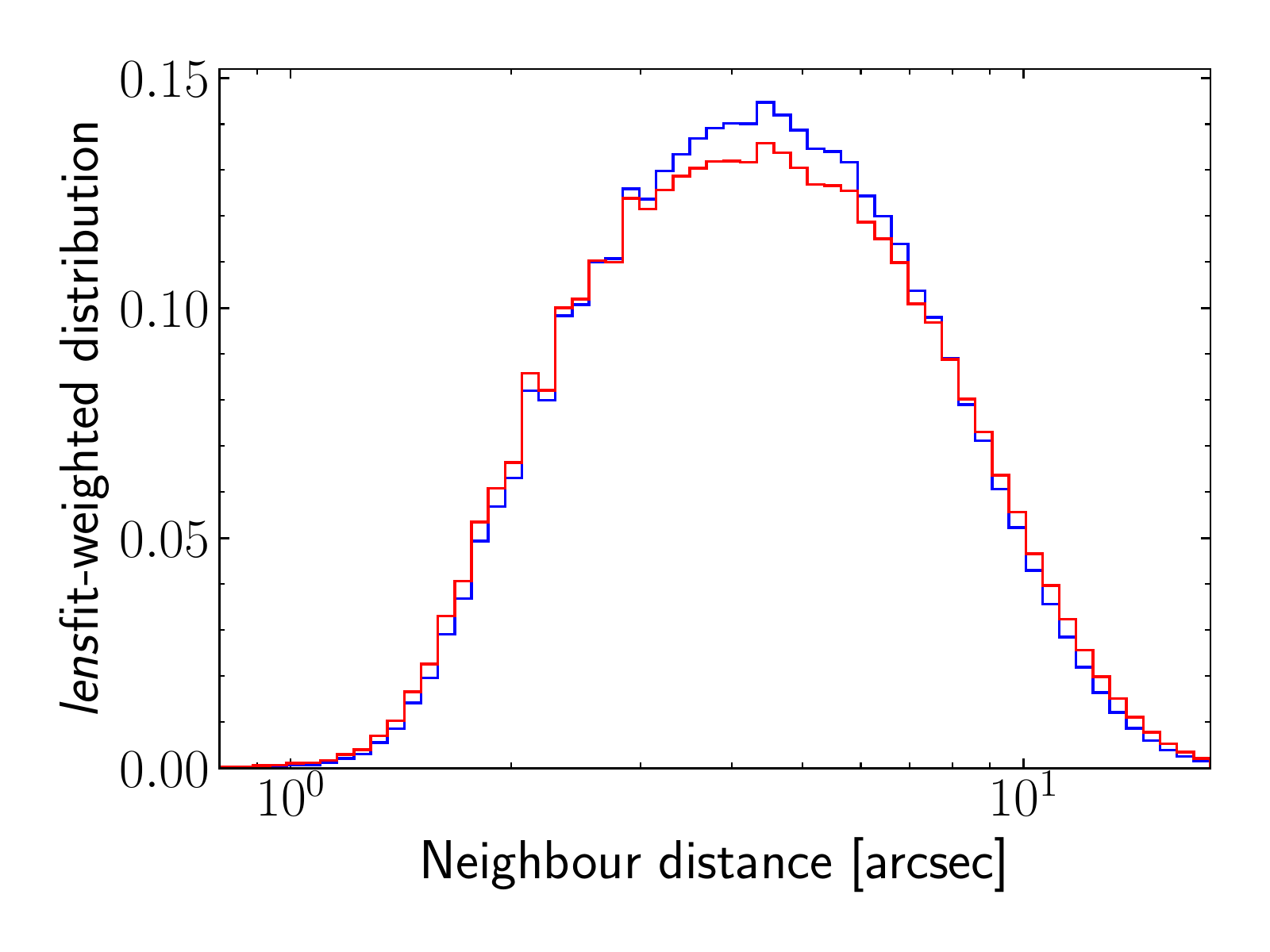}
      \caption{Comparison of the updated \textit{lens}fit measurements between KiDS (red) and SKiLLS (blue). All distributions are normalised with \textit{lens}fit weights, except for the distribution of \textit{lens}fit weight itself. The neighbour properties are based on the nearest neighbour found in the measured catalogue. The magnitude difference is defined as the neighbour magnitude minus the magnitude of the primary target. The lack of close pairs with distance below ${\sim}1$ arcsecond is due to the conservative blending cut used by KiDS (see Appendix~\ref{Sec:cuts}). This cut helps to mitigate the worst of the blending bias.}
         \label{fig:LFout}
   \end{figure*}

\subsection{Comparison between KiDS and SKiLLS}
\label{Sec:lensfitComp}

We applied the updated \textit{lens}fit code to KiDS-DR4 and SKiLLS $r$-band images. The object selections after the measurements are detailed in Appendix~\ref{Sec:cuts}. In short, we largely followed the selection criteria proposed in \citet{Hildebrandt2017MNRAS.465.1454H}, with an additional resolution cut introduced to mitigate the PSF contamination. We applied the same selections to the KiDS data and SKiLLS simulated catalogue to ensure a consistent selection effect, even though SKiLLS does not contain artefacts like asteroids and binary stars.

Figure~\ref{fig:LFout} compares the weighted distributions of some critical observables reported by the updated \textit{lens}fit. The SKiLLS results match the KiDS-DR4 data reasonably well. We also checked the properties of the close pairs. Specifically, we show the magnitude difference and the projected distance between close pairs in the measured catalogues. Both properties agree well between the data and simulations, implying SKiLLS has realistic clustering features. These realistic neighbouring properties are essential for an accurate shear calibration, especially when considering the shear interference between blended objects (see Sect.~\ref{Sec:shear} for details). 

%% file: Sec5_results.tex
The central task of image simulations is to quantify the average shear bias for a selected source sample. This is done by comparing the inferred shear ${\bm \gamma}_{\rm obs}$, to the input shear ${\bm \gamma}_{\rm input}$, which have a linear correlation to the first order~\citep{Heymans2006MNRAS3681323H}
\begin{equation}
    \label{eq:shearbias}
    {\bm \gamma}_{\rm obs} = (1+{\bm m})~{\bm \gamma}_{\rm input} + {\bm c}~,
\end{equation}
where ${\bm m}$ is known as the multiplicative bias, and ${\bm c}$ is the additive bias. The simulation-based calibration focuses on the multiplicative bias, as the additive bias is usually corrected empirically (for example, the correction scheme proposed in Sect.~\ref{Sec:lensfitUpdate}). So we use the term `shear bias' and `multiplicative bias' interchangeably throughout the paper. We note that all parameters in Eq.~(\ref{eq:shearbias}) are in complex forms, such as ${\bm m} = m_1+{\rm i}m_2$. However, we found $m_1$ and $m_2$ to be consistent in our analysis, so unless specified, we only report the amplitude $m$.

The shear calibration methodology keeps evolving as our understanding of systematics deepens. Early studies demonstrated that the shear bias correlated with galaxy properties and PSFs, especially the signal-to-noise ratio and resolution (e.g. \citealt{Miller2013MNRAS4292858M,Hoekstra2015MNRAS449685H,Mandelbaum2018MNRAS4813170M,Samuroff2018MNRAS4754524S}). So the first lesson is to avoid using one averaged result from the whole simulation as a scalar calibration to the entire data unless the simulations perfectly represent the data. A natural procedure then attempts to estimate the shear bias as a function of the galaxy and PSF properties (e.g. \citealt{Miller2013MNRAS4292858M,Jarvis2016MNRAS4602245J}). Nevertheless, we can only derive the relation of the bias to the noisy, measured properties, as the true properties are unknown in actual data. {\FC} found that the relation derived from the measured properties introduces biases because of the correlations between observed quantities, an effect referred to as the `calibration selection bias'. So the second lesson is that we should be cautious about object-based shear calibrations that rely on the relation to the noisy properties. That is why the recent simulations try to resemble the data and only provide a mean correction for an ensemble of galaxies~(e.g. \K). The latest lesson, stressed by \citet{MacCrann2022MNRAS5093371M}, is the interplay between shear estimates of blended objects at different redshifts, a higher-order effect that the traditional constant shear simulations cannot capture. It becomes more important as the precision of surveys improves.

Our shear calibration method builds on all these lessons. We created constant shear simulations following the previous KiDS tomographic calibration method but with improvements to the photo-$z$ estimates by taking advantage of the simulated multi-band images~(Sect.~\ref{Sec:shearBase}). Using additional blending-only variable shear simulations, we applied a correction to account for the interplay between blends containing different shears~(Sect.~\ref{Sec:shearVar}). When testing the PSF modelling algorithm in image simulations, we detected a small but noticeable change of shear bias, which was also corrected in our fiducial results~(Sect.~\ref{Sec:shearPSF}). 

\subsection{Results from the constant shear simulations}
\label{Sec:shearBase}
%
\begin{table*}
\caption{Differences between the COllege~(\K) and SKiLLS simulations.}             
\label{table:CompSim}      
\centering          
\begin{tabular}{p{0.1\textwidth} l p{0.33\textwidth} p{0.33\textwidth}}
\hline\hline\\       
\multicolumn{2}{c}{} & COllege~(\K) & SKiLLS (this work)\\ 
\hline\\                    
   Galaxies & Morphology & S\'ersic models with parameters taken directly from the HST-ACS measurements~\citep{Griffith2012ApJS} & S\'ersic models with parameters learned from the HST-ACS measurements~(Sect.~\ref{Sec:inputGalShape})\\  
   {} & Photometry & Single-band magnitudes from the Subaru $r^+$-band observations & Nine-band synthetic magnitudes based on a semi-analytic model~(Sect.~\ref{Sec:inputGalPhoto})\\
   {} & Depth & Limited by the HST-ACS measurements & Extending to $27$th magnitude in the $r$ band\\
   {} & Position & Based on the observed locations in the COSMOS field & Based on the \textsc{SURFS} $N$-body simulations~\citep{Elahi2018MNRAS}\\
\hline\\
   Stars & Photometry & Single-band synthetic magnitudes from the Besan{\c c}on model~\citep{Robin2003AA409523R,Czekaj2014AA564A102C} & Nine-band synthetic magnitudes from the \textsc{Trilegal} model~\citep{Girardi2005AA}\\
\hline\\
   Images & Band & the $r$-band images only & the full nine-band images\\
   {} & Layout & $32$ CCDs with even gaps in between & $32$ CCDs with variable gaps as in the actual camera (Fig.~\ref{fig:weightMap})\\
   {} & PSF & $13$ sets of spatially constant Moffat profiles, with each containing five different models corresponding to the five exposures & $108$ sets of spatially varying polynomial models, with each containing $5\times32$ different models\\
   {} & Noise & One fixed noise level for all tiles & $108$ different noise levels\\
   {} & Stack & Only \textsc{theli}-like stacks for shape measurements & Both  \textsc{theli}-like and \textsc{Astro}-WISE-like stacks for shape and photometric measurements, respectively\\  
\hline\\                  
   Measurements & Shape & From the self-calibration version of \textit{lens}fit with the weight bias correction of \FC & From the updated \textit{lens}fit with the AlphaRecal method detailed in Sect.~\ref{Sec:lensfitUpdate}\\
   {} & photo-$z$ & Assigned with the KiDS observations of the COSMOS field & Measured from the simulated nine-band images following the KiDS photometric processing steps (Sect.~\ref{Sec:simPhoto})\\
\hline\\
\multicolumn{2}{l}{Sample variance} & Identical input catalogues of galaxies and stars for all the $13$ realisations & Different galaxy catalogues for the $108$ realisations and six stellar catalogues for the selected sky blocks (Fig.~\ref{fig:footprint})\\
\multicolumn{2}{l}{Input shears\tablefootmark{a}} & Eight sets of constant shears & Four sets of constant shears in the baseline simulations and a variable shear field for the blended objects (Appendix~\ref{Sec:Zshear})\\
\multicolumn{2}{l}{Shape noise cancellation\tablefootmark{b}} & Each tile has three counterparts with galaxies rotated by $45$, $90$ and $135$ degrees & Each tile has one counterpart with galaxies rotated by $90$ degrees\\
\hline\\
\multicolumn{2}{l}{Total simulated area} & $416~{\rm deg}^2$ & $864~{\rm deg}^2$ in the constant shear simulations plus $7776~{\rm deg}^2$ of blending-only simulations for the correction of the `shear interplay' effect (Sect.~\ref{Sec:shearVar})\\
\hline
\end{tabular}
\tablefoot{\tablefoottext{a}{We verified that the four sets of input shears are sufficient to recover the previous results.}
\tablefoottext{b}{Although more rotations suppress shape noise more efficiently~(\FC), the selection effects diminish the actual performance of the shape noise cancellation~(\K).}
}

\end{table*}

Our constant shear simulations largely followed {\FC} and {\K} with some simplifications for better usage of computational resources. Table~\ref{table:CompSim} lists the main changes we made compared to our predecessor. Given the $108~{\rm deg}^2$ of unique synthetic galaxies we built in Sect.~\ref{Sec:input}, we mimicked $108$ KiDS pointings, where we vary the PSF, noise level and stellar density as detailed in Sect.~\ref{Sec:sim}. To reduce the shape noise, we copied each tile image with galaxies rotated by $90$ degrees. We created four sets of constant shear simulations with input shear: $(0.0283, 0.0283)$, $(0.0283, -0.0283)$, $(-0.0283, -0.0283)$, $(-0.0283, 0.0283)$. The total simulated area is $864~(=108\times 4 \times 2)~{\rm deg}^2$, which is equivalent to ${\sim}5170~{\rm deg}^2$ after accounting for the shape noise cancellation~($=864\times (\sigma_{\epsilon, {\rm raw}}/\sigma_{\epsilon, {\rm SNC}})^2$, where $\sigma_{\epsilon, {\rm raw}}$ and $\sigma_{\epsilon, {\rm SNC}}$ denote the weighted dispersion of the mean input ellipticities before and after the shape noise cancellation), which is roughly four times the final KiDS-Legacy area.

For a tomographic analysis, we need to estimate the bias for each redshift bin separately, given that the galaxy properties vary between bins. This requires photo-$z$ estimates for the simulated galaxies. For SKiLLS, we can follow the KiDS processing steps to directly measure photo-$z$s, thanks to the simulated nine-band images. We conducted the detection from the \textsc{theli}-like $r$-band images, the PSF Gaussianisation and forced multi-band photometry using the \textsc{GAaP} pipeline, and the photo-$z$ estimates with the \textsc{bpz} code (see Sect.~\ref{Sec:simPhoto} for details). This consistent data processing ensures that SKiLLS embraces realistic photometric properties, marking one of the most significant improvements over the previous image simulations.

As shown in Fig.~\ref{fig:LFout}, SKiLLS matches KiDS generally well but not perfectly. {\K} argued that an accurate estimate of the shear bias must account for any mismatches between the simulations and the target data. Therefore, we followed {\FC} and {\K} to reweight the simulation estimates using the \textit{lens}fit reported $\nu_{\rm SN}$ and resolution factor $\mathcal{R}$ (Eq.~\ref{eq:resolution}). Specifically, for each tomographic bin, we first divided simulated galaxies into $20\times20$ bins of $\nu_{\rm SN}$ and $\mathcal{R}$, each containing equal \textit{lens}fit weight. Then we estimated the multiplicative bias for each $\nu_{\rm SN}$-$\mathcal{R}$ bin using Eq.~(\ref{eq:shearbias}). Galaxies in the target data were assigned the bias based on the $\nu_{\rm SN}$-$\mathcal{R}$ bin they fall in, and the final bias for each tomographic bin was the \textit{lens}fit-weighted average of these individual assignments. This procedure ensures the estimated bias accounts for any $\nu_{\rm SN}$ and $\mathcal{R}$ differences between the simulations and the data while also minimising the impact of the calibration selection bias.

\begin{figure}
  \centering
  \includegraphics[width=\hsize]{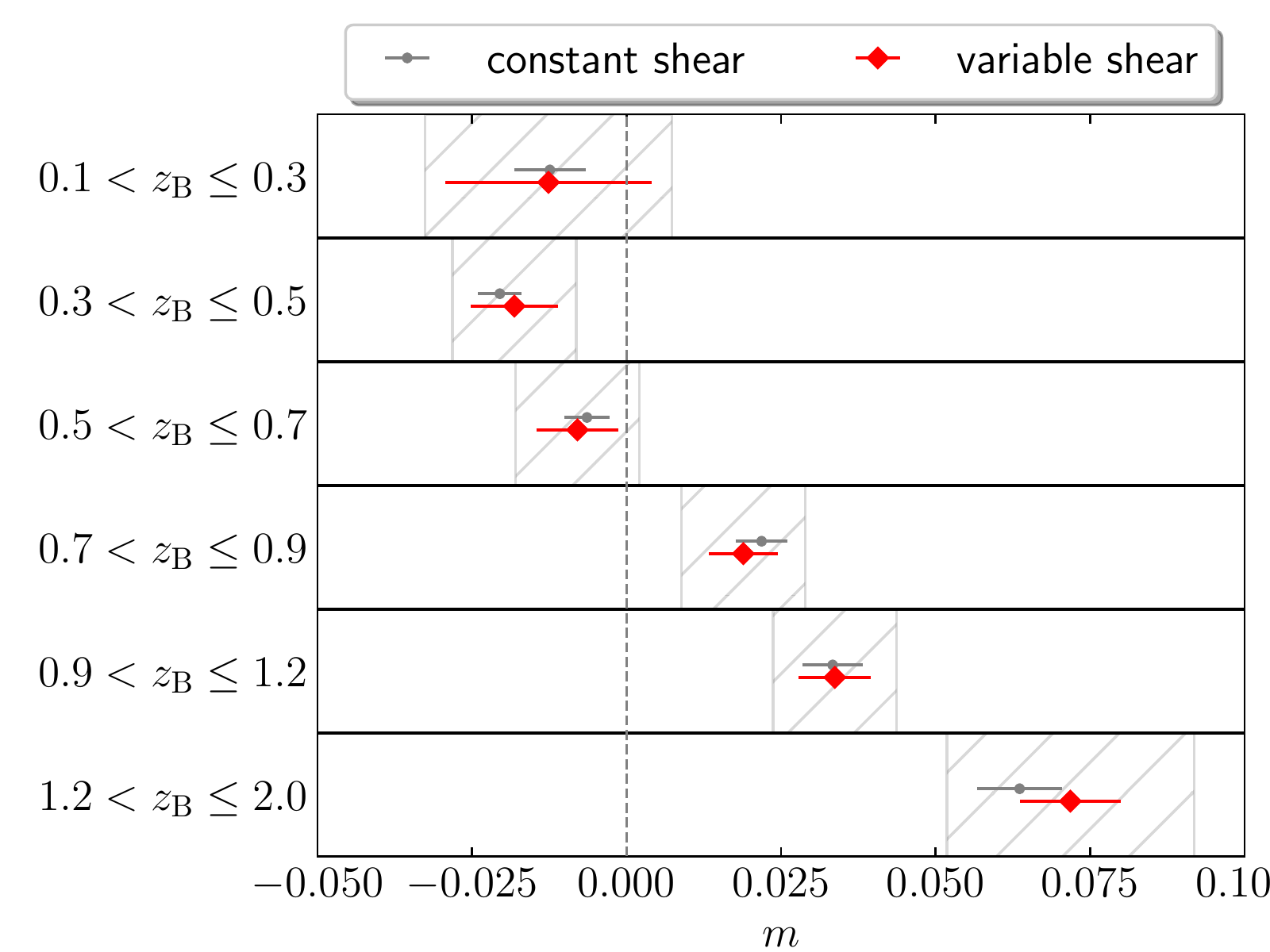}
      \caption{Multiplicative bias as a function of tomographic bins for KiDS-DR4 with the updated \textit{lens}fit. The red diamonds indicate our final results with the corrections for the shear-interplay effect (Sect.~\ref{Sec:shearVar}) and PSF modelling bias (Sect.~\ref{Sec:shearPSF}), whilst the grey points are the raw results from the idealised constant shear simulations (Sect.~\ref{Sec:shearBase}). The hatched regions indicate the nominal error budgets proposed for comparison (see Sect.~\ref{Sec:sensi} for details).}
         \label{fig:m}
  \end{figure}

Table~\ref{table:m} and Figure~\ref{fig:m} show the multiplicative bias estimates for the KiDS-DR4 re-run with the updated \textit{lens}fit from our constant shear simulations. The quoted errors only contain the statistical uncertainties from the linear fitting. Compared to Table 2 of {\K}, we reduced the statistical uncertainties by about half because of the larger sky area simulated. Direct comparisons between the calibration values quoted in Table~\ref{table:m}, cannot be made to those in {\K} and \citet{Giblin2021AA645A105G}. We updated the shape measurement algorithm \textit{lens}fit and calibrated the raw measurement against PSF contamination in our analysis (see Sect.~\ref{Sec:lensfitUpdate}). These changes modify the effective size and signal-to-noise ratio distribution of the samples and hence the overall calibration in each tomographic bin. Furthermore, \citet{Giblin2021AA645A105G} accounts for the \citet{Wright2020AA637A100W} `gold' selection for photometric redshifts, which reduces the effective number density by ${\sim}20\%$, compared to the sample simulated in this analysis.

%
\begin{table*}
\caption{Shear bias for the six tomographic bins.}             
\label{table:m}      
\centering          
\begin{tabular}{lcccccc}
\hline\hline\\       
$z_{\rm B}$ range & Ratio of $N_{\rm eff}$ & $\Delta \bar{m}_{\rm blending}$ & $\Delta m_{\rm PSF}$ & $m_{\rm raw}$ & $m_{\rm final}$\\
 & (blending / whole) &  &  &  &\\
\hline\\                    
$0.1<z_{\rm B}\leq 0.3$ & $0.345$ & $-0.012\pm 0.034$ & $+0.002\pm 0.001$ & $-0.012\pm 0.006$ & $-0.013\pm 0.017$\\
$0.3<z_{\rm B}\leq 0.5$ & $0.332$ & $-0.003\pm 0.014$ & $+0.004\pm 0.001$ & $-0.021\pm 0.004$ & $-0.018\pm 0.007$\\
$0.5<z_{\rm B}\leq 0.7$ & $0.365$ & $-0.021\pm 0.012$ & $+0.004\pm0.001$ & $-0.006\pm 0.004$ & $-0.008\pm 0.007$\\
$0.7<z_{\rm B}\leq 0.9$ & $0.366$ & $-0.018\pm 0.008$ & $+0.003\pm0.001$ & $+0.022\pm 0.004$ & $+0.019\pm 0.006$\\
$0.9<z_{\rm B}\leq 1.2$ & $0.370$ & $-0.013\pm 0.007$ & $+0.005\pm0.001$ & $+0.033\pm 0.005$ & $+0.034\pm 0.006$\\
$1.2<z_{\rm B}\leq 2.0$ & $0.358$ & $+0.000\pm 0.008$ & $+0.007\pm0.002$ & $+0.064\pm 0.007$ & $+0.072\pm 0.008$\\
\hline
\end{tabular}
\tablefoot{The ratio of $N_{\rm eff}$ between the blending-only simulation and the whole simulation is calculated from the measured catalogue with the \textit{lens}fit weight taken into account. The $\Delta \bar{m}_{\rm blending}$ is the mean residual bias introduced by the shear-interplay effect, estimated from the blending-only simulations (see Sect.~\ref{Sec:shearVar} for details). The correction to the whole sample should also account for the $N_{\rm eff}$ ratio and the correlation with the signal-to-ratio and resolution (see Sect.~\ref{Sec:shearVar} for details). The $\Delta m_{\rm PSF}$ is the residual bias introduced by the PSF modelling errors (see Sect.~\ref{Sec:shearPSF} for details). The $m_{\rm raw}$ results are derived from the idealised constant shear simulations (Sect.~\ref{Sec:shearBase}), and the $m_{\rm final}$ are our final estimates with the corrections for the shear-interplay effect and PSF modelling bias (Sect.~\ref{Sec:shearFinal}). The uncertainties quoted along with individual $m$ values are reported by the linear regression fitting, thus only reflecting the statistical power of SKiLLS simulations. All results are based on the KiDS-DR4 re-run with the updated \textit{lens}fit before any redshift calibration. They only indicate the general performance of the updated \textit{lens}fit.}
\end{table*}

%

\subsection{Impact of blends at different redshifts}
\label{Sec:shearVar}

\citet{MacCrann2022MNRAS5093371M} recently highlighted a complication that arises from blended objects at different redshifts, which are, therefore, sheared by different amounts. It stems from the fact that when objects are blended, a shear measurement of one object responds to the shear of the neighbouring object. This higher-order effect, which we refer to as `shear interplay' through this paper, cannot be captured by the aforementioned constant shear simulations. So, we built an extra suite of variable shear simulations to account for this effect.

\begin{figure}
  \centering
  \includegraphics[width=\hsize]{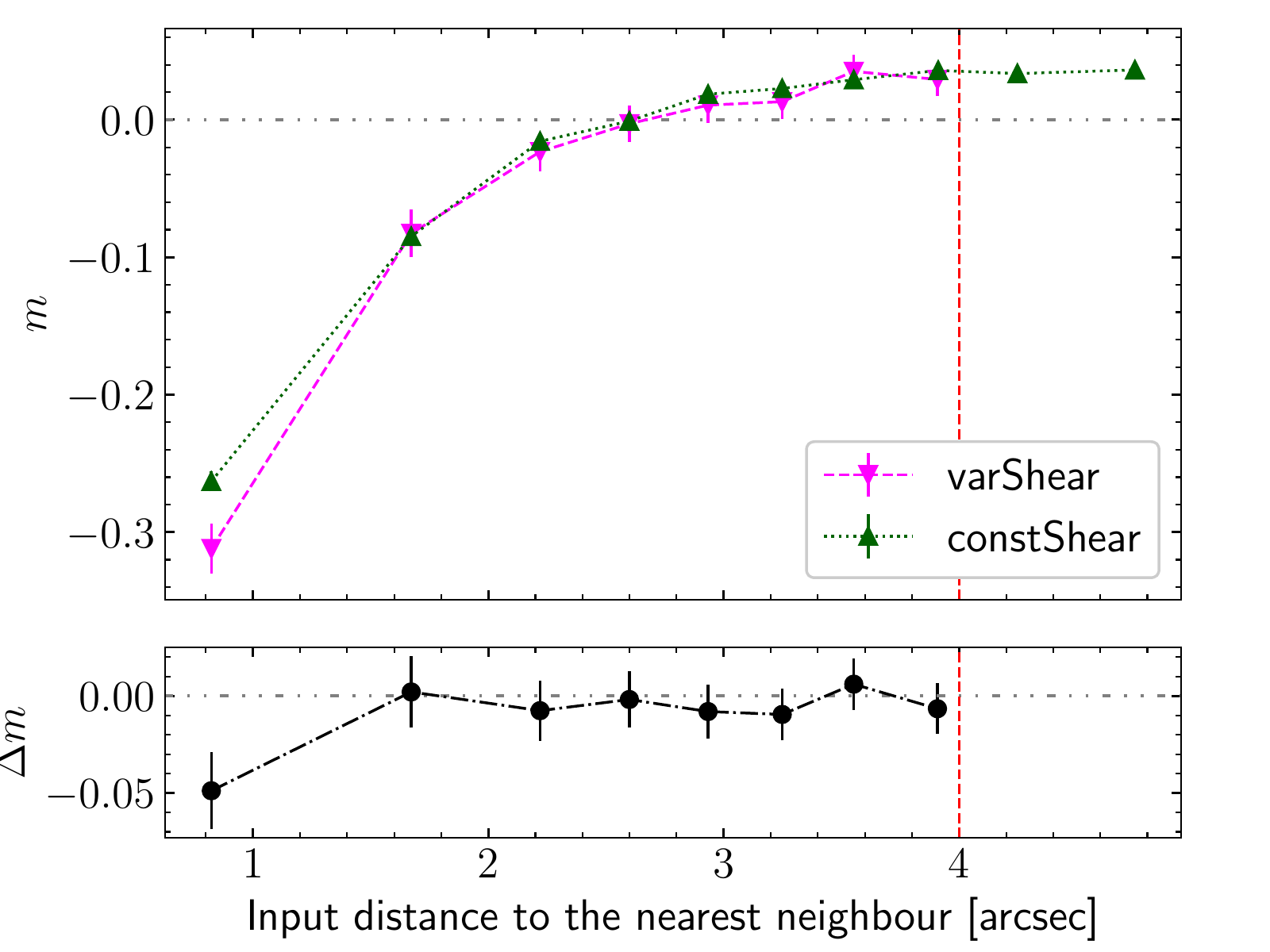}
      \caption{Multiplicative bias as a function of the nearest neighbour distance. The distance is measured in the input catalogue after removing faint galaxies whose $r$-band input magnitude ${>}25$. The $x$-axis values correspond to the weighted average of each sub-sample selected to estimate the multiplicative bias shown on the $y$-axis. The top panel shows the individual biases measured from the blending-only variable shear simulations (magenta points) and the blending-only constant shear simulations (dark green points). The vertical dashed lines show the threshold we set when building the blending-only simulations. Two extra dark green points beyond the threshold are calculated from the full constant shear simulations. The bottom panel shows the difference between these two estimates (varShear - constShear).}
         \label{fig:mBlend}
  \end{figure}

Since the shear interplay only happens when objects are blended, we built a blending-only input catalogue for these additional simulations to save some computing time. This blending-only catalogue only contains bright galaxies with bright neighbours, assuming that the blending effects caused by the faint objects are sufficiently accounted for by our main constant shear simulations, which include galaxies down to magnitude $27$. It means we only ignore the higher-order shear-interplay effect from the faint objects, which is valid as long as the excluded faint galaxies are below the measurement limit of the survey. In practice, we selected all galaxies with an input $r$-band magnitude $<25$. The choice of this magnitude cut meets the overall sensitivity of the KiDS survey. We further discarded those isolated galaxies whose nearest neighbour is $4\arcsec$ away based on their input positions (see Fig.~\ref{fig:mBlend}). The final selected sample covers ${\sim}10\%$ of the entire input catalogue. But after the \textit{lens}fit measurements, this blending-only simulation covers ${\sim}35\%$ of the objects measured in the whole simulation (see Table~\ref{table:m} for the exact values). The higher fraction in the measured catalogue is because most objects fainter than $25$ in the $r$-band magnitude are not measurable for KiDS.

\begin{figure}
  \centering
  \includegraphics[width=\hsize]{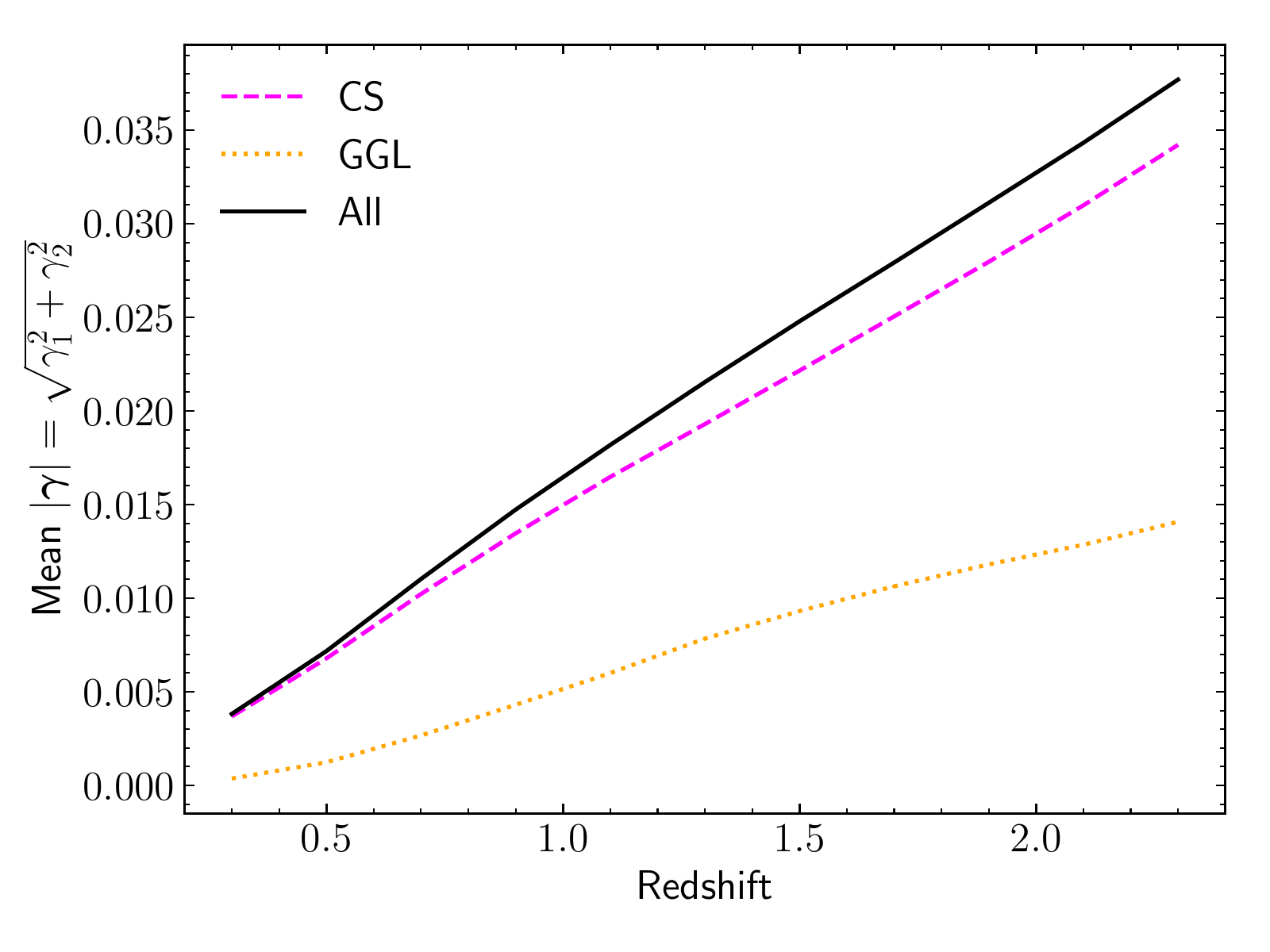}
      \caption{Variable shear field as a function of redshift. The solid black line shows the mean amplitude of the final used shears, which contain two components: the cosmic shear (dashed magenta line) and the tangential shear (dotted orange line). We refer to Appendix~\ref{Sec:Zshear} for details.}
         \label{fig:Zshear}
  \end{figure}

To properly account for the shear-interplay effect, we need realistic shear fields with proper correlations between the shear and the environment of galaxies. We refer to Appendix~\ref{Sec:Zshear} for technical details of our approach to creating such variable shear fields. In short, we considered two primary contributions to the weak lensing signal: the cosmic shear due to the large-scale structure and the tangential shear induced by the foreground objects (also known as the galaxy-galaxy lensing effect). The cosmic shear was learned from the MICE Grand Challenge (MICE-GC) simulation~\citep{Fosalba2015MNRAS4471319F}, whilst the tangential shear was calculated analytically by assuming Navarro-Frenk-White~\citep{Navarro1995MNRAS275720N} density profiles for the underlying dark matter halos. Figure~\ref{fig:Zshear} shows the average shear signals as a function of redshift. We see a roughly linear relationship between the mean signals and redshift. On average, the cosmic shear contributes more than the tangential shear. However, we note that the importance of the tangential shear varies between systems depending on the host halo mass of the foreground galaxies. 

To increase the constraining power, we used $32$ variable shear fields generated from the same learning algorithm but with different choices for the direction of the shear. Specifically, we created four variable shear fields with directions of the cosmic shear that differ by $90^{\circ}$. Then, we made eight copies for each shear field by rotating the final shear by $45^{\circ}$ each time. We also created an extra suite of blending-only constant shear simulations to serve as a reference. The final sky area of these additional simulations is \num[]{7776}~${\rm deg}^2 (=108\times36\times2)$. Except for the input shear, these blending-only simulations use the same pipeline, observational conditions and random seeds as the full simulations detailed in Sect.~\ref{Sec:shearBase} so that we can directly correct the constant shear results using the extra bias estimated from these additional simulations.

   \begin{figure*}
   \includegraphics[width=\hsize]{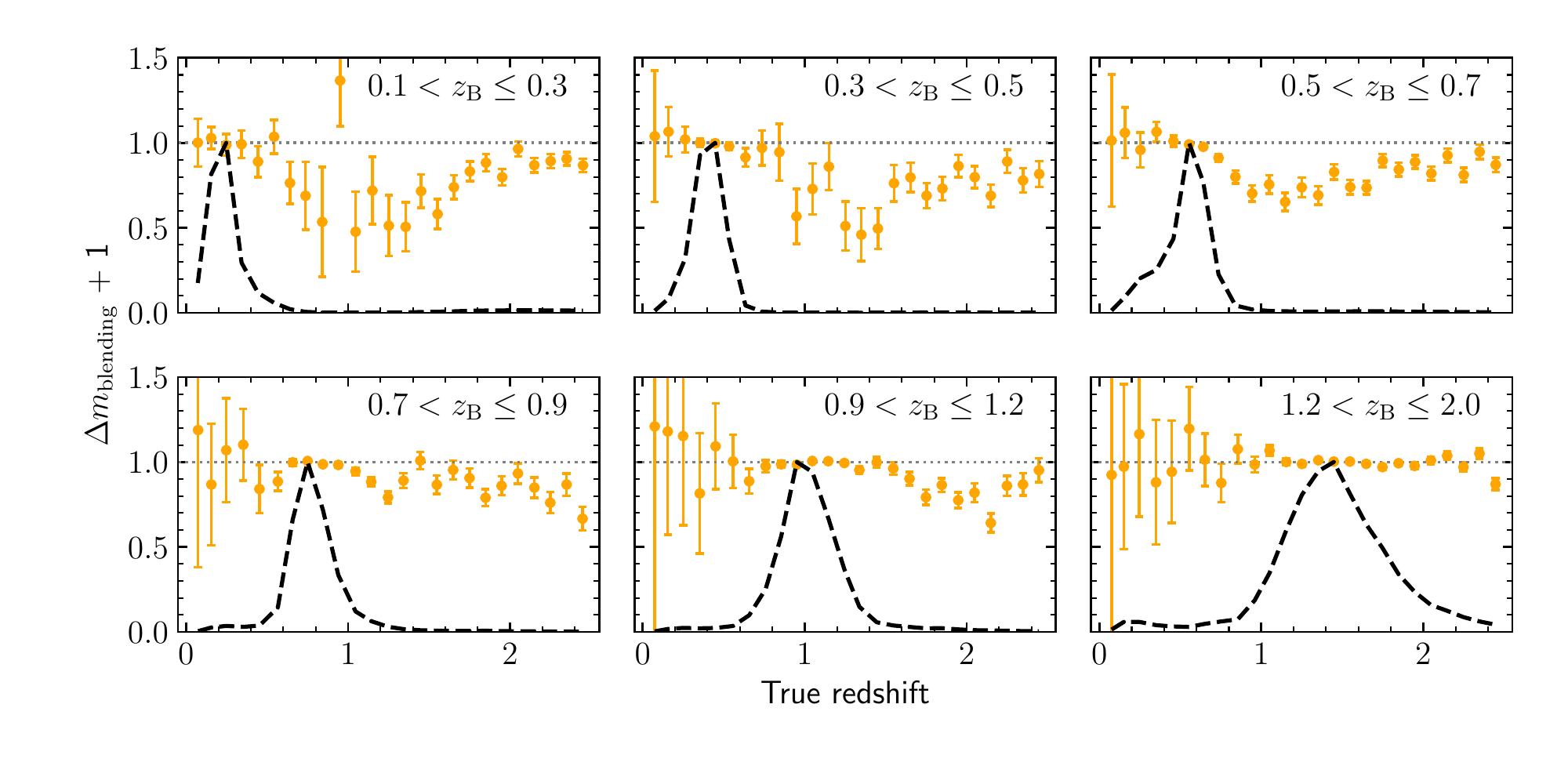}
     \caption{Residual shear bias introduced by the shear-interplay effect (orange points) as a function of the true redshift estimated from the blending-only simulations. The residuals are calculated from $\Delta m_{\rm blending}\equiv m_{\rm blending}^{\rm varShear}-m_{\rm blending}^{\rm constShear}$, with $m_{\rm blending}^{\rm varShear}$ the shear bias from the blending-only variable shear simulations and $m_{\rm blending}^{\rm constShear}$ the shear bias from the blending-only constant shear simulations. The error bars correspond to the fitting uncertainties reported by the linear regression. They are driven by two factors: the number of objects used by the fitting and the amplitude of the input shear value. The dashed lines show the normalised number density with respect to redshift. }
         \label{fig:dmZtrue}
   \end{figure*}
While estimating the shear bias for constant shear simulations is straightforward by directly conducting the linear least squares fitting to all measurements using Eq.~(\ref{eq:shearbias}), given that the input shear values do not depend on the underlying sample. The situation is more complicated for variable shear simulations. The crucial caveat is that the shear bias is now correlated with redshift [$m^{\rm varShear}_{\rm blending}(z_{\rm true})$] due to the shear-interplay effect. Owing to the realistic shear field we built, we can measure $m^{\rm varShear}_{\rm blending}(z_{\rm true})$ directly from simulations by performing the least squares fitting to sub-samples of galaxies split based on their true redshift. The same approach can also be applied to the blending-only constant shear simulations to get $m^{\rm constShear}_{\rm blending}(z_{\rm true})$; only, in that case, we would expect a negligible correlation with the true redshift, except for some fluctuations stemming from the different signal-to-noise ratios between true redshift bins. Figure~\ref{fig:dmZtrue} shows the difference $\Delta m_{\rm blending}(z_{\rm true}) \equiv m^{\rm varShear}_{\rm blending}(z_{\rm true}) - m^{\rm constShear}_{\rm blending}$, which is a direct measure of the impact of the shear-interplay effect, as the only difference between the simulations is the input shear value. It demonstrates evident residuals that correlate with redshift, indicating the non-trivial impact of the shear-interplay effect. Interestingly, the high-redshift outliers, which have an estimated photo-$z$ much lower than their true redshifts, show the most noticeable residuals across all tomographic bins, implying that the blends with objects from different redshifts are likely responsible for those outliers. This coupling between the photo-$z$ and shear biases in blended systems warrants a dedicated future study.

\begin{figure}
  \centering
  \includegraphics[width=\hsize]{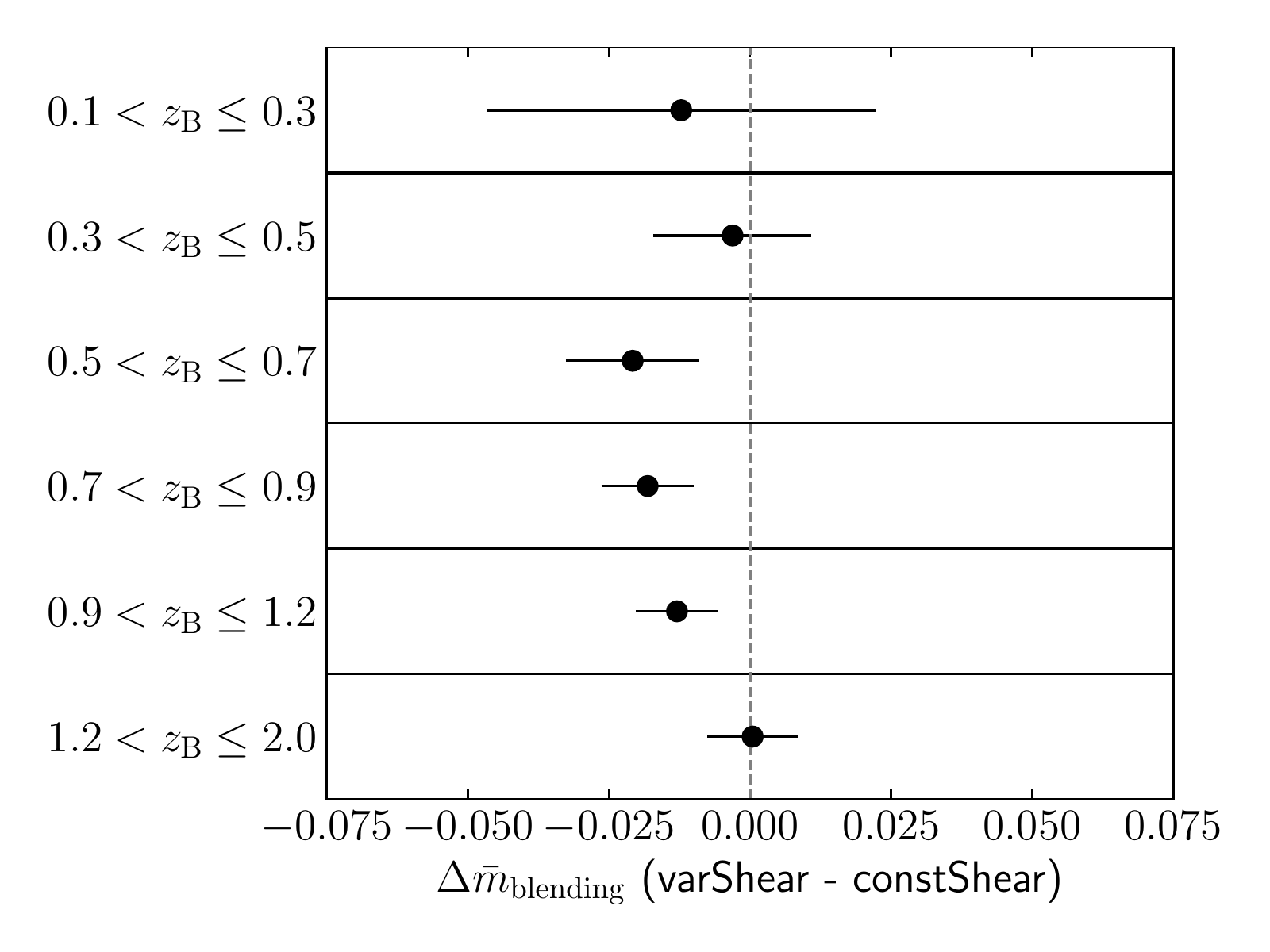}
      \caption{Mean residual multiplicative bias introduced by the shear-interplay effect. It is calculated from $\Delta \bar{m}_{\rm blending}=\int_{0}^{\infty}dz_{\rm true}~n(z_{\rm true})~\Delta m_{\rm blending}(z_{\rm true})$, with $n(z_{\rm true})$ and $\Delta m_{\rm blending}$ showing in Fig.~\ref{fig:dmZtrue}. We stress that the results are from the blending-only simulations. When applying to the whole sample, we must also consider the blending fraction (the third column of Table~\ref{table:m}).} 
         \label{fig:dmVar}
  \end{figure}

To correct the raw shear bias derived in Sect.~\ref{Sec:shearBase}, we need an average correction $\Delta \bar{m}_{\rm blending}$, which integrates over $z_{\rm true}$ as $\Delta \bar{m}_{\rm blending}=\int_{0}^{\infty}dz_{\rm true}~n(z_{\rm true})~\Delta m_{\rm blending}(z_{\rm true})$, where $n(z_{\rm true})$ is the weighted number density with respect to redshift (the dashed lines shown in Fig.~\ref{fig:dmZtrue}). The average results for individual tomographic bins are shown in Table~\ref{table:m} and Fig.~\ref{fig:dmVar}. In practice, we should also account for the blending fraction, which is correlated with the signal-to-noise ratio and resolution, as is the bias itself. Therefore, we perform the correction in each $\nu_{\rm SN}$-$\mathcal{R}$ bin, following the binning strategy proposed for reweighting the simulation (see Sect.~\ref{Sec:shearBase}). Specifically, for each $\nu_{\rm SN}$-$\mathcal{R}$ bin, we estimate the average correction $\Delta \bar{m}_{\rm blending}$ and the blending fraction. The blending fraction is estimated as the ratio of the effective number counts between the blending-only simulation and the whole simulation. Then, we shift the raw bias in each $\nu_{\rm SN}$-$\mathcal{R}$ bin with the product of $\Delta \bar{m}_{\rm blending}$ and blending fraction. The final corrected bias is the \textit{lens}fit-weighted average of these shifted biases. This correction process can be easily combined with the reweighting procedure, as they use the same binning strategy.

%

Another more direct way to inspect the blending effect is to check the relation between the shear bias and the nearest neighbour distance in the input catalogue. Figure~\ref{fig:mBlend} demonstrates such estimations for both constant shear and variable shear simulations. We see a clear correlation between the bias and the neighbour distance in both simulations, indicating the significant impact of the blending effect. It also confirms our choice of $4\arcsec$ to define blended systems, as we barely see any correlation after this threshold. The other important finding is that the traditional constant shear simulations can already capture the dominant contributions from the blending effect. The higher-order impact we study in this section, shown as the bias difference between the variable shear and constant shear simulations, contributes relatively minor except for the very close blends. The aggressive treatment of the blending in \textit{lens}fit can partially explain this finding, as it throws away most of the recognised blends~\citep{Hildebrandt2017MNRAS.465.1454H}.

%

We note that our variable shear simulations and the correction methodology differ from those of \citet{MacCrann2022MNRAS5093371M}. In their study, the simulated shear changes as a function of redshift, but, per redshift slice, it remains constant across the field of view. The chosen redshift intervals and adjusted shear have no physical meaning in their setups. But they built four sets of simulations by choosing different redshift intervals, so they were able to fit a smooth model to the simulated results, obtaining a continuous redshift-bias relation. In our approach, we computed the variable shear fields using a more physical model that accounts for the shear correlations to both the redshift and clustering of galaxies (see Appendix~\ref{Sec:Zshear}). Thanks to these realistic shear fields, we can measure the redshift-bias relation directly from the simulations without additional model fitting procedures. Our direct measurements confirmed the non-trivial impact of the shear-interplay effect (see Fig.~\ref{fig:dmZtrue}). By design, our method results in large uncertainties for low redshift bins due to the small input shear values. Fortunately, these low redshift bins carry little cosmic shear signals, making the overall downgrade tolerable. Albeit following a different approach, our final result is consistent with \citet{MacCrann2022MNRAS5093371M} finding that the overall correction due to the shear-interplay effect is negligible for the current weak lensing surveys. However, it will potentially impact the next-generation surveys.

\subsection{PSF modelling bias}
\label{Sec:shearPSF}

So far, we have ignored the PSF modelling errors, given the expected accuracy of PSF models relative to the requirement of the current weak lensing surveys (see e.g.~\citealt{Giblin2021AA645A105G}). We used the input PSF for shape measurements (i.e. assuming perfect PSF modelling). However, as the requirement of systematics becomes more stringent, it becomes necessary to check the impact of PSF modelling errors. This section quantifies this impact by including the PSF modelling procedure in the simulations.

The SKiLLS images have realistic stellar populations and variable PSFs across the field, so we can apply the PSF modelling code directly to the simulated images using similar setups as for the data. We refer to \citet{Kuijken2015MNRAS4543500K} for detailed descriptions of the PSF modelling algorithm used by KiDS. In short, it describes the position-dependent PSFs at the detector resolution using a set of amplitudes on a $48\times48$ pixel grid. The spatial variation of each pixel value is fitted with a two-dimensional polynomial of order $n$, with additional flexibility for allowing the lowest order coefficients to differ from CCD to CCD. This extra freedom allows for a more complex PSF variation between CCDs and, in principle, allows for discontinuities in the PSF between adjacent CCDs. When fitting to individual stars, the flux and centroid of each star are allowed to change, and a sinc function interpolation is used to align the PSF model with the star position. Following \citet{Giblin2021AA645A105G}, we set $n=4$ and allow the polynomial coefficients up to order $1$ to vary between CCDs. We skipped the complicated star-galaxy separation procedure with an implicit assumption that the point-source sample used by KiDS is sufficiently pure as verified using NIR colours in \citet{Giblin2021AA645A105G}. Instead, we built a perfect star sample by cross-matching the detected catalogue with the input star catalogue. However, we still applied the same magnitude and signal-to-noise ratio cuts as used in the data to ensure a similar noise level in the modelled stars. 

\begin{figure}
  \centering
  \includegraphics[width=\hsize]{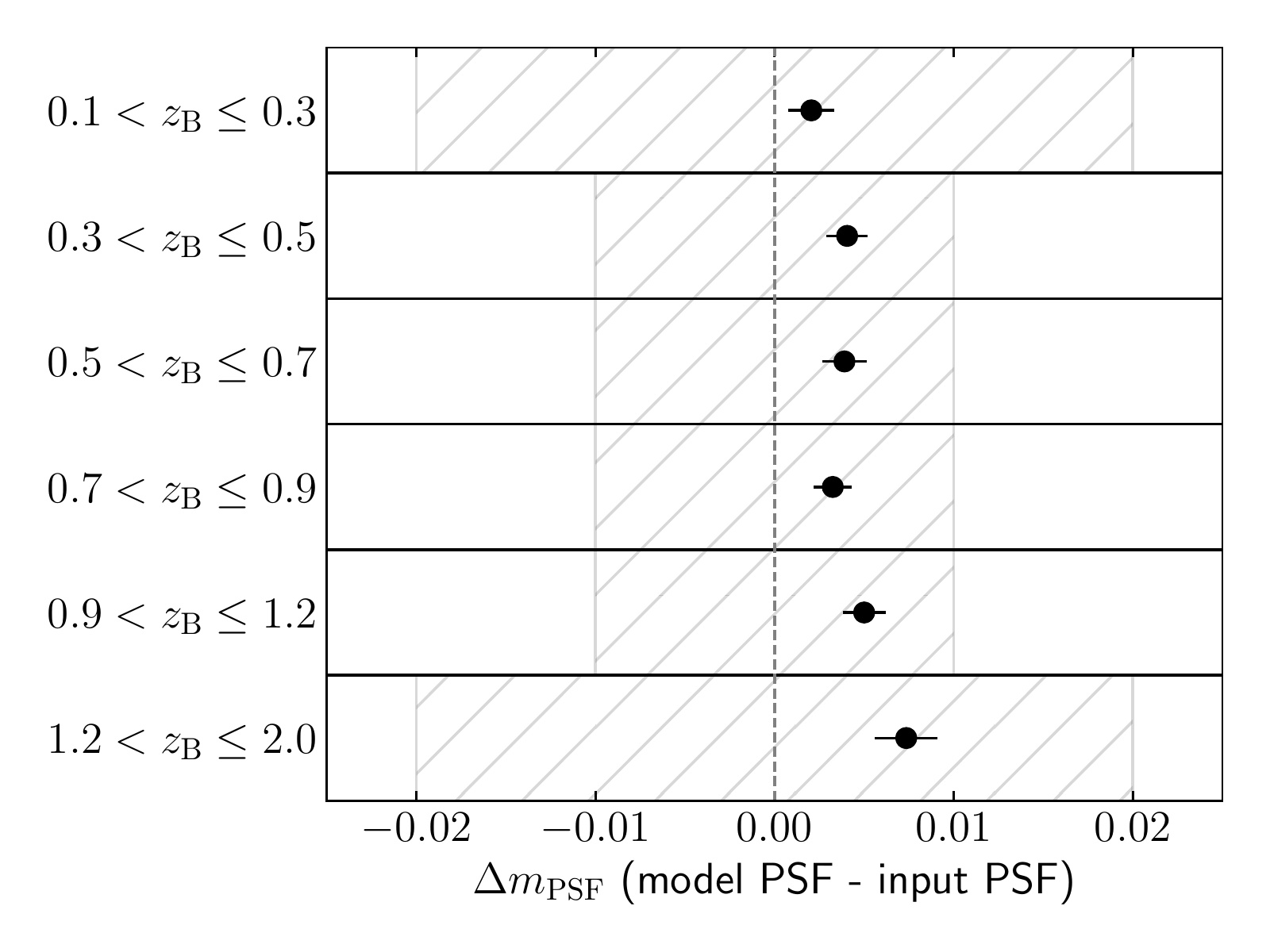}
      \caption{Changes in multiplicative bias when modelled PSFs are used instead of the input PSFs. The hatched regions indicate the nominal error budgets proposed for comparison (see Sect.~\ref{Sec:sensi} for details).} 
         \label{fig:dmPSF}
  \end{figure}

We selected $30$ tiles from the available $108$ fiducial tiles to test the influence of PSF modelling uncertainty on the multiplicative bias. These selected tiles cover the whole range of the PSF size, including the minimum and maximum. We performed the PSF modelling on the selected tiles and re-ran \textit{lens}fit using the modelled PSFs. Since all the images and detection catalogues are unchanged, the shift of the shear bias directly quantifies the contribution of the PSF modelling errors. Figure~\ref{fig:dmPSF} and Table~\ref{table:m} show the shifts for the six tomographic bins. We find the PSF modelling procedure does introduce small yet noticeable biases. Our fiducial results take these additional biases into account. 

%

\subsection{Results}
\label{Sec:shearFinal}

The final results after accounting for both the shear-interplay effect and PSF modelling errors are listed as $m_{\rm final}$ in Table~\ref{table:m} and shown as the red points in Fig.~\ref{fig:m}. Within the current statistical uncertainties, the average shifts due to the shear-interplay effect and PSF modelling errors are insignificant across all redshift bins, as indicated in Fig.~\ref{fig:m} between the grey points and the red points. A more noticeable change is the increased uncertainty introduced by the correction of the shear-interplay effect, especially in the low redshift bins where the input shear values are overall small in the variable shear simulations. Our proposed systematic error budgets account for these additional uncertainties (the hatched regions in Fig.~\ref{fig:m}).

%% file: Sec6_sensitivity.tex
Given the resemblance between the SKiLLS and KiDS images and the reweighting in the signal-to-noise ratio and $\mathcal{R}$ when estimating the shear biases, it is reasonable to assume that the estimates from SKiLLS can be used to correct the actual measurements. Nevertheless, it is still worth testing the robustness of SKiLLS results and accounting for any potential systematic uncertainties. We start with tests proposed by \FC\ and \K\ in Sect.~\ref{Sec:sensi1}. Thanks to the broad coverage of observational conditions in SKiLLS, we can quickly achieve these analyses without dedicated test runs. Additionally, we test how sensitive the \textit{lens}fit results are to the changes in the input galaxy morphology (Sect.~\ref{Sec:sensi3}). For comparison reasons, we propose some nominal error budgets based on the general performance of SKiLLS and the overall requirements of lensing analyses with KiDS. Specifically, we set an error budget of $0.02$ for the first and sixth tomographic bins and $0.01$ for the remaining bins. We found these nominal error budgets are conservative enough that our results are robust within them. Nevertheless, we note that these nominal error budgets can be over-conservative for cosmic shear analyses. In which case, we can estimate more accurate systematic uncertainties following other more aggressive approaches proposed by previous KiDS analyses~\citep{Giblin2021AA645A105G,Asgari2021AA...645A.104A}.

   \begin{figure*}
   \includegraphics[width=0.5\hsize]{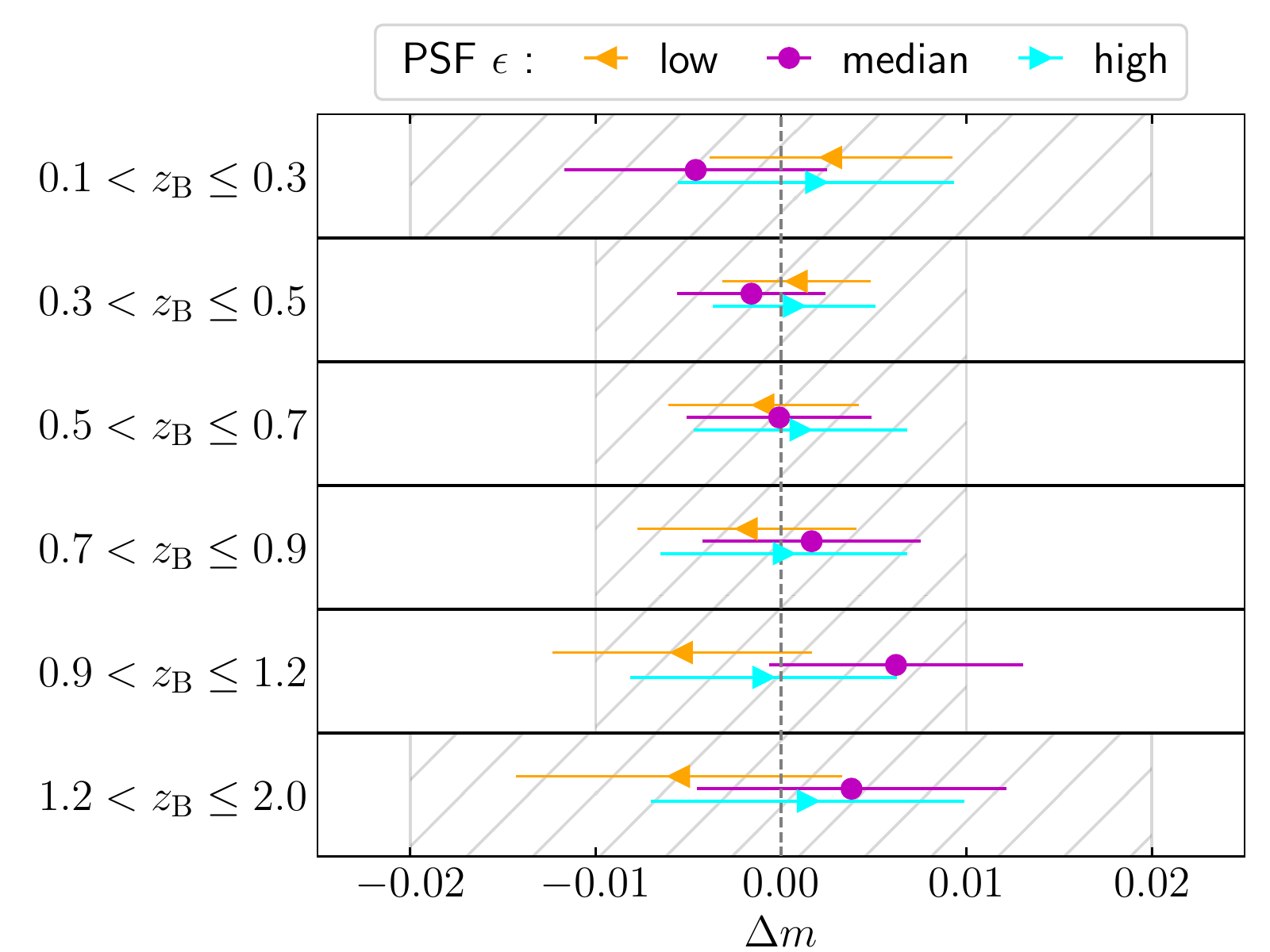}
   \includegraphics[width=0.5\hsize]{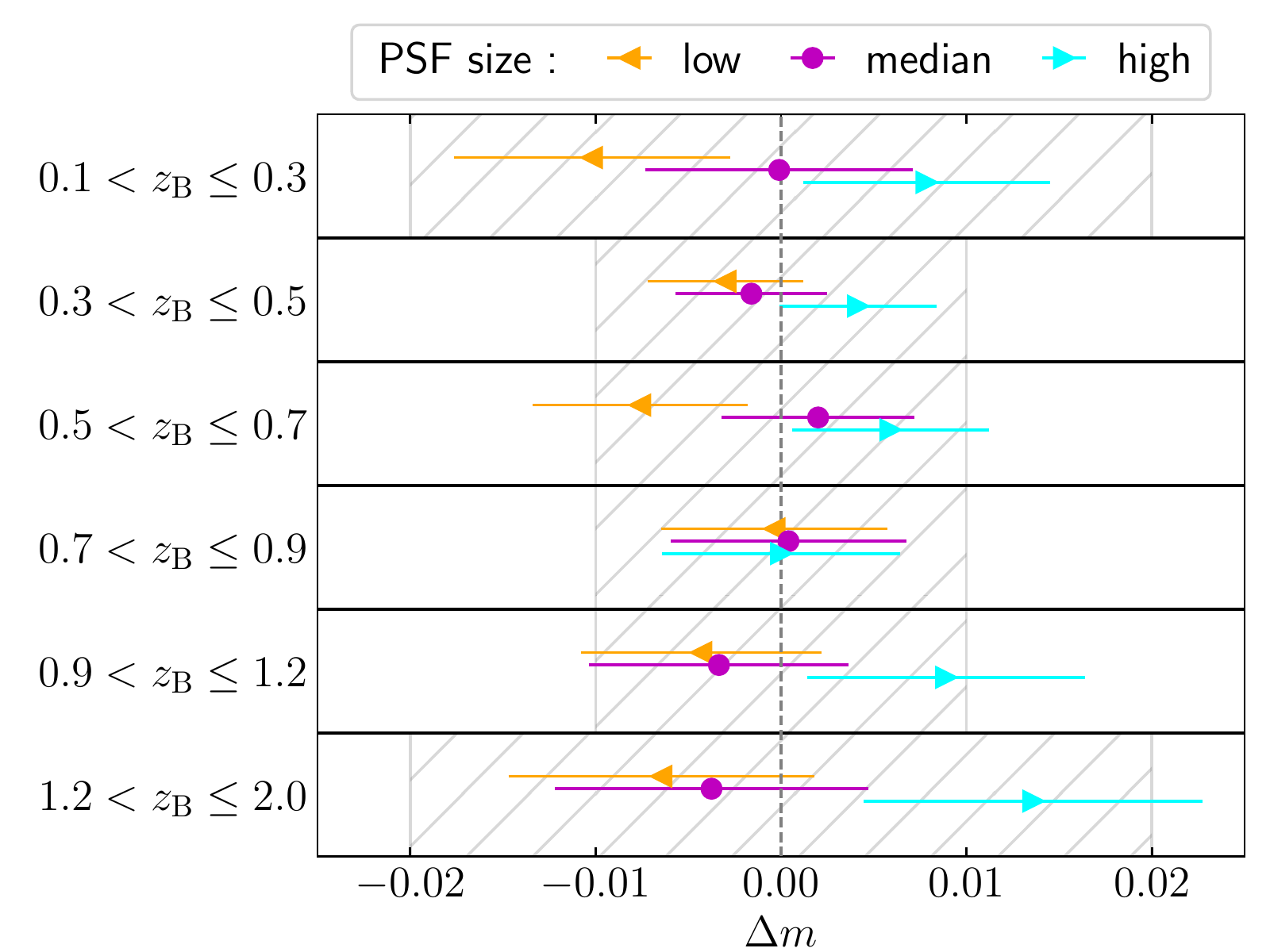}
  \includegraphics[width=0.5\hsize]{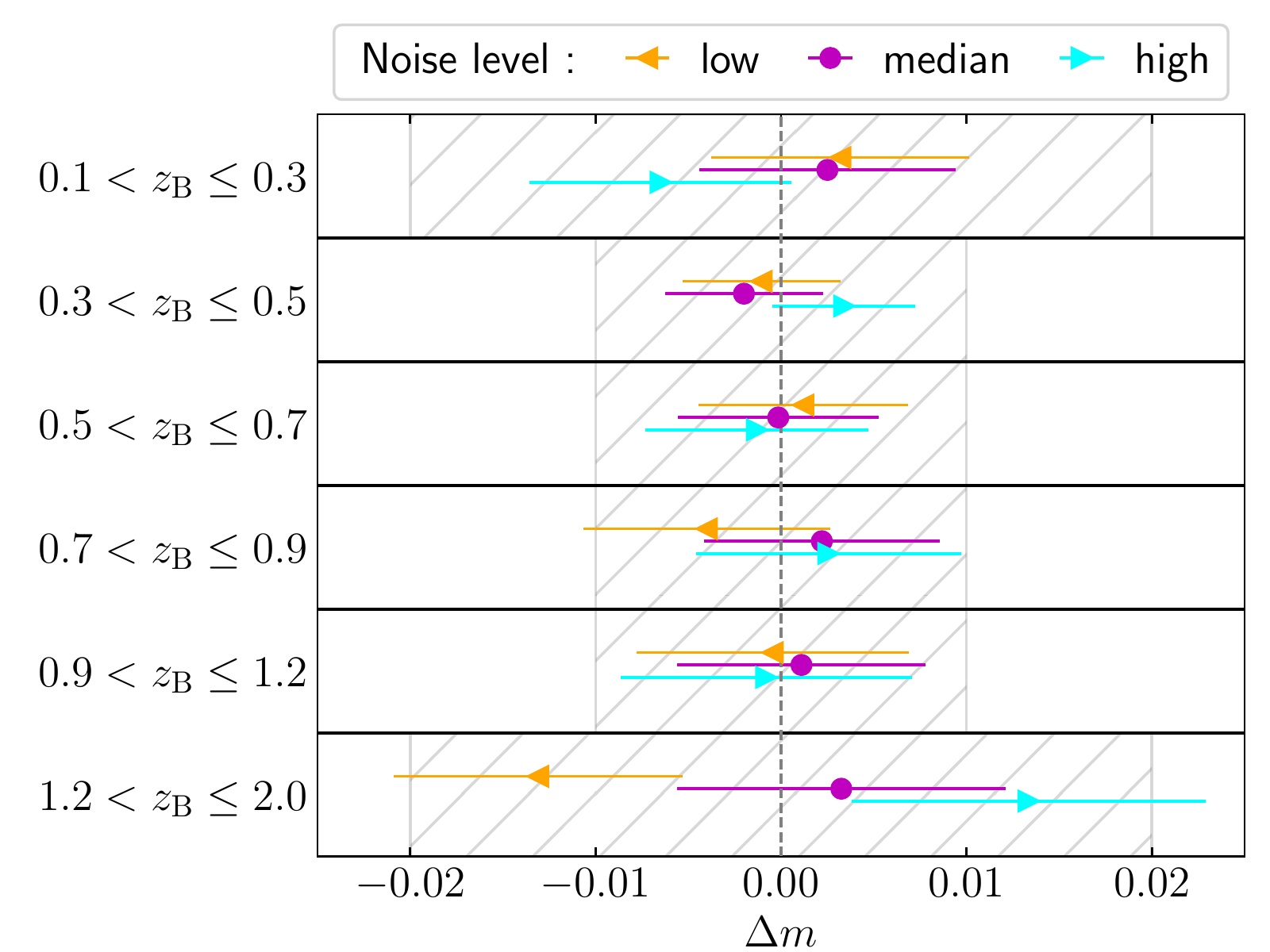}
   \includegraphics[width=0.5\hsize]{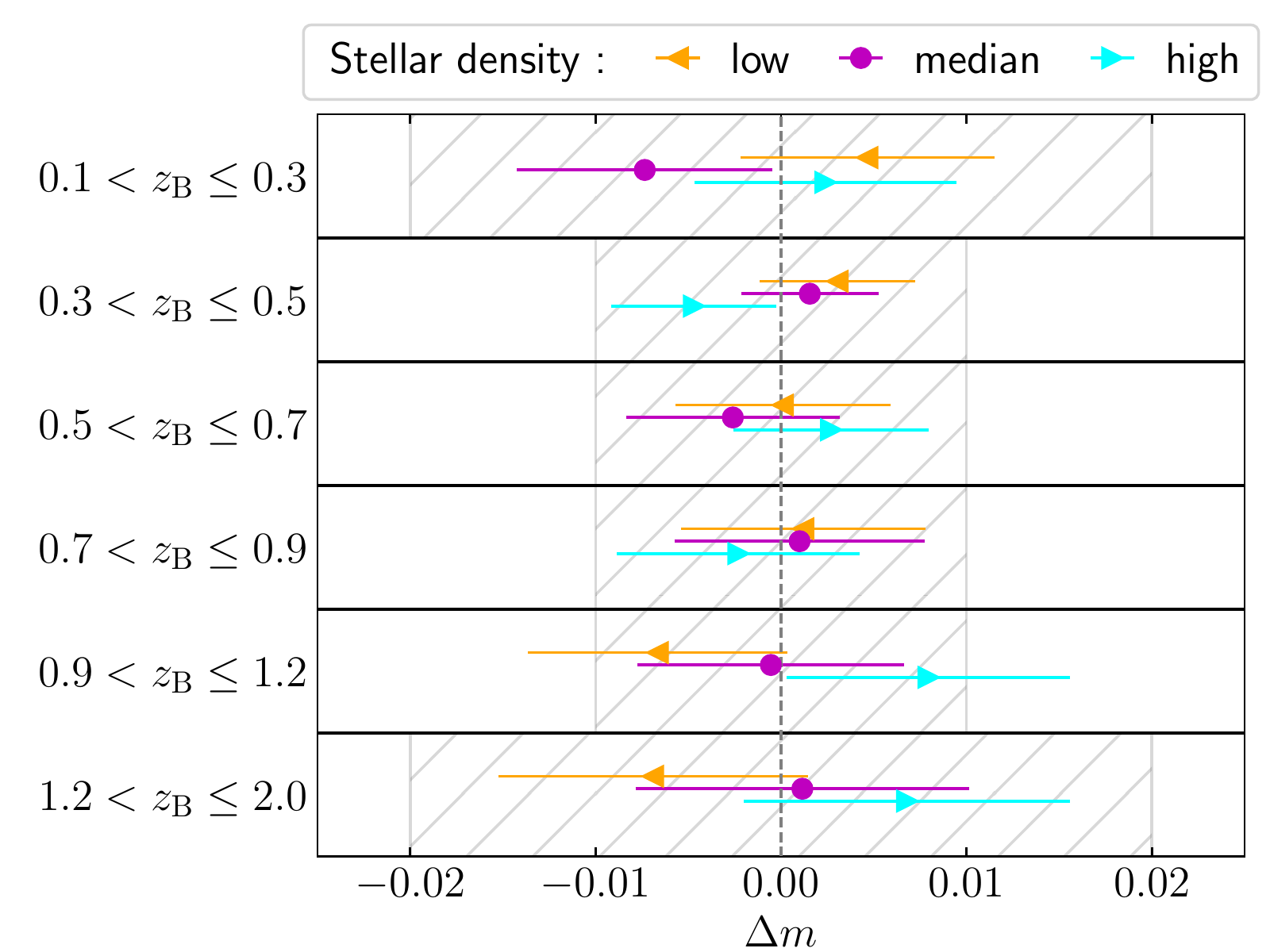}
      \caption{Changes in multiplicative bias when the fiducial simulations are divided into three sub-samples containing different observational conditions. From the left- to right-hand panels and top to bottom panels, the four panels show the results when splitting based on the PSF ellipticity, PSF size, background noise level in $r$-band images and stellar density. The hatched regions indicate the nominal error budgets proposed for comparison (see Sect.~\ref{Sec:sensi} for details). We note that the shifts correspond to the upper limits of potential systematic biases in our results (see Sect.~\ref{Sec:sensi1} for details).}
         \label{fig:dmSub}
   \end{figure*}
\subsection{Impact of observational conditions}
\label{Sec:sensi1}

When developing SKiLLS, we improved most of the critical sources of uncertainty in the previous KiDS simulations. For instance, we based our input galaxy catalogue on $N$-body simulations, so it has reasonable clustering features and is complete down to $27$ in the $r$-band magnitude. We learned realistic morphologies from observations using a powerful technique, dubbed vine copulas, which captures the multi-dimensional correlations between ellipticities and other galaxy properties. We included six stellar catalogues to account for the varying stellar densities across the survey sky. We covered more variations of the PSF models and background noise levels. Above all, we measured photo-$z$s directly from the simulated multi-band images to properly account for the correlation between the measurement uncertainties on the redshift and shear estimates. Consequently, most of the sensitivity analyses proposed by {\FC} and {\K} are either trivial or redundant for the SKiLLS results. 

Still, we examine the robustness of the \textit{lens}fit results against some crucial properties by comparing between sub-samples. The basic idea is to split the fiducial simulations into three sub-samples based on a targeted property and examine the consistency of their bias estimates to the fiducial results. These sub-samples contain roughly equal numbers of measured objects while covering different ranges of the targeted property. After applying the overall shear correction from the whole sample to the sub-samples, we calculate their residual biases to quantify the impact of the variations of the targeted property. We note that the estimated residuals are not systematic biases in our fiducial results, but they indicate the robustness of the shape measurement algorithm against the tested properties. Ideally, if the simulations fully match the data in the distributions of the targeted property, we would still expect an accurate bias estimate even if the estimated residuals are large. For that account, the estimated residuals are conservative upper limits of the systematic biases in our results.

Figure~\ref{fig:dmSub} shows the estimated residuals for the variations in four critical properties of the simulated images: the PSF ellipticity, PSF size, background noise level in $r$-band images, and stellar density. It indicates that our fiducial results are robust within the nominal error budgets, considering the shifts shown in the plots are the upper limits of possible deviations. 

\subsection{Impact of the input galaxy morphology}
\label{Sec:sensi3}

We learned the galaxy morphology from \citet{Griffith2012ApJS} based on S\'ersic models fitted to the HST observations. We have shown that our copula-based learning algorithm captures the properties of the reference sample (see Sect.~\ref{Sec:inputGalShape}). However, the reference sample itself contains measurement errors. This section examines how sensitive the \textit{lens}fit measurements are to the changes in the input galaxy morphology.

We focus on the three morphological parameters used to describe the S\'ersic profile: the half-light radius, axis ratio and S\'ersic index. To get some indication of the overall accuracy of the reference sample, we first checked the fitting uncertainties. We found that the median relative uncertainties for these three parameters are $\lesssim 5\%$, $\lesssim 5\%$ and $\lesssim 10\%$, respectively. We took these values (quoted as $\sigma$ below) as the benchmark for changing the input galaxy morphology. We built new input catalogues by increasing a certain parameter with $1\sigma$, $2\sigma$ and $3\sigma$ each time while keeping the other parameters unchanged. We generated test simulations using these new input catalogues and measured the bias difference with respect to the fiducial simulations.

\begin{figure*}
   \includegraphics[width=\hsize]{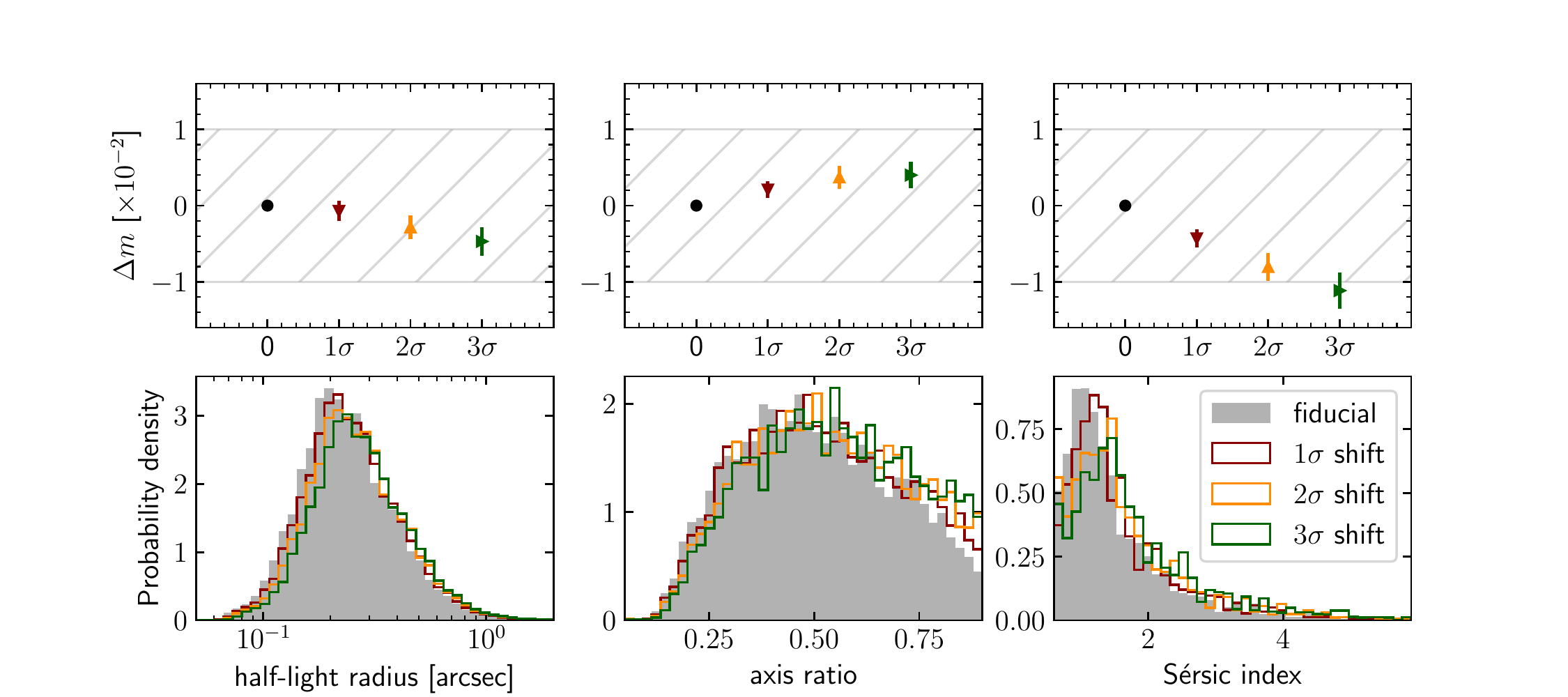}
    \caption{Changes in multiplicative bias when the morphological parameters of all input galaxies are increased by a certain fraction. The top panels show the shifts of the multiplicative bias caused by changing morphological parameter values. The three shift points correspond to the increased factor of $1+1\sigma$ (dark red), $1+2\sigma$ (dark orange) and $1+3\sigma$ (dark green), where $\sigma$ denotes the median relative uncertainties reported by \citet{Griffith2012ApJS} ($\sigma=5\%, 5\%, 10\%$ for the half-light radius, the axis ratio and S\'ersic index, respectively). The hatched regions indicate the nominal $0.01$ error budget for comparison. The bottom panels show the normalised histograms comparing before and after changing morphological parameter values. We note that we shifted all galaxies by the same fraction, resulting in an overall shift of the whole distribution, which corresponds to the most extreme cases, as the uncertainty on the entire distribution is much smaller than on individual values.}
         \label{fig:dmGal}
   \end{figure*}

Figure~\ref{fig:dmGal} presents the test results from $10$ tiles of simulations. We find minor residuals in most cases, with the most significant shifts seen when changing the S\'ersic index. We note that we shifted all galaxies with the same amount of fractions, resulting in an overall shift of the whole distribution, as shown in the bottom panels of Fig.~\ref{fig:dmGal}. Given that the entire distribution's uncertainty is much smaller than the individual measurement uncertainties, we are testing the most extreme cases. Hence, the measured residuals only indicate the sensitivity of \textit{lens}fit towards the input galaxy properties but cannot be seen as systematics in our fiducial results. To achieve tighter requirements for future surveys, we will need a shape measurement method that is less susceptible to the galaxy properties, as the fidelity of image simulations will always be limited by the realism of the input galaxy catalogue. For the upcoming KiDS-Legacy analysis, we will, therefore, also explore an alternative method based on the \textsc{Metacalibration} technique~\citep{Huff2017arXiv170202600H,Sheldon2017ApJ84124S}, which is expected to be more robust against the galaxy properties (Yoon et al., in prep.).

%% file: Sec7_conclusion.tex
An unbiased measurement of the ensemble shear signal is the backbone of reliable precision cosmology with weak lensing surveys. The state-of-the-art shape measurement methods have already reached a percent, if not a sub-percent, level of accuracy. But meanwhile, the statistical powers of weak lensing surveys keep growing, thus putting more stringent requirements on the systematics. Higher-order effects distinct from the shape estimation bias start drawing more and more attention, including the selection bias, PSF modelling errors and shear-interplay bias, which are challenging to eliminate by only improving the shape measurement algorithms. On the other hand, image simulations show promising performance in calibrating these higher-order effects.

This paper presents the third-generation image simulations for the KiDS survey, dubbed SKiLLS, after SCHOo\textit{l}~(\FC) and COllege~(\K). The current image simulations implemented several significant developments to meet the calibration requirement of the KiDS-Legacy analysis, which uses an updated \textit{lens}fit. First and foremost, we simulated the full nine-band images and produced a self-consistent joint shear-redshift mock catalogue. We combined the cosmological simulations with deep imaging observations as input to balance the sample volume and the realism of the galaxy morphology. We also improved the realism of images by varying PSFs between CCDs, including stellar density variations and varying noise levels between pointings. We followed the whole KiDS procedure for the photometric measurements, including the $r$-band detection, PSF Gaussianisation, forced multi-band photometry and photo-$z$ estimates. Given the large volume of simulated galaxies and their realistic photometric properties, the joint shear-redshift mock catalogue not only improves the shear calibration but also benefits the redshift calibration (van den Busch et al., in prep). 

We further explored the impact of blends of galaxies at different redshifts by building realistic shear fields accounting for the redshift and clustering of galaxies. We accounted for the PSF modelling errors by conducting the PSF modelling procedures on the image simulations. Finally, we performed a series of sensitivity tests, including changing the input galaxy properties, demonstrating robustness in the SKiLLS measured calibration values for future lensing studies with KiDS. The final shear calibration results for the updated \textit{lens}fit are summarised in Table~\ref{table:m} and shown in Fig.~\ref{fig:m}. Our statistical uncertainties and sensitivity tests suggest that the shear bias estimated from SKiLLS is robust within the nominal error budget of $0.02$ for the first and sixth tomographic bins and $0.01$ for the remaining bins. Besides, we share some lessons and findings that can be instructive for calibrating future weak lensing surveys.

The fidelity of image simulations relies heavily on the realism of the input galaxy population in terms of photometry, morphology, and clustering. Therefore, the latest image simulations have used high-quality imaging observations as the input. But, this observation-based input is limited by its sample volume and depth, which will soon be inadequate for the next generation of weak lensing surveys. An alternative is to acquire the input galaxy population from cosmological simulations. However, the cosmological simulations still cannot fully reproduce the observed galaxy morphology --- the first-order feature that cannot be compromised in image simulations. In our work, we show that it is possible to keep the merits from both sides by integrating cosmological simulations with high-quality imaging observations. We proposed a copula-based learning algorithm that can mimic and link the observed morphology to synthetic galaxies from cosmological simulations. Our results suggest that this hybrid approach is promising for future image simulations that require a large volume of galaxies.

Recent studies have already pointed out that the shear calibration must consider redshift-related selections, which requires simulating multi-band observations to account for the measurement of photo-$z$s~(e.g. {\K}, \citealt{MacCrann2022MNRAS5093371M}). We further show that multi-band image simulations with a sufficiently large volume of galaxies benefit not only the shear calibration but also the redshift calibration. It allows us to perform the whole procedure for photometric measurements, ensuring realistic photometric properties in the mock catalogue. This end-to-end approach is a significant improvement compared to previous catalogue-level simulations~(e.g. \citealt{Hoyle2018MNRAS.478..592H,Busch2020AA642A200V,DeRose2022PhRvD.105l3520D}). Moreover, image simulations allow us to study the blending effect in redshift estimates, which are otherwise hard to consider at the catalogue level. Given the importance of blending, we argue that unifying the shear and redshift calibrations with multi-band image simulations will be crucial for future high-accuracy tomographic analysis.

\citet{MacCrann2022MNRAS5093371M} recently studied the impact of blended systems that contain galaxies experiencing different shears, an effect we referred to as `shear interplay' throughout the paper. We extended their study by building realistic variable shear fields accounting for both redshift and clustering of galaxies. We also explicitly included the contribution from galaxy-galaxy lensing. Our final results confirmed its overall minor impact on current weak lensing surveys (see Fig.~\ref{fig:m}). However, we measured an evident redshift-bias correlation from our blending-only variable shear simulations, proving the presence of the shear-interplay effect and its non-trivial contributions (see Fig.~\ref{fig:dmZtrue}). We also found that the photo-$z$ outliers showcase the most significant shear interplay, implying a common cause of the shear and redshift biases. A dedicated study is warranted to further explore this coupling in blended systems, as it will soon be relevant for the next-generation weak lensing surveys.

Image simulations usually skip the PSF modelling process, given the PSF validation conducted in data~(see e.g. \citealt{Giblin2021AA645A105G}). Thanks to the realistic SKiLLS images, we can test the impact of the PSF modelling errors by directly running the PSF modelling code in simulated images. By comparing the shear biases measured from runs with and without PSF modelling, we identified sub-percent residual biases from the PSF modelling errors. Although this is insignificant for the current requirement, it will concern future weak lensing surveys. Therefore, we stress the importance of improving or including the PSF modelling algorithm in image simulations for future surveys.

Finally, we tested the sensitivity to the properties of the input galaxy population. By changing the input values of morphological parameters, we found that our current fiducial shape measurement method, \textit{lens}fit, is sensitive to the input galaxy shapes but within a tolerable level for KiDS analysis. Still, we will develop an alternative method based on the \textsc{Metacalibration} technique~\citep{Huff2017arXiv170202600H,Sheldon2017ApJ84124S} for KiDS-Legacy analysis, which is more robust against the galaxy properties (Yoon et al., in prep.). It will be essential for future weak lensing surveys to develop such a method that is less sensitive to the galaxy properties, as the image simulations will never fully represent the observed galaxy population given the limits of its input catalogue.

%% file: Sec999_appendix.tex
\section{An empirical modification to the synthetic photometry}
\label{Sec:modifPhoto}

We detail the proposed empirical modification of the \textsc{Shark} photometry in this appendix. It intends to improve the agreement of the magnitude counts between the simulations and observations, which is critical for the redshift and shear calibrations. 

We took the COSMOS2015 catalogue as the benchmark under an implicit assumption that the COSMOS field is representative. The COSMOS2015 catalogue is a near-infrared-selected photometric catalogue containing 30-band photometry, precise photometric redshifts and stellar masses for more than half a million objects~\citep{Laigle2016ApJS}. We note that measurement uncertainties and modelling errors are inevitable for observations, especially for faint objects. Therefore, the COSMOS2015 catalogue cannot, in principle, be treated as the truth. Nevertheless, these uncertainties are tolerable for calibrating a KiDS-like sample. Following this reasoning, we tuned the simulated properties solely based on the COSMOS2015 measurements for the sake of simplicity, but caution any physical interpretation of our modified results. 

\begin{figure}
\FloatBarrier
\centering
\includegraphics[width=\hsize]{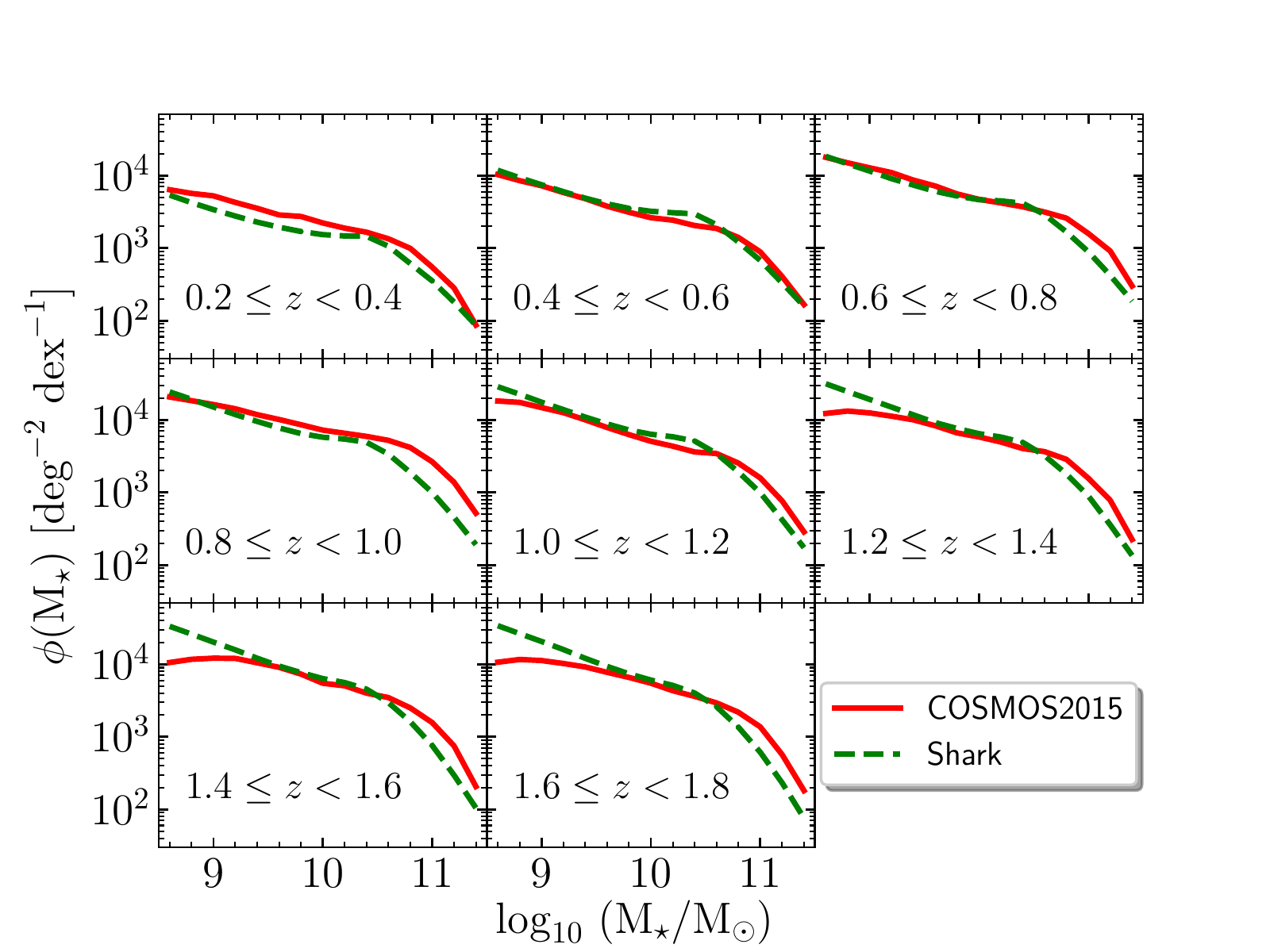}
  \caption{Comparison of the stellar mass functions. For the COSMOS2015 catalogue (red solid lines), we use the median values of the marginalised likelihood distributions. For the \textsc{Shark} catalogue (green dashed lines), we assume that the total stellar mass equals the sum of the stellar masses in the bulge and the disc.}
  \label{fig:mass}
\end{figure}

\begin{figure}
\centering
\includegraphics[width=\hsize]{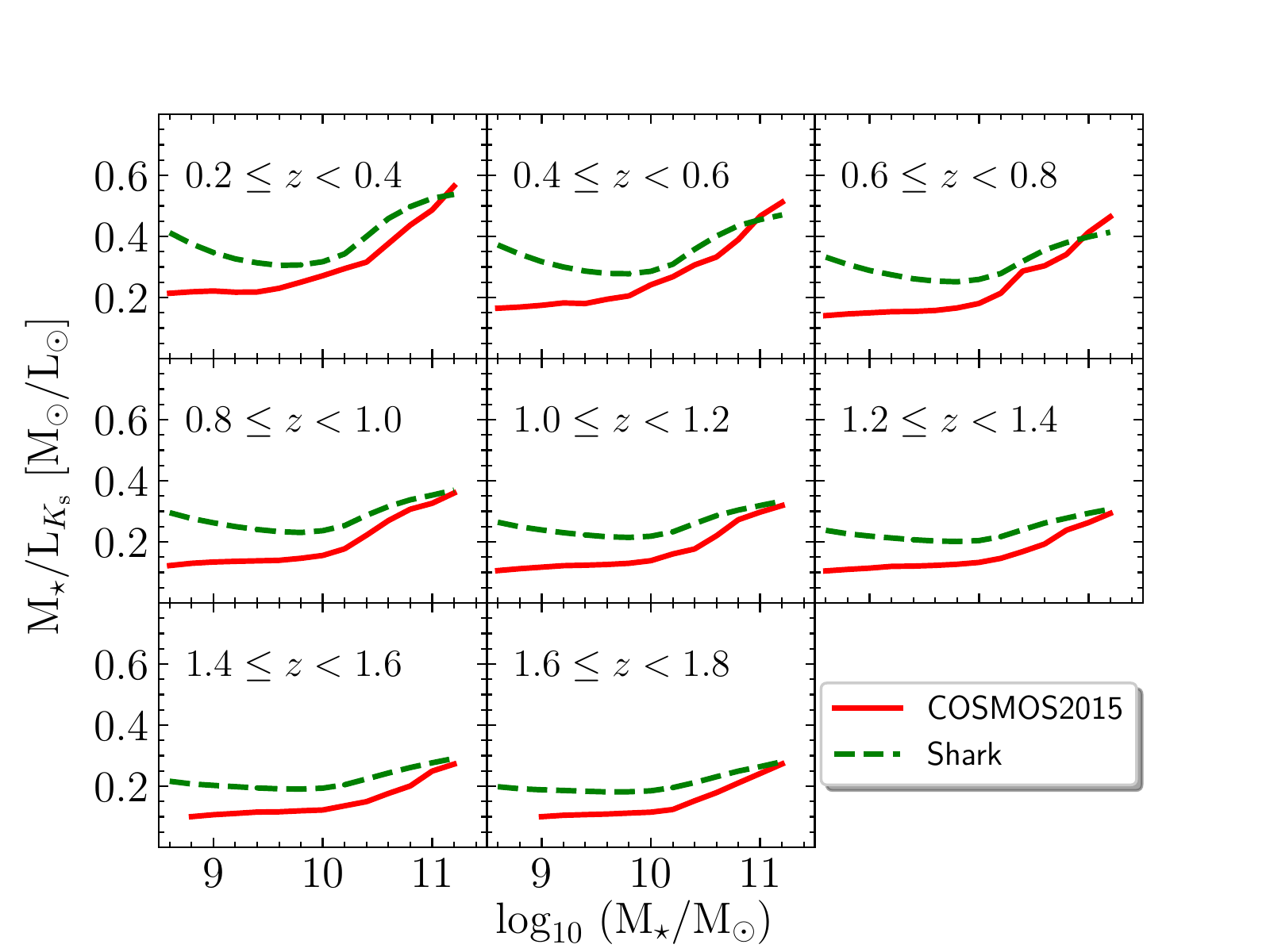}
  \caption{$K_s$-band stellar mass-to-light ratio as a function of stellar mass. The red and green lines correspond to the COSMOS2015 and \textsc{Shark} galaxies, respectively.}
\label{fig:massratio}
\end{figure}

First of all, we must locate the cause of the discrepancy. As the \textsc{Shark} free parameters were tuned using the observed stellar mass functions, we would expect the number density of the \textsc{Shark} galaxies is realistic. This is confirmed by Figure~\ref{fig:mass}, where we see a good agreement of the stellar mass distributions between the data and simulations. As a next step, we inspected the stellar mass-to-light ratio ($\Upsilon_{\star}$), for which took the $K_s$-band photometry as an indicator of the total luminosity as it is least affected by the dust extinctions. Figure~\ref{fig:massratio} shows the comparing results as a function of the stellar mass in several redshift bins. Noticeably, the \textsc{Shark} $\Upsilon_{\star}$ is systematically higher than the COSMOS2015 one, especially in the low stellar mass and low redshift ranges. It can, at least partially, explain the discrepancy seen in the magnitude distributions. Fortunately, this $\Upsilon_{\star}$ difference is easy to calibrate without changing other intrinsic properties, such as the colours, redshifts, and positions.

\begin{figure}
\centering
\includegraphics[width=\hsize]{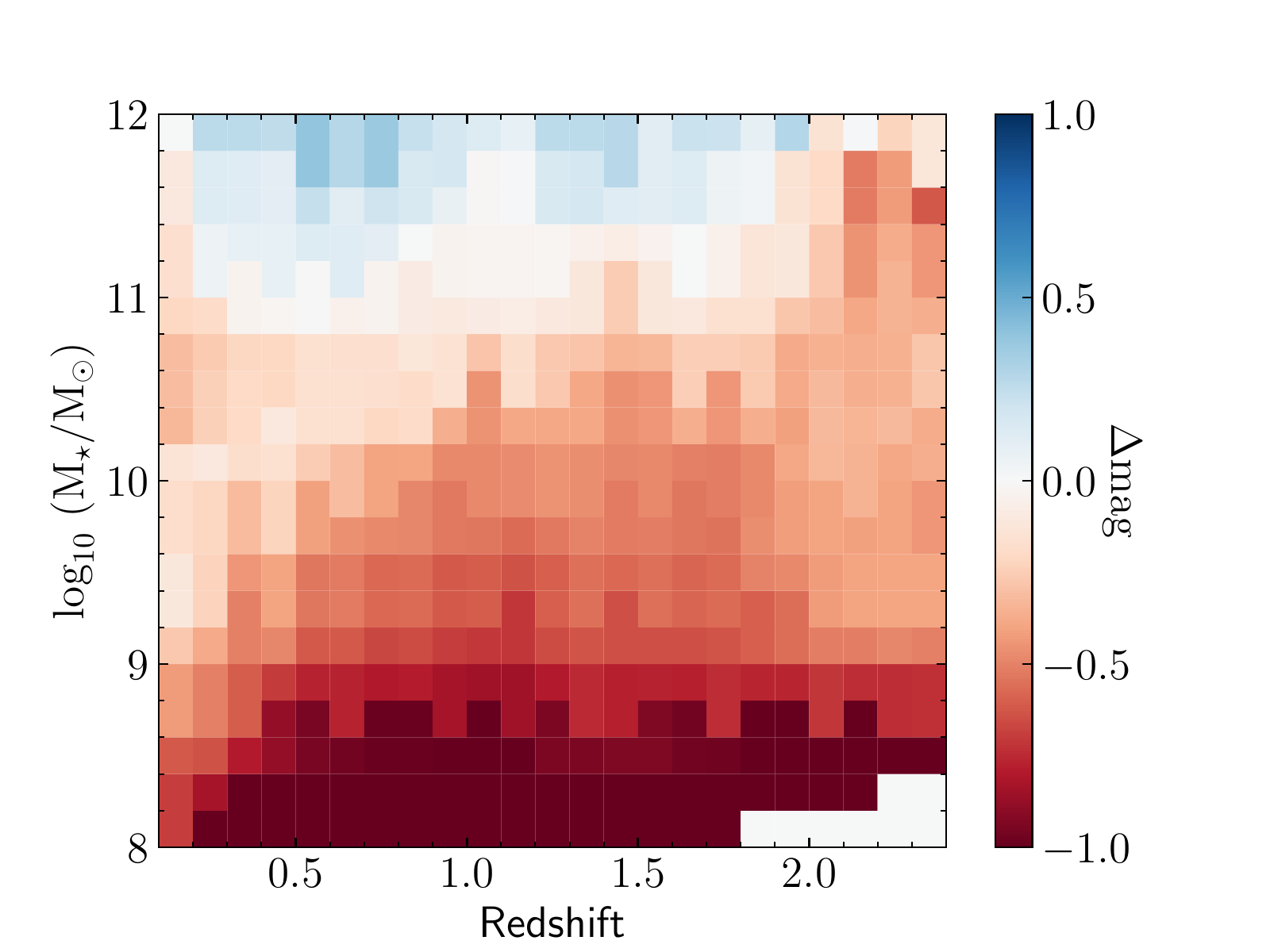}
  \caption{Distribution of the magnitude modification factor $\Delta {\rm mag}$ in the redshift-stellar mass plane. The red colour denotes negative values, whilst the blue colour denotes positive values. The definition of $\Delta {\rm mag}$ is shown in Eq.~\ref{eq:Upsilon}. For a given galaxy, the same $\Delta {\rm mag}$ value is added to the apparent magnitudes for all available bands.}
     \label{fig:Upsilon}
\end{figure}

We, therefore, conducted an empirical modification of the simulated magnitudes to account for the $\Upsilon_{\star}$ difference. We divided \textsc{Shark} and COSMOS2015 galaxies into $24\times23$ evenly spaced small bins based on their redshifts and stellar masses. In each bin, we calculated the median $\Upsilon_{\star}$ for the \textsc{Shark} and COSMOS2015 galaxies, separately. To mitigate the observational uncertainties, we only used the COSMOS2015 galaxies with good stellar mass estimations ($\delta {\rm M}_{\star}<0.15{\rm M}_{\star}$). For bins that lack observations, we extrapolated $\Upsilon_{\star,~{\rm obs}}$ as a function of ${\rm M}_{\star}$ for each redshift slice. After inspecting the general trend, we found a good fit by combining an exponential descending function in the low ${\rm M}_{\star}$ end and a linear ascending function in the high ${\rm M}_{\star}$ end. From these estimates, we constructed a magnitude modification factor $\Delta {\rm mag}$ as
\begin{equation}
\label{eq:Upsilon}
    \Delta {\rm mag} = -2.5\log_{10}\left(\frac{{\rm median}[\Upsilon_{\star,~{\rm \textsc{Shark}}}]}{{\rm median}[\Upsilon_{\star,~{\rm obs}}]}\right)~.
\end{equation}
Figure~\ref{fig:Upsilon} demonstrates the estimated $\Delta {\rm mag}$ values in the 2D redshift-stellar mass plane. Following the difference seen in Fig.~\ref{fig:massratio}, substantial modifications happen in the low mass and low redshift bins. Therefore, the magnitude modification reduces the range of magnitudes of \textsc{Shark} galaxies. We note that the different bands share the same $\Delta {\rm mag}$ values, so the colours of individual galaxies are preserved. 

\section{Modelling multivariate distributions with vine copulas}
\label{Sec:vineCopulas}

We outline some necessary background on the vine-copula modelling in this appendix. For a comprehensive introduction, we refer to \citet{joe2014dependence} and \citet{CzadoClaudia2019ADDw}.

A copula is simply a multivariate cumulative distribution function (CDF) with uniformly distributed margins. The \citet{sklar1959functions} theorem states that any $d$-dimensional CDF $F(\bm{x})$, with univariate margins $F_1(x_1), ..., F_d(x_d)$, can be described as $F(\bm{x})=C_{1,...,d}(F_1(x_1), ..., F_d(x_d))$, where $C_{1,...,d}$ is the corresponding copula function. Therefore, given a joint probability distribution function (PDF) $f(\bm{x})$ with $d$-dimensional variables $\bm{x}=(x_1,..., x_d)$, we can always find a copula density $c_{1,...,d}$ that is the partial differentiation of the copula $C_{1,...,d}$, such that
\begin{equation}\label{eq:copula}
    f(\bm{x}) = c_{1,...,d}(F_1(x_1), ..., F_d(x_d))\cdot f_1(x_1)\cdot\cdot\cdot f_d(x_d)~.
\end{equation}
It means we can divide the modelling of any joint multi-dimensional PDF into two parts: one for the independent distributions of the individual random variables $\{f_i(x_i)\}$, and the other for their mutual dependence captured by the copula density $c_{1,...,d}(F_1(x_1), ..., F_d(x_d))$.

The restriction of the classical copula method is that most of the flexible copula families available in the literature are bivariate, making it tricky to deal with high-dimensional distributions. In this aspect, the vine copula method stands out as an effective approach~\citep{bedford2002vines,aas2009pair}. A vine copula is a graphical model organising a set of bivariate copulas, called pair-copulas. The chain rule states that any PDF $f(\bm{x})$ can be decomposed as
\begin{equation}\label{eq:chain}
    f(\bm{x}) = f(x_d)\cdot f(x_{d-1}|x_d)\cdot f(x_{d-2}|x_{d-1},x_{d})\cdot\cdot\cdot f(x_1|x_2,...,x_{d})~,
\end{equation}
with $f(.|.)$ being the conditional PDF. \citet{aas2009pair} further states that each term in Eq.~(\ref{eq:chain}) can be decomposed into an appropriate pair-copula times a conditional marginal density as described by the following general formula
\begin{equation}\label{eq:conditional}
    f(x|\bm{v}) = c_{xv_j|\bm{v}_{-j}}(F(x|\bm{v}_{-j}), F(v_j|\bm{v}_{-j}))\cdot f(x|\bm{v}_{-j})~,
\end{equation}
where $\bm{v}$ stands for a $d$-dimensional vector, $v_j$ is an arbitrary component of $\bm{v}$, and $\bm{v}_{-j}$ denotes the $v$-vector excluding this component. Therefore, the multiple dependence can be captured by a product of pair-copulas acting on underlying conditional probability distributions. Since the decomposition shown in Eq.~(\ref{eq:chain}) is not unique, there is a significant number of possible pair-copula constructions. These possibilities are organised by the graphical models, that is the vines. 

\section{Transformation of the SDSS filters to the KiDS/VIKING filters}
\label{Sec:FilterTransform}

\begin{figure}
\centering
\includegraphics[width=\hsize]{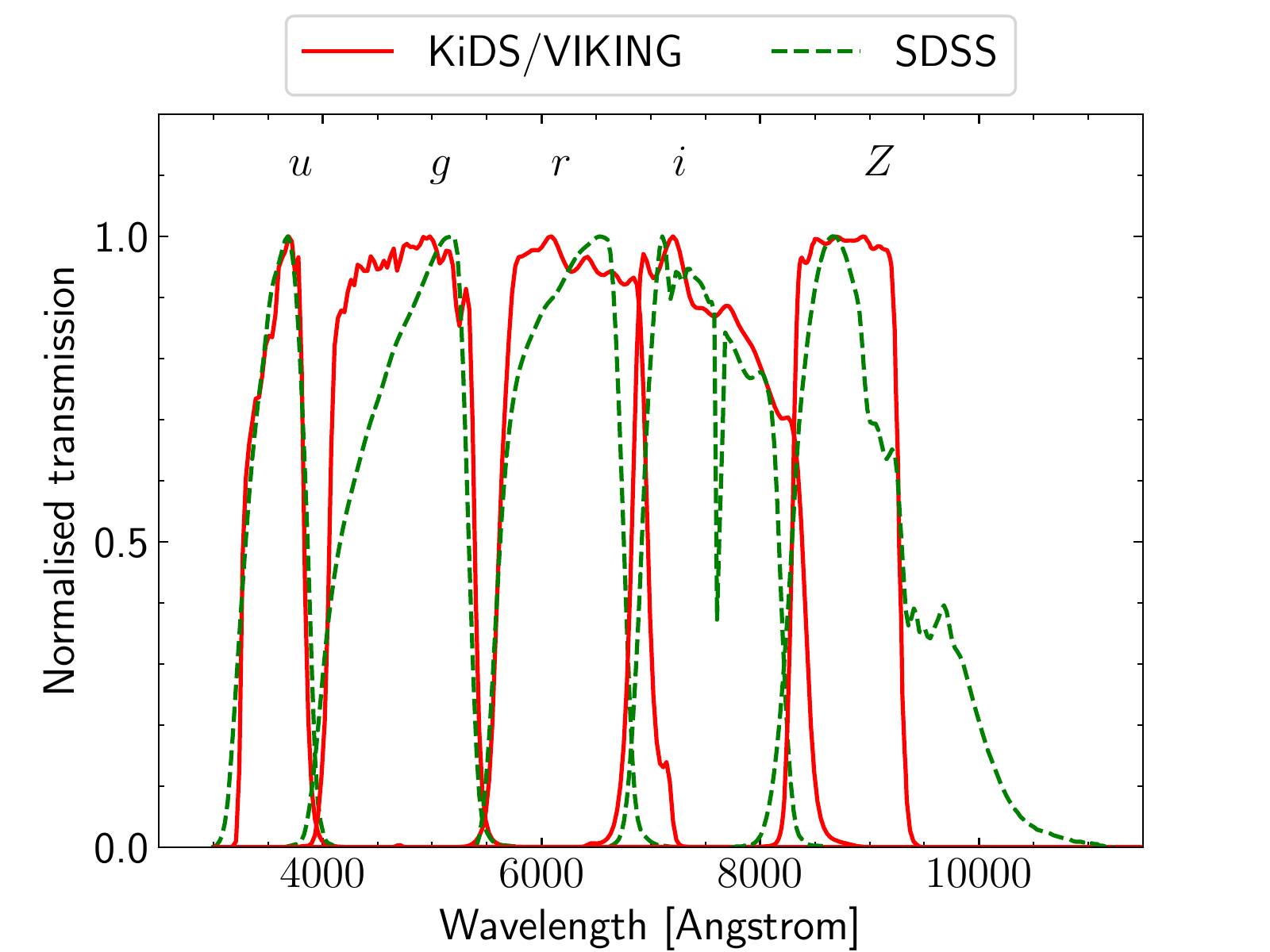}
  \caption{Comparison of the normalised transmission curves of the $ugriZ$ filters in the SDSS photometric system (red solid lines) and the KiDS/VIKING system (green dashed lines).}
     \label{fig:FiterComp}
\end{figure}

This appendix details the transformation of the Sloan Digital Sky Survey (SDSS) photometric system to the KiDS/VIKING system. The SDSS photometric system comprises five colour bands ($u$, $g$, $r$, $i$, $z$) that cover wavelengths ranging from ultra-violet at \num[]{3000}~$\AA$ to near-infrared at \num[]{11000}~$\AA$~\citep{Fukugita1996AJ....111.1748F}, whilst the KiDS/VIKING system contains optical filters ($u$, $g$, $r$, $i$) mounted on the VST OmegaCAM camera~\citep{Kuijken2011Msngr1468K} and near-infrared filters ($Z$, $Y$, $J$, $H$, $K_s$) mounted on the VISTA infrared camera~\citep{Gonz2018MNRAS.474.5459G}. Figure~\ref{fig:FiterComp} compares the filter curves from these two systems. The differences are noticeable, especially for the $Z$ filter, where the KiDS/VIKING system cuts the tail towards long wavelengths. We used the following relation to correct these differences:
\begin{equation}
    \label{eq:filterTransform}
    X_{\rm KiDS/VIKING} = X_{\rm SDSS} + j(z_{\rm true})~(X_{\rm SDSS} - W_{\rm SDSS}) + h(z_{\rm true})~,
\end{equation}
where $X$ corresponds to the target filter, whilst $W$ is another filter, helping to define the colour. Given the superior depth of the $r$-band measurement, we picked it as the $Y$ filter whenever possible. When the $r$ band is the target filter, we chose the $g$ band as the $Y$ filter. The coefficients $j(z_{\rm true})$ and $h(z_{\rm true})$ are correlated with the redshift, for which we took values from the \textsc{ProSpect} web-portal\footnote{\url{https://transformcalc.icrar.org/}}. For the redshift, we used the true redshift from the input \textsc{SURFS}-\textsc{Shark} simulations.

\begin{figure}
\centering
\includegraphics[width=\hsize]{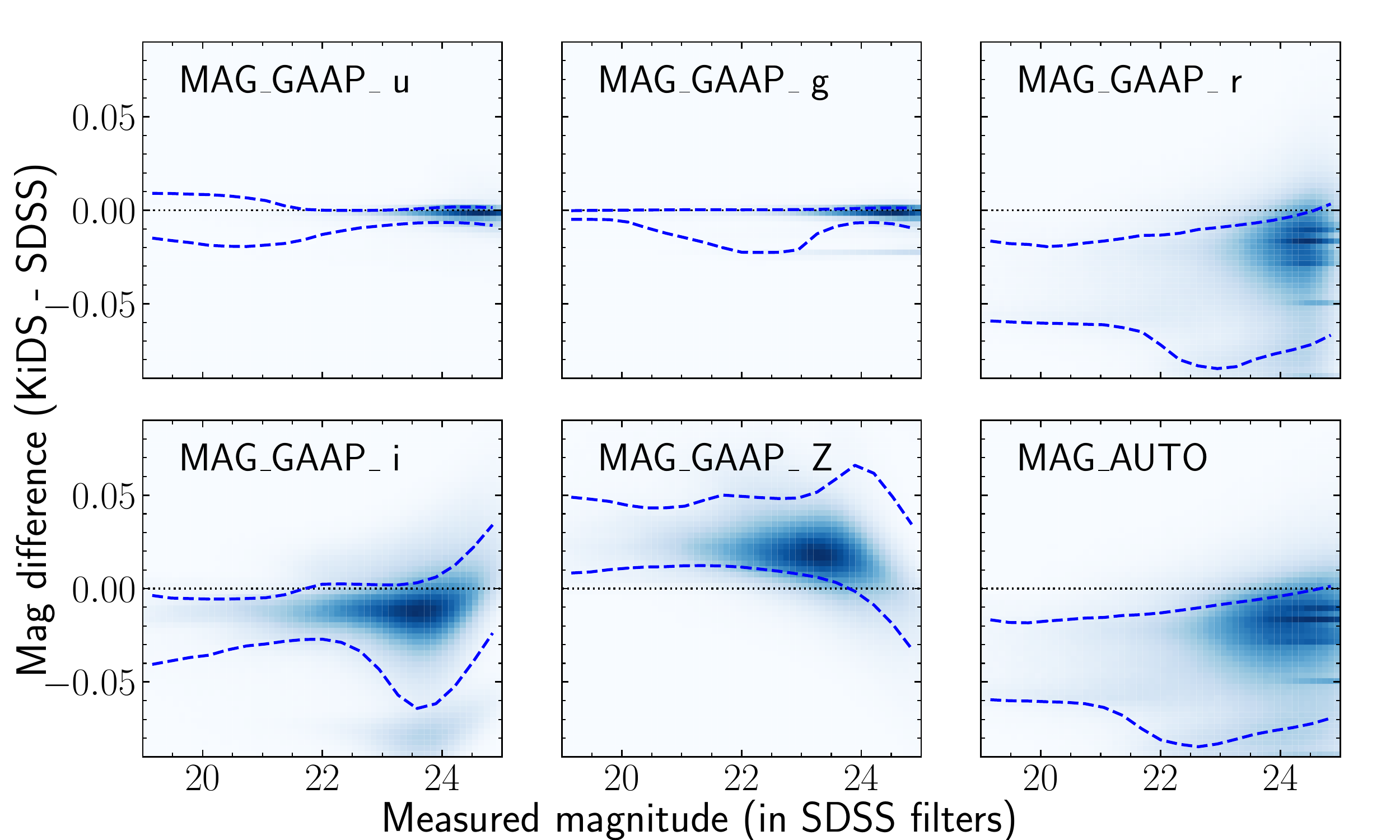}
\caption{Joint distributions of the initially measured magnitude and the magnitude modifications. The dashed lines show the $16$ and $84$ percentiles. The `MAG\_GAAP\_X' magnitudes correspond to those measured by \textsc{GAaP} in the \textsc{Astro}-WISE images, whilst the `MAG\_AUTO' is measured by \textsc{SExtractor} in the $r$-band \textsc{theli} images (see Sect.~\ref{Sec:sim}).}
     \label{fig:magDiff}
\end{figure}

\begin{figure}
\centering
\includegraphics[width=\hsize]{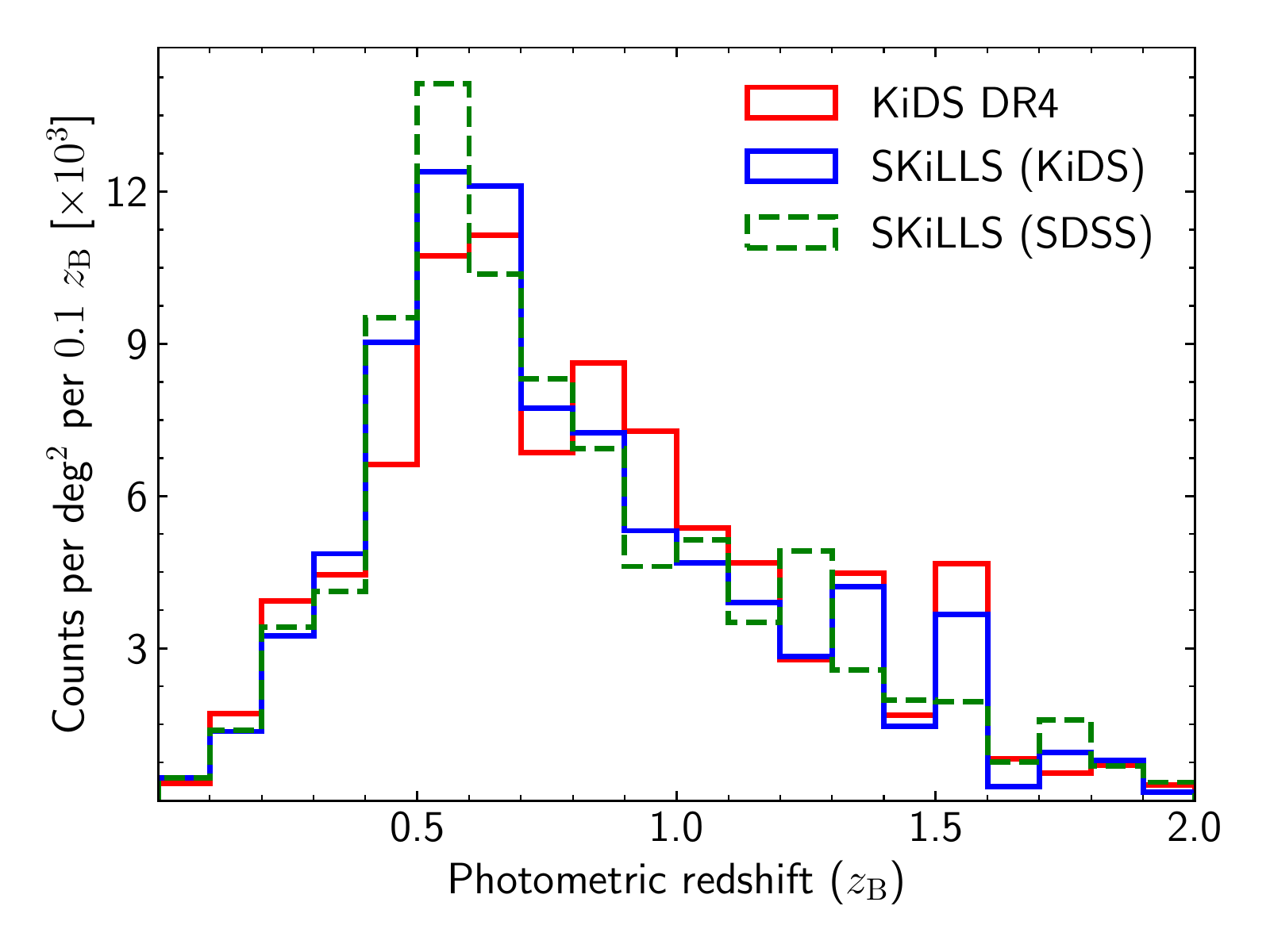}
  \caption{Distributions of the photo-$z$ estimates. The red histogram shows the KiDS-DR4 results. The green histogram is from the initial measurements in the SDSS filters, whilst the blue histogram uses results corrected to the KiDS/VIKING filters. The improvement mostly shows in the peaks around the $z_{\rm B}{\sim} 0.55$ and $1.55$. }
     \label{fig:ZBcomp}
\end{figure}

As for the SKiLLS measured photometry, we need to correct six measurements: the five $u$-, $g$-, $r$-, $i$-, $Z$-band magnitudes measured in the \textsc{Astro}-WISE images (MAG\_GAAP\_X) and the $r$-band magnitudes measured in the \textsc{theli} images (MAG\_AUTO). There is no need to correct the remaining $YJHK_s$ bands as SKiLLS also uses VISTA filters for them. Figure~\ref{fig:magDiff} shows the distributions of the magnitude modification as a function of the initially measured magnitude. The modifications are generally small, especially for the $u$ and $g$ bands. Even for the $r$ and $Z$ bands with the most significant differences, the majority of objects has a modification $\lesssim 0.05$. Accordingly, the changes in the overall magnitude and colour distributions are negligible. Still, we get a better agreement with the data in the photo-$z$ distributions after transforming to the KiDS/VIKING filters, as shown in Fig.~\ref{fig:ZBcomp}. 

\section{Selection criteria for the updated \textit{lens}fit catalogue}
\label{Sec:cuts}

This appendix details all selections we propose to the updated \textit{lens}fit shear catalogue. Most of the selection criteria were taken from earlier KiDS analyses, documented in \citet{Hildebrandt2017MNRAS.465.1454H}. These include:
\begin{enumerate}
    \item Several \textit{lens}fit $\mathtt{fitclass}$ cuts to discard:
        \begin{enumerate}
            \item objects without sufficient data, for example, those fall near the image edge or a defect ($\mathtt{fitclass}=-1$),
            \item objects classified as duplicates ($\mathtt{fitclass}=-10$),
            \item objects poorly fitted by the given bulge plus disc galaxy model ($\mathtt{fitclass}=-4$),
            \item objects identified as stars and star-like point sources ($\mathtt{fitclass}=1$ and $2$),
            \item objects whose fitted centroid is more than $4$ pixels away from the input centroid ($\mathtt{fitclass}=-7$),
            \item objects that are unmeasurable, usually because of being too faint ($\mathtt{fitclass}=-3$).
        \end{enumerate}
    \item A magnitude cut to remove bright objects ($\mathtt{MAG\_AUTO}>20$).
    \item A contamination radius cut to mitigate blending effects ($\mathtt{contamination\_radius}>4.25$ pixels)
    \item Removing asteroids based on the object colours ($\mathtt{MAG\_GAAP\_g} - \mathtt{MAG\_GAAP\_r}\leq 1.5$ or $\mathtt{MAG\_GAAP\_i} - \mathtt{MAG\_GAAP\_r}\leq 1.5$).
    \item Removing unresolved binary stars by requiring objects with ellipticity $>0.8$ to have a measured scalelength 
    \begin{equation*}
        \geq 0.5\times 10^{(24.2-\mathtt{MAG\_GAAP\_r})/3.5} ~{\rm pixels}~.
    \end{equation*}
    \item A non-zero weight cut using the weight bias corrected weight (Sect.~\ref{Sec:weightCorr}).
    \item A resolution cut to remove poorly resolved objects ($\mathcal{R}<0.9$).
\end{enumerate}

\begin{figure}
  \centering
  \includegraphics[width=\hsize]{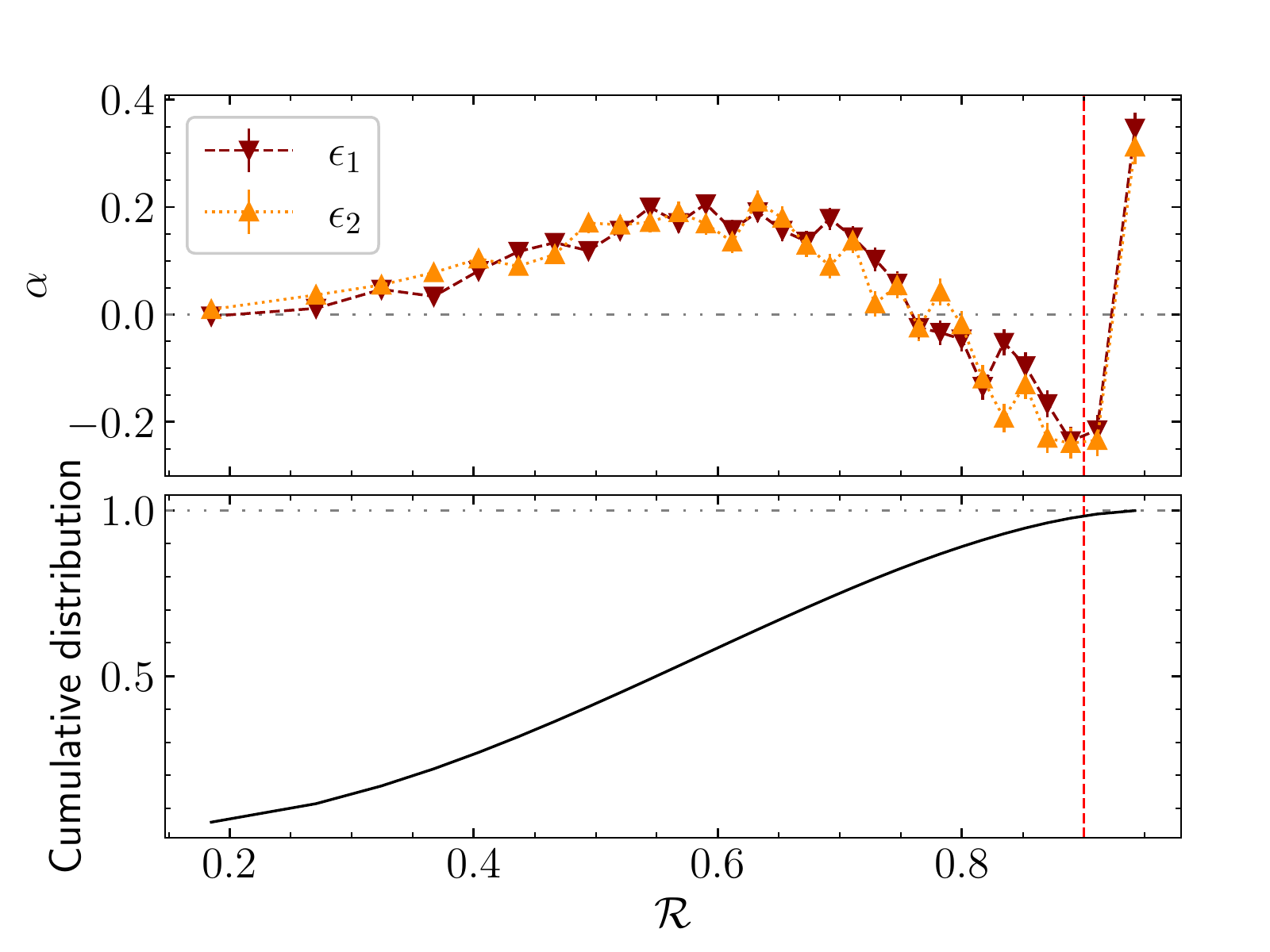}
      \caption{PSF leakage and effective number density as a function of the resolution factor. The upper panel shows the measured PSF leakage, whilst the lower panel shows the effective cumulative distribution. The resolution factor $\mathcal{R}$ is defined in Eq.~(\ref{eq:resolution}), and the PSF leakage factor $\alpha$ is measured from the linear regression with Eq.~(\ref{eq:alpha}). We perform the measurement to the weighted average ellipticity $\epsilon_1$ (dark-red down-pointing triangle) and $\epsilon_2$ (dark-orange up-pointing triangle) using the \textit{lens}fit measurements before the correction of PSF contamination. The vertical red dashed line indicates the proposed resolution cut of $\mathcal{R}<0.9$. The loss of effective number density due to this resolution cut is ${\sim}2$ per cent.}
         \label{fig:Rcut}
  \end{figure}

The resolution cut is a new criterion proposed in this work. When developing our empirical correction method for the PSF contamination (Sect.~\ref{Sec:lensfitUpdate}), we noticed that objects with poor resolution contain very high PSF leakages, as demonstrated in Fig.~\ref{fig:Rcut}. These poor-resolution outliers contribute little to the effective number density but introduce significant bias. So we propose a new selection using the resolution factor defined in Eq.~(\ref{eq:resolution}). We found the proposed cut of $\mathcal{R}<0.9$ can remove most outliers while only decreasing the effective number density by ${\sim}2$ per cent.

\section{Building the variable shear field}
\label{Sec:Zshear}

In this appendix, we detail the creation of a realistic shear field accounting for the shear dependence on the redshift and clustering of galaxies. We considered the two main contributions to the weak lensing signals: the cosmic shear from the large-scale structure, and the tangential shear from the foreground objects (also known as the galaxy-galaxy lensing effect). 

We split the blending-only sample into two classes based on their relative line-of-sight distances to their brightest neighbours. Those more distant than their brightest neighbours are referred as the background galaxies, whilst the remaining are the foreground galaxies. This classification is necessary to quantify the shear correlations within the blended systems. We found a roughly equal number of foreground and background galaxies in our sample.

For the cosmic shear effect, we learned it from the galaxy lensing mocks associated with the MICE Grand Challenge (MICE-GC) simulation~\citep{Fosalba2015MNRAS4471319F}. The MICE-GC simulation is a large volume $N$-body light-cone simulation developed by the Marenostrum Institut de Ci\`{e}ncies de l’Espai (MICE) collaboration~\citep{Fosalba2015MNRAS4482987F}. It contains ${\sim}6.9\times10^{10}$ dark matter particles with a mass of ${\sim}2.9\times10^{10}~h^{-1}{\rm M_{\sun}}$ and a softening length of $50~h^{-1}{\rm kpc}$, in a box of $3072~h^{-1}{\rm Mpc}$ aside. The simulation starts at $z_{\rm i}=100$ and produces the light-cone in 265 steps from $z=1.4$ to $0$. It builds halo catalogues using the Friends-of-Friends algorithm~\citep{Crocce2015MNRAS4531513C}, and subsequently populates galaxies using halo occupation distribution recipes along with the subhalo abundance matching technique~\citep{Carretero2015MNRAS447646C}. The construction of all-sky lensing maps follows the Onion Universe approach, which reaches a sub-arcminute spatial resolution up to $z=1.4$~\citep{Fosalba2015MNRAS4471319F}. Here we used the second version of the catalogue, named MICECAT2, from the CosmoHub web-portal~\citep{Carretero2017ehepconfE488C,Tallada2020AC3200391T}\footnote{\url{https://cosmohub.pic.es/}}.

\begin{figure}
  \centering
  \includegraphics[width=\hsize]{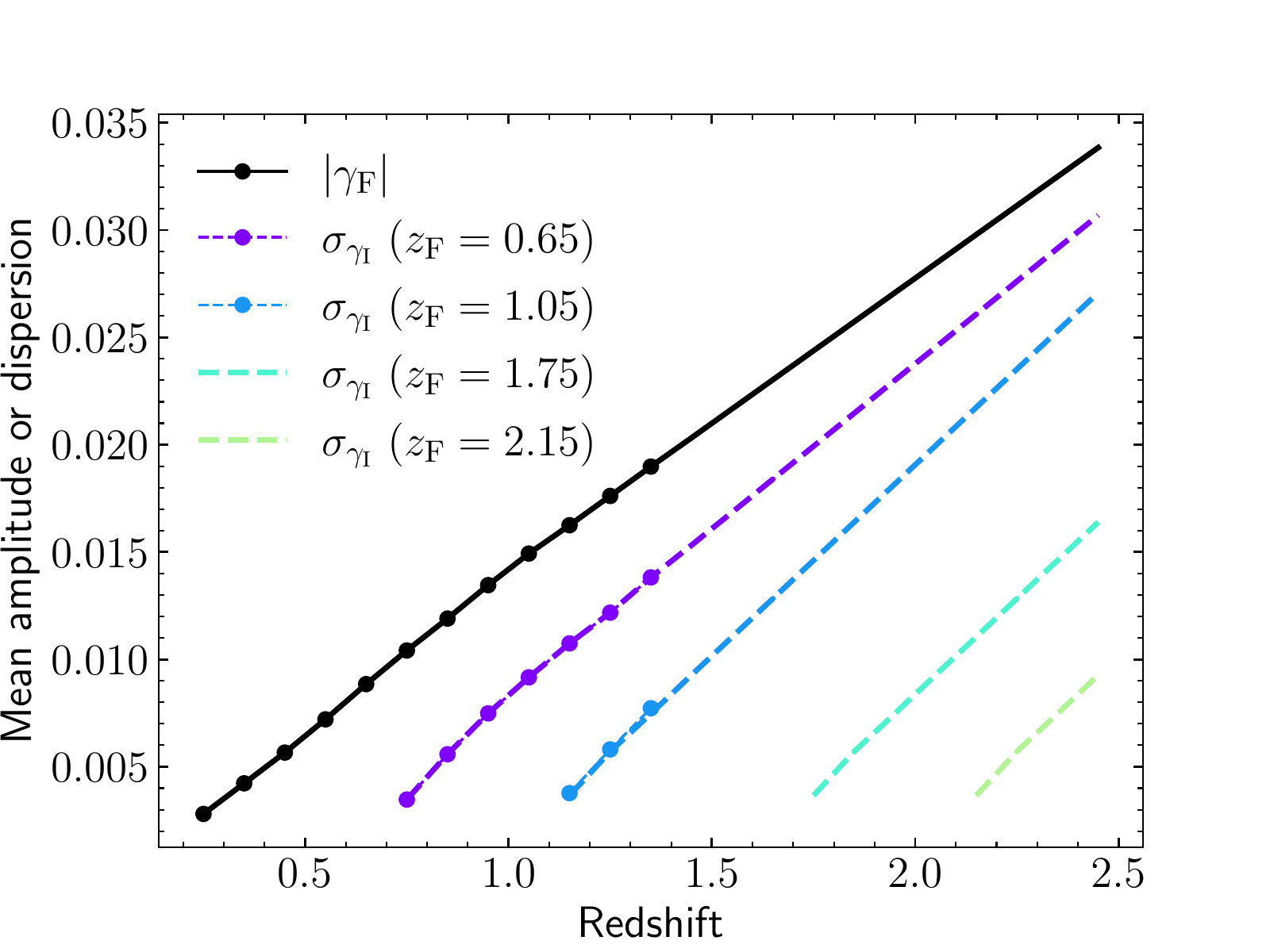}
      \caption{Cosmic shear signals learned from the MICECAT2 (Eq.~\ref{equ:gammaFB}). The black solid line and points indicate the mean amplitude of the ${\bm \gamma}_{\rm F}$ component, whilst the coloured lines and points show the ${\bm \gamma}_{\rm I}$ dispersion for several redshifts of the foreground galaxies. The points are direct measurements from the MICECAT2, while the lines are linear extrapolations.}
         \label{fig:ZshearCS}
  \end{figure}

Following the building of the blending-only sample for SKiLLS, we selected blended objects and classified foreground and background galaxies for MICECAT2 under the same conditions expect for the magnitude cut. We first estimated the relationship between the mean cosmic shear amplitude and redshifts by averaging individual shear values of galaxies in redshift bins defined with a width of $0.1$. These redshift-dependent mean amplitudes are good approximations for cosmic shears experienced by the foreground galaxies. It is more intricate to get proper cosmic shears for the background galaxies. Because of the overlapping line-of-sights of the blended objects, we expect the cosmic shear experienced by the background galaxy (${\bm \gamma}_{\rm B}$) to correlate with that in its neighbour (${\bm \gamma}_{\rm F}$). Based on our tests, the correlation can be described by a linear formula
\begin{equation}\label{equ:gammaFB}
{\bm \gamma}_{\rm B}(z_{\rm B}, z_{\rm F}) = A(z_{\rm B}, z_{\rm F})\cdot{\bm \gamma}_{\rm F} + {\bm \gamma}_{\rm I}(z_{\rm B}, z_{\rm F})~,
\end{equation}
with the scaling factor
\begin{equation}\label{equ:A}
A(z_{\rm B}, z_{\rm F})\equiv\frac{D_{\rm c, B}-0.5D_{\rm c, F}}{D_{\rm c, B}}\cdot\frac{D_{\rm c, F}}{D_{\rm c, F}-0.5D_{\rm c, F}}~,
\end{equation}
and an offset ${\bm \gamma}_{\rm I}(z_{\rm B}, z_{\rm F})=\mathcal{N}\left[0,~\sigma_{\rm I}(z_{\rm B}, z_{\rm F})\right]$ following the Gaussian distribution with a mean of zero and variance depending on redshifts of both galaxies. The $D_{\rm c, B}$ and $D_{\rm c, F}$ denote the comoving distances to the  background galaxy and its neighbour, respectively. The scaling factor $A$ reflects the geometrical relation between the blended objects; whilst the offset ${\bm \gamma}_{\rm I}$ specifies contributions from the intermediate structures between blended galaxies. We estimated the redshift-dependent variance of ${\bm \gamma}_{\rm I}$ again from MICECAT2 by measuring the dispersion of ${\bm \gamma}_{\rm B}-A\cdot{\bm \gamma}_{\rm F}$ in each redshift bin. Because the MICECAT2 stops at $z=1.4$, we linearly extrapolated measured values to $z=2.5$, which is the limit of SKiLLS. Figure~\ref{fig:ZshearCS} shows the learned cosmic shear as a function of redshift. The black solid line indicates the mean amplitude of the ${\bm \gamma}_{\rm F}$ component; whilst the coloured lines present the dispersion of the ${\bm \gamma}_{\rm I}$ component. It illustrates that the linear extrapolation captures the general trends towards the high redshift for both components.

We note that MICECAT2 assumes a $\Lambda$CDM cosmology with parameters from the Wilkinson Microwave Anisotropy Probe five-year data (WMAP5, \citealt{Dunkley2009ApJS..180..306D}), whilst our base \textsc{SURFS}-\textsc{Shark} simulation uses cosmological parameters from \citet{Planck2016AA}. Therefore, the cosmic shear field we learned from MICECAT2 does not necessarily match the galaxy mock we are using. But, since the current calibration still adopts one-point statistics (see Eq.~\ref{eq:shearbias}), our calibration results are robust against detailed galaxy populations or underlying cosmologies and even more so to the higher-order correlation between galaxy populations and cosmology. We defer the proper treatment using a ray-tracing approach with consistent properties from the underlying cosmological simulations to future studies.

Besides the cosmic shear, a background galaxy also suffers from the tangential shear induced by the host dark matter halo of its neighbour. We calculated this effect analytically by assuming Navarro-Frenk-White (NFW) density profiles for dark matter halos presented in the \textsc{SURFS}-\textsc{Shark} simulation. The NFW profile, proposed by \citet{Navarro1995MNRAS275720N}, is the most popular analytical model for dark matter halos, given its ability to describe the radial matter distribution of dark matter halos over a wide range of masses~\citep{Navarro1996ApJ462563N,Navarro1997ApJ490493N}. Its mass density is described by the formula
\begin{equation}\label{equ:nfw}
    \rho(r) = \frac{\rho_{\rm cr}~\delta_c}{(r/r_{\rm s})(1+r/r_{\rm s})^2}~,
\end{equation}
where $\delta_c$ and $r_{\rm s}$ are two free parameters known as the characteristic overdensity and the scale radius, respectively. We set the normalisation to the critical density at the redshift of the halo $\rho_{\rm cr}\equiv 3H^2(z)/(8\pi G)$ with $H(z)$ the Hubble parameter at that same redshift and $G$ the gravitational constant. With the definition of the virial radius, $r_{\rm 200c}$, the radius inside which the mean mass density of the halo equals $200\rho_{\rm cr}$, we can construct a so-called concentration parameter $c\equiv r_{\rm 200c}/r_{\rm s}$ and relate it to $\delta_c$ through 
\begin{equation}\label{equ:deltaC}
    \delta_c = \frac{200}{3}\frac{c^3}{\ln(1+c)-c/(1+c)}~.
\end{equation}

In practice, we used \texttt{mvir\_subhalo}, the virial mass of the subhalo from the \textsc{SURFS}-\textsc{Shark} simulation\footnote{\url{https://shark-sam.readthedocs.io/en/latest/output_files.html}}, to calculate the virial radius for each lens. For the concentration parameter, we adopted the concentration–mass relation from \citet{Duffy2008MNRAS390L64D}
\begin{equation}\label{equ:c}
    c = 7.85~\left(\frac{M_{\rm vir}}{2\times 10^{12}~h^{-1}{\rm M}_{\sun}}\right)^{-0.081}(1+z)^{-0.71}~.
\end{equation}
We note that Eq.~(\ref{equ:c}) is estimated from $N$-body simulations based on a WMAP5 cosmology~\citep{Komatsu2009ApJS..180..330K}, which has slightly different parameter values from the \citet{Planck2016AA} cosmology used by the \textsc{SURFS} simulations. Nevertheless, the weak-lensing shear amplitude is dominated by the enclosed mass of the lens but has minor sensitivity to the concentration (e.g., \citealt{Viola2015MNRAS.452.3529V}). Therefore, we ignored any potential WMAP5-to-Planck cosmology correction to Eq.~(\ref{equ:c}).

Recognising the spherically symmetric feature of the NFW profile, we can derive the radial-dependent tangential shear as~\citep{Bartelmann1996AA313697B,Wright2000ApJ53434W}:
\begin{equation}\label{equ:Tshear}
\gamma_{\rm t}(x) = \frac{\rho_{\rm cr}~\delta_c~r_{\rm s}}{\Sigma_{\rm cr}}g(x)~,
\end{equation}
where $x\equiv R_{\rm FB}/r_{\rm s}$ is a dimensionless radial distance factor defined as the ratio of $R_{\rm FB}$, the projected radial separation between the lens and the source, to the scale radius of the lens. The critical surface mass density
\begin{equation}\label{equ:Sigmacr}
    \Sigma_{\rm cr}\equiv \frac{c^2}{4\pi G}\frac{D_{\rm a, B}}{D_{\rm a, F}~D_{\rm a, FB}}
\end{equation}
is a geometric term depending on the angular diameter distances to the source $D_{\rm a, B}$, to the lens $D_{\rm a, F}$ and between the lens and the source $D_{\rm a, FB}$. The radial dependence of the shear is captured by the function $g(x)$ as
\begin{equation*}\label{equ:gx}
\begin{aligned}
g(x) =& ~\frac{4}{x^2}\ln\left(\frac{x}{2}\right)\\
&+\begin{dcases*}
      \frac{2}{1-x^2} + \frac{8-12x^2}{x^2(1-x^2)^{3/2}}~{\rm arctanh}\sqrt{\frac{1-x}{1+x}}~ & ($x<1$) \\
      \frac{10}{3}~ &($x=1$)\\
      \frac{2}{1-x^2}+\frac{12x^2-8}{x^2(x^2-1)^{3/2}}~{\rm arctan}\sqrt{\frac{x-1}{1+x}}~ & ($x>1$)
\end{dcases*}~.
\end{aligned}
\end{equation*}

With all these ingredients in hand, we can now assign galaxy a specific shear value based on its redshift and neighbouring conditions. In summary, those identified as foreground galaxies only contain the redshift-dependent mean amplitude $\gamma_{\rm F}(z_{\rm F})$, whilst the background galaxies combine the cosmic shear from Eq.~(\ref{equ:gammaFB}) and the tangential shear from Eq.~(\ref{equ:Tshear}). This treatment accounts for not only the redshift-shear dependence but also the correlations between the blended objects.

%% file: main.bbl
\begin{thebibliography}{132}
\expandafter\ifx\csname natexlab\endcsname\relax\def\natexlab#1{#1}\fi

\bibitem[{Aas {et~al.}(2009)Aas, Czado, Frigessi, \& Bakken}]{aas2009pair}
Aas, K., Czado, C., Frigessi, A., \& Bakken, H. 2009, Insurance: Mathematics
  and economics, 44, 182

\bibitem[{{Abbott} {et~al.}(2022){Abbott}, {Aguena}, {Alarcon}, {Allam},
  {Alves}, {Amon}, {Andrade-Oliveira}, {Annis}, {Avila}, {Bacon}, {Baxter},
  {Bechtol}, {Becker}, {Bernstein}, {Bhargava}, {Birrer}, {Blazek},
  {Brandao-Souza}, {Bridle}, {Brooks}, {Buckley-Geer}, {Burke}, {Camacho},
  {Campos}, {Carnero Rosell}, {Carrasco Kind}, {Carretero}, {Castander},
  {Cawthon}, {Chang}, {Chen}, {Chen}, {Choi}, {Conselice}, {Cordero},
  {Costanzi}, {Crocce}, {da Costa}, {da Silva Pereira}, {Davis}, {Davis}, {De
  Vicente}, {DeRose}, {Desai}, {Di Valentino}, {Diehl}, {Dietrich}, {Dodelson},
  {Doel}, {Doux}, {Drlica-Wagner}, {Eckert}, {Eifler}, {Elsner}, {Elvin-Poole},
  {Everett}, {Evrard}, {Fang}, {Farahi}, {Fernandez}, {Ferrero}, {Fert{\'e}},
  {Fosalba}, {Friedrich}, {Frieman}, {Garc{\'\i}a-Bellido}, {Gatti},
  {Gaztanaga}, {Gerdes}, {Giannantonio}, {Giannini}, {Gruen}, {Gruendl},
  {Gschwend}, {Gutierrez}, {Harrison}, {Hartley}, {Herner}, {Hinton},
  {Hollowood}, {Honscheid}, {Hoyle}, {Huff}, {Huterer}, {Jain}, {James},
  {Jarvis}, {Jeffrey}, {Jeltema}, {Kovacs}, {Krause}, {Kron}, {Kuehn},
  {Kuropatkin}, {Lahav}, {Leget}, {Lemos}, {Liddle}, {Lidman}, {Lima}, {Lin},
  {MacCrann}, {Maia}, {Marshall}, {Martini}, {McCullough}, {Melchior},
  {Mena-Fern{\'a}ndez}, {Menanteau}, {Miquel}, {Mohr}, {Morgan}, {Muir},
  {Myles}, {Nadathur}, {Navarro-Alsina}, {Nichol}, {Ogando}, {Omori},
  {Palmese}, {Pandey}, {Park}, {Paz-Chinch{\'o}n}, {Petravick}, {Pieres},
  {Plazas Malag{\'o}n}, {Porredon}, {Prat}, {Raveri}, {Rodriguez-Monroy},
  {Rollins}, {Romer}, {Roodman}, {Rosenfeld}, {Ross}, {Rykoff}, {Samuroff},
  {S{\'a}nchez}, {Sanchez}, {Sanchez}, {Sanchez Cid}, {Scarpine}, {Schubnell},
  {Scolnic}, {Secco}, {Serrano}, {Sevilla-Noarbe}, {Sheldon}, {Shin}, {Smith},
  {Soares-Santos}, {Suchyta}, {Swanson}, {Tabbutt}, {Tarle}, {Thomas}, {To},
  {Troja}, {Troxel}, {Tucker}, {Tutusaus}, {Varga}, {Walker}, {Weaverdyck},
  {Wechsler}, {Weller}, {Yanny}, {Yin}, {Zhang}, {Zuntz}, \& {DES
  Collaboration}}]{Abbott2022PhRvD105b3520A}
{Abbott}, T.~M.~C., {Aguena}, M., {Alarcon}, A., {et~al.} 2022, \prd, 105,
  023520

\bibitem[{{Aihara} {et~al.}(2018){Aihara}, {Arimoto}, {Armstrong}, {Arnouts},
  {Bahcall}, {Bickerton}, {Bosch}, {Bundy}, {Capak}, {Chan}, {Chiba}, {Coupon},
  {Egami}, {Enoki}, {Finet}, {Fujimori}, {Fujimoto}, {Furusawa}, {Furusawa},
  {Goto}, {Goulding}, {Greco}, {Greene}, {Gunn}, {Hamana}, {Harikane},
  {Hashimoto}, {Hattori}, {Hayashi}, {Hayashi}, {He{\l}miniak}, {Higuchi},
  {Hikage}, {Ho}, {Hsieh}, {Huang}, {Huang}, {Ikeda}, {Imanishi}, {Inoue},
  {Iwasawa}, {Iwata}, {Jaelani}, {Jian}, {Kamata}, {Karoji}, {Kashikawa},
  {Katayama}, {Kawanomoto}, {Kayo}, {Koda}, {Koike}, {Kojima}, {Komiyama},
  {Konno}, {Koshida}, {Koyama}, {Kusakabe}, {Leauthaud}, {Lee}, {Lin}, {Lin},
  {Lupton}, {Mandelbaum}, {Matsuoka}, {Medezinski}, {Mineo}, {Miyama},
  {Miyatake}, {Miyazaki}, {Momose}, {More}, {More}, {Moritani}, {Moriya},
  {Morokuma}, {Mukae}, {Murata}, {Murayama}, {Nagao}, {Nakata}, {Niida},
  {Niikura}, {Nishizawa}, {Obuchi}, {Oguri}, {Oishi}, {Okabe}, {Okamoto},
  {Okura}, {Ono}, {Onodera}, {Onoue}, {Osato}, {Ouchi}, {Price}, {Pyo}, {Sako},
  {Sawicki}, {Shibuya}, {Shimasaku}, {Shimono}, {Shirasaki}, {Silverman},
  {Simet}, {Speagle}, {Spergel}, {Strauss}, {Sugahara}, {Sugiyama}, {Suto},
  {Suyu}, {Suzuki}, {Tait}, {Takada}, {Takata}, {Tamura}, {Tanaka}, {Tanaka},
  {Tanaka}, {Tanaka}, {Terai}, {Terashima}, {Toba}, {Tominaga}, {Toshikawa},
  {Turner}, {Uchida}, {Uchiyama}, {Umetsu}, {Uraguchi}, {Urata}, {Usuda},
  {Utsumi}, {Wang}, {Wang}, {Wong}, {Yabe}, {Yamada}, {Yamanoi}, {Yasuda},
  {Yeh}, {Yonehara}, \& {Yuma}}]{Aihara2018PASJ70S4A}
{Aihara}, H., {Arimoto}, N., {Armstrong}, R., {et~al.} 2018, \pasj, 70, S4

\bibitem[{{Alarcon} {et~al.}(2021){Alarcon}, {Gaztanaga}, {Eriksen}, {Baugh},
  {Cabayol}, {Casas}, {Carretero}, {Castander}, {De Vicente}, {Fernandez},
  {Garcia-Bellido}, {Hildebrandt}, {Hoekstra}, {Joachimi}, {Manzoni}, {Miquel},
  {Norberg}, {Padilla}, {Renard}, {Sanchez}, {Serrano}, {Sevilla-Noarbe},
  {Siudek}, \& {Tallada-Cresp{\'\i}}}]{Alarcon2021MNRAS}
{Alarcon}, A., {Gaztanaga}, E., {Eriksen}, M., {et~al.} 2021, \mnras, 501, 6103

\bibitem[{{Asgari} {et~al.}(2021){Asgari}, {Lin}, {Joachimi}, {Giblin},
  {Heymans}, {Hildebrandt}, {Kannawadi}, {St{\"o}lzner}, {Tr{\"o}ster}, {van
  den Busch}, {Wright}, {Bilicki}, {Blake}, {de Jong}, {Dvornik}, {Erben},
  {Getman}, {Hoekstra}, {K{\"o}hlinger}, {Kuijken}, {Miller}, {Radovich},
  {Schneider}, {Shan}, \& {Valentijn}}]{Asgari2021AA...645A.104A}
{Asgari}, M., {Lin}, C.-A., {Joachimi}, B., {et~al.} 2021, \aap, 645, A104

\bibitem[{{Bartelmann}(1996)}]{Bartelmann1996AA313697B}
{Bartelmann}, M. 1996, \aap, 313, 697

\bibitem[{{Bartelmann} \& {Schneider}(2001)}]{Bartelmann2001PhR340291B}
{Bartelmann}, M. \& {Schneider}, P. 2001, \physrep, 340, 291

\bibitem[{Bedford \& Cooke(2002)}]{bedford2002vines}
Bedford, T. \& Cooke, R.~M. 2002, The Annals of Statistics, 30, 1031

\bibitem[{{Ben{\'\i}tez}(2000)}]{Benitez2000ApJ}
{Ben{\'\i}tez}, N. 2000, \apj, 536, 571

\bibitem[{{Bernstein} \& {Armstrong}(2014)}]{Bernstein2014MNRAS4381880B}
{Bernstein}, G.~M. \& {Armstrong}, R. 2014, \mnras, 438, 1880

\bibitem[{{Bernstein} \& {Jarvis}(2002)}]{Bernstein2002AJ123583B}
{Bernstein}, G.~M. \& {Jarvis}, M. 2002, \aj, 123, 583

\bibitem[{{Bertin}(2010)}]{Bertin2010asclsoft10068B}
{Bertin}, E. 2010, {SWarp: Resampling and Co-adding FITS Images Together}

\bibitem[{{Bertin} \& {Arnouts}(1996)}]{Bertin1996AAS117393B}
{Bertin}, E. \& {Arnouts}, S. 1996, \aaps, 117, 393

\bibitem[{{Bravo} {et~al.}(2020){Bravo}, {Lagos}, {Robotham}, {Bellstedt}, \&
  {Obreschkow}}]{Bravo2020MNRAS4973026B}
{Bravo}, M., {Lagos}, C. d.~P., {Robotham}, A. S.~G., {Bellstedt}, S., \&
  {Obreschkow}, D. 2020, \mnras, 497, 3026

\bibitem[{{Bridle} {et~al.}(2010){Bridle}, {Balan}, {Bethge}, {Gentile},
  {Harmeling}, {Heymans}, {Hirsch}, {Hosseini}, {Jarvis}, {Kirk}, {Kitching},
  {Kuijken}, {Lewis}, {Paulin-Henriksson}, {Sch{\"o}lkopf}, {Velander},
  {Voigt}, {Witherick}, {Amara}, {Bernstein}, {Courbin}, {Gill}, {Heavens},
  {Mandelbaum}, {Massey}, {Moghaddam}, {Rassat}, {R{\'e}fr{\'e}gier}, {Rhodes},
  {Schrabback}, {Shawe-Taylor}, {Shmakova}, {van Waerbeke}, \&
  {Wittman}}]{Bridle2010MNRAS4052044B}
{Bridle}, S., {Balan}, S.~T., {Bethge}, M., {et~al.} 2010, \mnras, 405, 2044

\bibitem[{{Ca{\~n}as} {et~al.}(2019){Ca{\~n}as}, {Elahi}, {Welker}, {del P
  Lagos}, {Power}, {Dubois}, \& {Pichon}}]{Canas2019MNRAS4822039C}
{Ca{\~n}as}, R., {Elahi}, P.~J., {Welker}, C., {et~al.} 2019, \mnras, 482, 2039

\bibitem[{{Capak}(2004)}]{Capak2004PhDT}
{Capak}, P.~L. 2004, PhD thesis, UNIVERSITY OF HAWAI'I

\bibitem[{{Cappellaro} {et~al.}(2015){Cappellaro}, {Botticella}, {Pignata},
  {Grado}, {Greggio}, {Limatola}, {Vaccari}, {Baruffolo}, {Benetti}, {Bufano},
  {Capaccioli}, {Cascone}, {Covone}, {De Cicco}, {Falocco}, {Della Valle},
  {Jarvis}, {Marchetti}, {Napolitano}, {Paolillo}, {Pastorello}, {Radovich},
  {Schipani}, {Spiro}, {Tomasella}, \&
  {Turatto}}]{Cappellaro2015AA...584A..62C}
{Cappellaro}, E., {Botticella}, M.~T., {Pignata}, G., {et~al.} 2015, \aap, 584,
  A62

\bibitem[{{Carretero} {et~al.}(2015){Carretero}, {Castander}, {Gazta{\~n}aga},
  {Crocce}, \& {Fosalba}}]{Carretero2015MNRAS447646C}
{Carretero}, J., {Castander}, F.~J., {Gazta{\~n}aga}, E., {Crocce}, M., \&
  {Fosalba}, P. 2015, \mnras, 447, 646

\bibitem[{{Carretero} {et~al.}(2017){Carretero}, {Tallada}, {Casals}, {Caubet},
  {Castander}, {Blot}, {Alarc{\'o}n}, {Serrano}, {Fosalba}, {Acosta-Silva},
  {Tonello}, {Torradeflot}, {Eriksen}, {Neissner}, \&
  {Delfino}}]{Carretero2017ehepconfE488C}
{Carretero}, J., {Tallada}, P., {Casals}, J., {et~al.} 2017, in Proceedings of
  the European Physical Society Conference on High Energy Physics. 5-12 July,
  488

\bibitem[{{Chang} {et~al.}(2013){Chang}, {Jarvis}, {Jain}, {Kahn}, {Kirkby},
  {Connolly}, {Krughoff}, {Peng}, \& {Peterson}}]{Chang2013MNRAS4342121C}
{Chang}, C., {Jarvis}, M., {Jain}, B., {et~al.} 2013, \mnras, 434, 2121

\bibitem[{{Chauhan} {et~al.}(2019){Chauhan}, {Lagos}, {Obreschkow}, {Power},
  {Oman}, \& {Elahi}}]{Chauhan2019MNRAS.488.5898C}
{Chauhan}, G., {Lagos}, C. d.~P., {Obreschkow}, D., {et~al.} 2019, \mnras, 488,
  5898

\bibitem[{{Coe} {et~al.}(2006){Coe}, {Ben{\'\i}tez}, {S{\'a}nchez}, {Jee},
  {Bouwens}, \& {Ford}}]{Coe2006AJ132926C}
{Coe}, D., {Ben{\'\i}tez}, N., {S{\'a}nchez}, S.~F., {et~al.} 2006, \aj, 132,
  926

\bibitem[{{Crocce} {et~al.}(2015){Crocce}, {Castander}, {Gazta{\~n}aga},
  {Fosalba}, \& {Carretero}}]{Crocce2015MNRAS4531513C}
{Crocce}, M., {Castander}, F.~J., {Gazta{\~n}aga}, E., {Fosalba}, P., \&
  {Carretero}, J. 2015, \mnras, 453, 1513

\bibitem[{Czado(2019)}]{CzadoClaudia2019ADDw}
Czado, C. 2019, Lecture notes in statistics, Vol. 222, Analyzing Dependent Data
  with Vine Copulas: A Practical Guide with R (Cham: Springer International
  Publishing AG)

\bibitem[{{Czekaj} {et~al.}(2014){Czekaj}, {Robin}, {Figueras}, {Luri}, \&
  {Haywood}}]{Czekaj2014AA564A102C}
{Czekaj}, M.~A., {Robin}, A.~C., {Figueras}, F., {Luri}, X., \& {Haywood}, M.
  2014, \aap, 564, A102

\bibitem[{{Damjanov} {et~al.}(2018){Damjanov}, {Zahid}, {Geller}, {Fabricant},
  \& {Hwang}}]{Damjanov2018ApJS}
{Damjanov}, I., {Zahid}, H.~J., {Geller}, M.~J., {Fabricant}, D.~G., \&
  {Hwang}, H.~S. 2018, \apjs, 234, 21

\bibitem[{{Dark Energy Survey Collaboration} {et~al.}(2016){Dark Energy Survey
  Collaboration}, {Abbott}, {Abdalla}, {Aleksi{\'c}}, {Allam}, {Amara},
  {Bacon}, {Balbinot}, {Banerji}, {Bechtol}, {Benoit-L{\'e}vy}, {Bernstein},
  {Bertin}, {Blazek}, {Bonnett}, {Bridle}, {Brooks}, {Brunner}, {Buckley-Geer},
  {Burke}, {Caminha}, {Capozzi}, {Carlsen}, {Carnero-Rosell}, {Carollo},
  {Carrasco-Kind}, {Carretero}, {Castander}, {Clerkin}, {Collett}, {Conselice},
  {Crocce}, {Cunha}, {D'Andrea}, {da Costa}, {Davis}, {Desai}, {Diehl},
  {Dietrich}, {Dodelson}, {Doel}, {Drlica-Wagner}, {Estrada}, {Etherington},
  {Evrard}, {Fabbri}, {Finley}, {Flaugher}, {Foley}, {Fosalba}, {Frieman},
  {Garc{\'\i}a-Bellido}, {Gaztanaga}, {Gerdes}, {Giannantonio}, {Goldstein},
  {Gruen}, {Gruendl}, {Guarnieri}, {Gutierrez}, {Hartley}, {Honscheid}, {Jain},
  {James}, {Jeltema}, {Jouvel}, {Kessler}, {King}, {Kirk}, {Kron}, {Kuehn},
  {Kuropatkin}, {Lahav}, {Li}, {Lima}, {Lin}, {Maia}, {Makler}, {Manera},
  {Maraston}, {Marshall}, {Martini}, {McMahon}, {Melchior}, {Merson}, {Miller},
  {Miquel}, {Mohr}, {Morice-Atkinson}, {Naidoo}, {Neilsen}, {Nichol}, {Nord},
  {Ogando}, {Ostrovski}, {Palmese}, {Papadopoulos}, {Peiris}, {Peoples},
  {Percival}, {Plazas}, {Reed}, {Refregier}, {Romer}, {Roodman}, {Ross},
  {Rozo}, {Rykoff}, {Sadeh}, {Sako}, {S{\'a}nchez}, {Sanchez}, {Santiago},
  {Scarpine}, {Schubnell}, {Sevilla-Noarbe}, {Sheldon}, {Smith}, {Smith},
  {Soares-Santos}, {Sobreira}, {Soumagnac}, {Suchyta}, {Sullivan}, {Swanson},
  {Tarle}, {Thaler}, {Thomas}, {Thomas}, {Tucker}, {Vieira}, {Vikram},
  {Walker}, {Wechsler}, {Weller}, {Wester}, {Whiteway}, {Wilcox}, {Yanny},
  {Zhang}, \& {Zuntz}}]{Dark2016MNRAS4601270D}
{Dark Energy Survey Collaboration}, {Abbott}, T., {Abdalla}, F.~B., {et~al.}
  2016, \mnras, 460, 1270

\bibitem[{{Davies} {et~al.}(2015){Davies}, {Driver}, {Robotham}, {Baldry},
  {Lange}, {Liske}, {Meyer}, {Popping}, {Wilkins}, \&
  {Wright}}]{Davies2015MNRAS}
{Davies}, L.~J.~M., {Driver}, S.~P., {Robotham}, A.~S.~G., {et~al.} 2015,
  \mnras, 447, 1014

\bibitem[{{Davies} {et~al.}(2018){Davies}, {Robotham}, {Driver}, {Lagos},
  {Cortese}, {Mannering}, {Foster}, {Lidman}, {Hashemizadeh}, {Koushan},
  {O'Toole}, {Baldry}, {Bilicki}, {Bland-Hawthorn}, {Bremer}, {Brown},
  {Bryant}, {Catinella}, {Croom}, {Grootes}, {Holwerda}, {Jarvis}, {Maddox},
  {Meyer}, {Moffett}, {Phillipps}, {Taylor}, {Windhorst}, \&
  {Wolf}}]{Davies2018MNRAS}
{Davies}, L.~J.~M., {Robotham}, A.~S.~G., {Driver}, S.~P., {et~al.} 2018,
  \mnras, 480, 768

\bibitem[{{Dawson} {et~al.}(2016){Dawson}, {Schneider}, {Tyson}, \&
  {Jee}}]{Dawson2016ApJ81611D}
{Dawson}, W.~A., {Schneider}, M.~D., {Tyson}, J.~A., \& {Jee}, M.~J. 2016,
  \apj, 816, 11

\bibitem[{{De Cicco} {et~al.}(2019){De Cicco}, {Paolillo}, {Falocco},
  {Poulain}, {Brandt}, {Bauer}, {Vagnetti}, {Longo}, {Grado}, {Ragosta},
  {Botticella}, {Pignata}, {Vaccari}, {Radovich}, {Salvato}, {Covone},
  {Napolitano}, {Marchetti}, \& {Schipani}}]{Cicco2019AA...627A..33D}
{De Cicco}, D., {Paolillo}, M., {Falocco}, S., {et~al.} 2019, \aap, 627, A33

\bibitem[{{de Jong} {et~al.}(2015){de Jong}, {Verdoes Kleijn}, {Boxhoorn},
  {Buddelmeijer}, {Capaccioli}, {Getman}, {Grado}, {Helmich}, {Huang},
  {Irisarri}, {Kuijken}, {La Barbera}, {McFarland}, {Napolitano}, {Radovich},
  {Sikkema}, {Valentijn}, {Begeman}, {Brescia}, {Cavuoti}, {Choi}, {Cordes},
  {Covone}, {Dall'Ora}, {Hildebrandt}, {Longo}, {Nakajima}, {Paolillo},
  {Puddu}, {Rifatto}, {Tortora}, {van Uitert}, {Buddendiek},
  {Harnois-D{\'e}raps}, {Erben}, {Eriksen}, {Heymans}, {Hoekstra}, {Joachimi},
  {Kitching}, {Klaes}, {Koopmans}, {K{\"o}hlinger}, {Roy}, {Sif{\'o}n},
  {Schneider}, {Sutherland}, {Viola}, \& {Vriend}}]{Jong2015AA582A62D}
{de Jong}, J. T.~A., {Verdoes Kleijn}, G.~A., {Boxhoorn}, D.~R., {et~al.} 2015,
  \aap, 582, A62

\bibitem[{{de Jong} {et~al.}(2013){de Jong}, {Verdoes Kleijn}, {Kuijken}, \&
  {Valentijn}}]{Jong2013ExA3525D}
{de Jong}, J. T.~A., {Verdoes Kleijn}, G.~A., {Kuijken}, K.~H., \& {Valentijn},
  E.~A. 2013, Experimental Astronomy, 35, 25

\bibitem[{{DeRose} {et~al.}(2022){DeRose}, {Wechsler}, {Becker}, {Rykoff},
  {Pandey}, {MacCrann}, {Amon}, {Myles}, {Krause}, {Gruen}, {Jain}, {Troxel},
  {Prat}, {Alarcon}, {S{\'a}nchez}, {Blazek}, {Crocce}, {Giannini}, {Gatti},
  {Bernstein}, {Zuntz}, {Dodelson}, {Fang}, {Friedrich}, {Secco},
  {Elvin-Poole}, {Porredon}, {Everett}, {Choi}, {Harrison}, {Cordero},
  {Rodriguez-Monroy}, {McCullough}, {Cawthon}, {Chen}, {Alves},
  {Andrade-Oliveira}, {Bechtol}, {Camacho}, {Campos}, {Rosell}, {Kind},
  {Diehl}, {Drlica-Wagner}, {Eckert}, {Eifler}, {Gruendl}, {Hartley}, {Huang},
  {Huff}, {Kuropatkin}, {Raveri}, {Rosenfeld}, {Ross}, {Sanchez},
  {Sevilla-Noarbe}, {Sheldon}, {Yanny}, {Yin}, {Zhang}, {Fosalba}, {Aguena},
  {Allam}, {Annis}, {Avila}, {Bacon}, {Bhargava}, {Brooks}, {Buckley-Geer},
  {Burke}, {Carretero}, {Castander}, {Chang}, {Costanzi}, {da Costa},
  {Pereira}, {De Vicente}, {Desai}, {Dietrich}, {Doel}, {Evrard}, {Ferrero},
  {Fert{\'e}}, {Flaugher}, {Frieman}, {Garc{\'\i}a-Bellido}, {Gaztanaga},
  {Giannantonio}, {Gschwend}, {Gutierrez}, {Hinton}, {Hollowood}, {Honscheid},
  {Huterer}, {James}, {Kuehn}, {Lahav}, {Lima}, {Maia}, {Marshall}, {Melchior},
  {Menanteau}, {Miquel}, {Mohr}, {Morgan}, {Palmese}, {Paz-Chinch{\'o}n},
  {Petravick}, {Pieres}, {Malag{\'o}n}, {Sanchez}, {Scarpine}, {Serrano},
  {Smith}, {Soares-Santos}, {Suchyta}, {Tarle}, {Thomas}, {To}, {Varga}, \&
  {DES Collaboration}}]{DeRose2022PhRvD.105l3520D}
{DeRose}, J., {Wechsler}, R.~H., {Becker}, M.~R., {et~al.} 2022, \prd, 105,
  123520

\bibitem[{{Duffy} {et~al.}(2008){Duffy}, {Schaye}, {Kay}, \& {Dalla
  Vecchia}}]{Duffy2008MNRAS390L64D}
{Duffy}, A.~R., {Schaye}, J., {Kay}, S.~T., \& {Dalla Vecchia}, C. 2008,
  \mnras, 390, L64

\bibitem[{{Dunkley} {et~al.}(2009){Dunkley}, {Komatsu}, {Nolta}, {Spergel},
  {Larson}, {Hinshaw}, {Page}, {Bennett}, {Gold}, {Jarosik}, {Weiland},
  {Halpern}, {Hill}, {Kogut}, {Limon}, {Meyer}, {Tucker}, {Wollack}, \&
  {Wright}}]{Dunkley2009ApJS..180..306D}
{Dunkley}, J., {Komatsu}, E., {Nolta}, M.~R., {et~al.} 2009, \apjs, 180, 306

\bibitem[{{Edge} {et~al.}(2013){Edge}, {Sutherland}, {Kuijken}, {Driver},
  {McMahon}, {Eales}, \& {Emerson}}]{Edge2013Msngr15432E}
{Edge}, A., {Sutherland}, W., {Kuijken}, K., {et~al.} 2013, The Messenger, 154,
  32

\bibitem[{{Elahi} {et~al.}(2019{\natexlab{a}}){Elahi}, {Ca{\~n}as}, {Poulton},
  {Tobar}, {Willis}, {Lagos}, {Power}, \& {Robotham}}]{Elahi2019PASA3621E}
{Elahi}, P.~J., {Ca{\~n}as}, R., {Poulton}, R. J.~J., {et~al.}
  2019{\natexlab{a}}, \pasa, 36, e021

\bibitem[{{Elahi} {et~al.}(2019{\natexlab{b}}){Elahi}, {Poulton}, {Tobar},
  {Ca{\~n}as}, {Lagos}, {Power}, \& {Robotham}}]{Elahi2019PASA3628E}
{Elahi}, P.~J., {Poulton}, R. J.~J., {Tobar}, R.~J., {et~al.}
  2019{\natexlab{b}}, \pasa, 36, e028

\bibitem[{{Elahi} {et~al.}(2018){Elahi}, {Welker}, {Power}, {Lagos},
  {Robotham}, {Ca{\~n}as}, \& {Poulton}}]{Elahi2018MNRAS}
{Elahi}, P.~J., {Welker}, C., {Power}, C., {et~al.} 2018, \mnras, 475, 5338

\bibitem[{{Erben} {et~al.}(2005){Erben}, {Schirmer}, {Dietrich}, {Cordes},
  {Haberzettl}, {Hetterscheidt}, {Hildebrandt}, {Schmithuesen}, {Schneider},
  {Simon}, {Deul}, {Hook}, {Kaiser}, {Radovich}, {Benoist}, {Nonino}, {Olsen},
  {Prandoni}, {Wichmann}, {Zaggia}, {Bomans}, {Dettmar}, \&
  {Miralles}}]{Erben2005AN326432E}
{Erben}, T., {Schirmer}, M., {Dietrich}, J.~P., {et~al.} 2005, Astronomische
  Nachrichten, 326, 432

\bibitem[{{Euclid Collaboration} {et~al.}(2020){Euclid Collaboration},
  {Guglielmo}, {Saglia}, {Castander}, {Galametz}, {Paltani}, {Bender},
  {Bolzonella}, {Capak}, {Ilbert}, {Masters}, {Stern}, {Andreon}, {Auricchio},
  {Balaguera-Antol{\'\i}nez}, {Baldi}, {Bardelli}, {Biviano}, {Bodendorf},
  {Bonino}, {Bozzo}, {Branchini}, {Brau-Nogue}, {Brescia}, {Burigana},
  {Cabanac}, {Camera}, {Capobianco}, {Cappi}, {Carbone}, {Carretero},
  {Carvalho}, {Casas}, {Casas}, {Castellano}, {Castignani}, {Cavuoti},
  {Cimatti}, {Cledassou}, {Colodro-Conde}, {Congedo}, {Conselice}, {Conversi},
  {Copin}, {Corcione}, {Costille}, {Coupon}, {Courtois}, {Cropper}, {Da Silva},
  {de la Torre}, {Di Ferdinando}, {Dubath}, {Duncan}, {Dupac}, {Dusini},
  {Fabricius}, {Farrens}, {Ferreira}, {Fotopoulou}, {Frailis}, {Franceschi},
  {Fumana}, {Galeotta}, {Garilli}, {Gillis}, {Giocoli}, {Gozaliasl},
  {Graci{\'a}-Carpio}, {Grupp}, {Guzzo}, {Hildebrandt}, {Hoekstra}, {Hormuth},
  {Israel}, {Jahnke}, {Keihanen}, {Kermiche}, {Kilbinger}, {Kirkpatrick},
  {Kitching}, {Kubik}, {Kunz}, {Kurki-Suonio}, {Laureijs}, {Ligori}, {Lilje},
  {Lloro}, {Maino}, {Maiorano}, {Maraston}, {Marggraf}, {Martinet}, {Marulli},
  {Massey}, {Maurogordato}, {Medinaceli}, {Mei}, {Meneghetti}, {Metcalf},
  {Meylan}, {Moresco}, {Moscardini}, {Munari}, {Nakajima}, {Neissner}, {Niemi},
  {Nucita}, {Padilla}, {Pasian}, {Patrizii}, {Pocino}, {Poncet}, {Pozzetti},
  {Raison}, {Renzi}, {Rhodes}, {Riccio}, {Romelli}, {Roncarelli}, {Rossetti},
  {S{\'a}nchez}, {Sapone}, {Schneider}, {Scottez}, {Secroun}, {Serrano},
  {Sirignano}, {Sirri}, {Sureau}, {Tallada-Cresp{\'\i}}, {Tavagnacco},
  {Taylor}, {Tenti}, {Tereno}, {Toledo-Moreo}, {Torradeflot}, {Tramacere},
  {Valenziano}, {Vassallo}, {Wang}, {Welikala}, {Wetzstein}, {Whittaker},
  {Zacchei}, {Zamorani}, {Zoubian}, \& {Zucca}}]{Euclid2020AA}
{Euclid Collaboration}, {Guglielmo}, V., {Saglia}, R., {et~al.} 2020, \aap,
  642, A192

\bibitem[{{Euclid Collaboration} {et~al.}(2019){Euclid Collaboration},
  {Martinet}, {Schrabback}, {Hoekstra}, {Tewes}, {Herbonnet}, {Schneider},
  {Hernandez-Martin}, {Taylor}, {Brinchmann}, {Carvalho}, {Castellano},
  {Congedo}, {Gillis}, {Jullo}, {K{\"u}mmel}, {Ligori}, {Lilje}, {Padilla},
  {Paris}, {Peacock}, {Pilo}, {Pujol}, {Scott}, \&
  {Toledo-Moreo}}]{Martinet2019AA627A59E}
{Euclid Collaboration}, {Martinet}, N., {Schrabback}, T., {et~al.} 2019, \aap,
  627, A59

\bibitem[{{Fenech Conti} {et~al.}(2017){Fenech Conti}, {Herbonnet}, {Hoekstra},
  {Merten}, {Miller}, \& {Viola}}]{Conti2017MNRAS}
{Fenech Conti}, I., {Herbonnet}, R., {Hoekstra}, H., {et~al.} 2017, \mnras,
  467, 1627

\bibitem[{{Fosalba} {et~al.}(2015{\natexlab{a}}){Fosalba}, {Crocce},
  {Gazta{\~n}aga}, \& {Castander}}]{Fosalba2015MNRAS4482987F}
{Fosalba}, P., {Crocce}, M., {Gazta{\~n}aga}, E., \& {Castander}, F.~J.
  2015{\natexlab{a}}, \mnras, 448, 2987

\bibitem[{{Fosalba} {et~al.}(2015{\natexlab{b}}){Fosalba}, {Gazta{\~n}aga},
  {Castander}, \& {Crocce}}]{Fosalba2015MNRAS4471319F}
{Fosalba}, P., {Gazta{\~n}aga}, E., {Castander}, F.~J., \& {Crocce}, M.
  2015{\natexlab{b}}, \mnras, 447, 1319

\bibitem[{{Fukugita} {et~al.}(1996){Fukugita}, {Ichikawa}, {Gunn}, {Doi},
  {Shimasaku}, \& {Schneider}}]{Fukugita1996AJ....111.1748F}
{Fukugita}, M., {Ichikawa}, T., {Gunn}, J.~E., {et~al.} 1996, \aj, 111, 1748

\bibitem[{{Giblin} {et~al.}(2021){Giblin}, {Heymans}, {Asgari}, {Hildebrandt},
  {Hoekstra}, {Joachimi}, {Kannawadi}, {Kuijken}, {Lin}, {Miller},
  {Tr{\"o}ster}, {van den Busch}, {Wright}, {Bilicki}, {Blake}, {de Jong},
  {Dvornik}, {Erben}, {Getman}, {Napolitano}, {Schneider}, {Shan}, \&
  {Valentijn}}]{Giblin2021AA645A105G}
{Giblin}, B., {Heymans}, C., {Asgari}, M., {et~al.} 2021, \aap, 645, A105

\bibitem[{{Girardi} {et~al.}(2005){Girardi}, {Groenewegen}, {Hatziminaoglou},
  \& {da Costa}}]{Girardi2005AA}
{Girardi}, L., {Groenewegen}, M.~A.~T., {Hatziminaoglou}, E., \& {da Costa}, L.
  2005, \aap, 436, 895

\bibitem[{{Gonz{\'a}lez-Fern{\'a}ndez}
  {et~al.}(2018){Gonz{\'a}lez-Fern{\'a}ndez}, {Hodgkin}, {Irwin},
  {Gonz{\'a}lez-Solares}, {Koposov}, {Lewis}, {Emerson}, {Hewett},
  {Yolda{\c{s}}}, \& {Riello}}]{Gonz2018MNRAS.474.5459G}
{Gonz{\'a}lez-Fern{\'a}ndez}, C., {Hodgkin}, S.~T., {Irwin}, M.~J., {et~al.}
  2018, \mnras, 474, 5459

\bibitem[{{Griffith} {et~al.}(2012){Griffith}, {Cooper}, {Newman}, {Moustakas},
  {Stern}, {Comerford}, {Davis}, {Lotz}, {Barden}, {Conselice}, {Capak},
  {Faber}, {Kirkpatrick}, {Koekemoer}, {Koo}, {Noeske}, {Scoville}, {Sheth},
  {Shopbell}, {Willmer}, \& {Weiner}}]{Griffith2012ApJS}
{Griffith}, R.~L., {Cooper}, M.~C., {Newman}, J.~A., {et~al.} 2012, \apjs, 200,
  9

\bibitem[{{Hamana} {et~al.}(2020){Hamana}, {Shirasaki}, {Miyazaki}, {Hikage},
  {Oguri}, {More}, {Armstrong}, {Leauthaud}, {Mandelbaum}, {Miyatake},
  {Nishizawa}, {Simet}, {Takada}, {Aihara}, {Bosch}, {Komiyama}, {Lupton},
  {Murayama}, {Strauss}, \& {Tanaka}}]{Hamana2020PASJ7216H}
{Hamana}, T., {Shirasaki}, M., {Miyazaki}, S., {et~al.} 2020, \pasj, 72, 16

\bibitem[{{Hartlap} {et~al.}(2011){Hartlap}, {Hilbert}, {Schneider}, \&
  {Hildebrandt}}]{Hartlap2011AA528A51H}
{Hartlap}, J., {Hilbert}, S., {Schneider}, P., \& {Hildebrandt}, H. 2011, \aap,
  528, A51

\bibitem[{{Hasinger} {et~al.}(2018){Hasinger}, {Capak}, {Salvato}, {Barger},
  {Cowie}, {Faisst}, {Hemmati}, {Kakazu}, {Kartaltepe}, {Masters}, {Mobasher},
  {Nayyeri}, {Sanders}, {Scoville}, {Suh}, {Steinhardt}, \&
  {Yang}}]{Hasinger2018ApJ}
{Hasinger}, G., {Capak}, P., {Salvato}, M., {et~al.} 2018, \apj, 858, 77

\bibitem[{{Heymans} {et~al.}(2021){Heymans}, {Tr{\"o}ster}, {Asgari}, {Blake},
  {Hildebrandt}, {Joachimi}, {Kuijken}, {Lin}, {S{\'a}nchez}, {van den Busch},
  {Wright}, {Amon}, {Bilicki}, {de Jong}, {Crocce}, {Dvornik}, {Erben},
  {Fortuna}, {Getman}, {Giblin}, {Glazebrook}, {Hoekstra}, {Joudaki},
  {Kannawadi}, {K{\"o}hlinger}, {Lidman}, {Miller}, {Napolitano}, {Parkinson},
  {Schneider}, {Shan}, {Valentijn}, {Verdoes Kleijn}, \&
  {Wolf}}]{Heymans2021AA646A140H}
{Heymans}, C., {Tr{\"o}ster}, T., {Asgari}, M., {et~al.} 2021, \aap, 646, A140

\bibitem[{{Heymans} {et~al.}(2006){Heymans}, {Van Waerbeke}, {Bacon}, {Berge},
  {Bernstein}, {Bertin}, {Bridle}, {Brown}, {Clowe}, {Dahle}, {Erben}, {Gray},
  {Hetterscheidt}, {Hoekstra}, {Hudelot}, {Jarvis}, {Kuijken}, {Margoniner},
  {Massey}, {Mellier}, {Nakajima}, {Refregier}, {Rhodes}, {Schrabback}, \&
  {Wittman}}]{Heymans2006MNRAS3681323H}
{Heymans}, C., {Van Waerbeke}, L., {Bacon}, D., {et~al.} 2006, \mnras, 368,
  1323

\bibitem[{{Heymans} {et~al.}(2012){Heymans}, {Van Waerbeke}, {Miller}, {Erben},
  {Hildebrandt}, {Hoekstra}, {Kitching}, {Mellier}, {Simon}, {Bonnett},
  {Coupon}, {Fu}, {Harnois D{\'e}raps}, {Hudson}, {Kilbinger}, {Kuijken},
  {Rowe}, {Schrabback}, {Semboloni}, {van Uitert}, {Vafaei}, \&
  {Velander}}]{Heymans2012MNRAS427146H}
{Heymans}, C., {Van Waerbeke}, L., {Miller}, L., {et~al.} 2012, \mnras, 427,
  146

\bibitem[{{Hildebrandt} {et~al.}(2016){Hildebrandt}, {Choi}, {Heymans},
  {Blake}, {Erben}, {Miller}, {Nakajima}, {van Waerbeke}, {Viola},
  {Buddendiek}, {Harnois-D{\'e}raps}, {Hojjati}, {Joachimi}, {Joudaki},
  {Kitching}, {Wolf}, {Gwyn}, {Johnson}, {Kuijken}, {Sheikhbahaee}, {Tudorica},
  \& {Yee}}]{Hildebrandt2016MNRAS.463..635H}
{Hildebrandt}, H., {Choi}, A., {Heymans}, C., {et~al.} 2016, \mnras, 463, 635

\bibitem[{{Hildebrandt} {et~al.}(2017){Hildebrandt}, {Viola}, {Heymans},
  {Joudaki}, {Kuijken}, {Blake}, {Erben}, {Joachimi}, {Klaes}, {Miller},
  {Morrison}, {Nakajima}, {Verdoes Kleijn}, {Amon}, {Choi}, {Covone}, {de
  Jong}, {Dvornik}, {Fenech Conti}, {Grado}, {Harnois-D{\'e}raps}, {Herbonnet},
  {Hoekstra}, {K{\"o}hlinger}, {McFarland}, {Mead}, {Merten}, {Napolitano},
  {Peacock}, {Radovich}, {Schneider}, {Simon}, {Valentijn}, {van den Busch},
  {van Uitert}, \& {Van Waerbeke}}]{Hildebrandt2017MNRAS.465.1454H}
{Hildebrandt}, H., {Viola}, M., {Heymans}, C., {et~al.} 2017, \mnras, 465, 1454

\bibitem[{{Hirata} \& {Seljak}(2003)}]{Hirata2003MNRAS343459H}
{Hirata}, C. \& {Seljak}, U. 2003, \mnras, 343, 459

\bibitem[{{Hoekstra} {et~al.}(2015){Hoekstra}, {Herbonnet}, {Muzzin}, {Babul},
  {Mahdavi}, {Viola}, \& {Cacciato}}]{Hoekstra2015MNRAS449685H}
{Hoekstra}, H., {Herbonnet}, R., {Muzzin}, A., {et~al.} 2015, \mnras, 449, 685

\bibitem[{{Hoekstra} \& {Jain}(2008)}]{Hoekstra2008ARNPS5899H}
{Hoekstra}, H. \& {Jain}, B. 2008, Annual Review of Nuclear and Particle
  Science, 58, 99

\bibitem[{{Hoekstra} {et~al.}(2017){Hoekstra}, {Viola}, \&
  {Herbonnet}}]{Hoekstra2017MNRAS4683295H}
{Hoekstra}, H., {Viola}, M., \& {Herbonnet}, R. 2017, \mnras, 468, 3295

\bibitem[{{Hoyle} {et~al.}(2018){Hoyle}, {Gruen}, {Bernstein}, {Rau}, {De
  Vicente}, {Hartley}, {Gaztanaga}, {DeRose}, {Troxel}, {Davis}, {Alarcon},
  {MacCrann}, {Prat}, {S{\'a}nchez}, {Sheldon}, {Wechsler}, {Asorey}, {Becker},
  {Bonnett}, {Carnero Rosell}, {Carollo}, {Carrasco Kind}, {Castander},
  {Cawthon}, {Chang}, {Childress}, {Davis}, {Drlica-Wagner}, {Gatti},
  {Glazebrook}, {Gschwend}, {Hinton}, {Hoormann}, {Kim}, {King}, {Kuehn},
  {Lewis}, {Lidman}, {Lin}, {Macaulay}, {Maia}, {Martini}, {Mudd},
  {M{\"o}ller}, {Nichol}, {Ogando}, {Rollins}, {Roodman}, {Ross}, {Rozo},
  {Rykoff}, {Samuroff}, {Sevilla-Noarbe}, {Sharp}, {Sommer}, {Tucker}, {Uddin},
  {Varga}, {Vielzeuf}, {Yuan}, {Zhang}, {Abbott}, {Abdalla}, {Allam}, {Annis},
  {Bechtol}, {Benoit-L{\'e}vy}, {Bertin}, {Brooks}, {Buckley-Geer}, {Burke},
  {Busha}, {Capozzi}, {Carretero}, {Crocce}, {D'Andrea}, {da Costa}, {DePoy},
  {Desai}, {Diehl}, {Doel}, {Eifler}, {Estrada}, {Evrard}, {Fernandez},
  {Flaugher}, {Fosalba}, {Frieman}, {Garc{\'\i}a-Bellido}, {Gerdes},
  {Giannantonio}, {Goldstein}, {Gruendl}, {Gutierrez}, {Honscheid}, {James},
  {Jarvis}, {Jeltema}, {Johnson}, {Johnson}, {Kirk}, {Krause}, {Kuhlmann},
  {Kuropatkin}, {Lahav}, {Li}, {Lima}, {March}, {Marshall}, {Melchior},
  {Menanteau}, {Miquel}, {Nord}, {O'Neill}, {Plazas}, {Romer}, {Sako},
  {Sanchez}, {Santiago}, {Scarpine}, {Schindler}, {Schubnell}, {Smith},
  {Smith}, {Soares-Santos}, {Sobreira}, {Suchyta}, {Swanson}, {Tarle},
  {Thomas}, {Tucker}, {Vikram}, {Walker}, {Weller}, {Wester}, {Wolf}, {Yanny},
  {Zuntz}, \& {DES Collaboration}}]{Hoyle2018MNRAS.478..592H}
{Hoyle}, B., {Gruen}, D., {Bernstein}, G.~M., {et~al.} 2018, \mnras, 478, 592

\bibitem[{{Hu}(1999)}]{Hu1999ApJ522L21H}
{Hu}, W. 1999, \apjl, 522, L21

\bibitem[{{Huff} \& {Mandelbaum}(2017)}]{Huff2017arXiv170202600H}
{Huff}, E. \& {Mandelbaum}, R. 2017, arXiv e-prints, arXiv:1702.02600

\bibitem[{{Huterer}(2002)}]{Huterer2002PhRvD65f3001H}
{Huterer}, D. 2002, \prd, 65, 063001

\bibitem[{{Ivezi{\'c}} {et~al.}(2019){Ivezi{\'c}}, {Kahn}, {Tyson}, {Abel},
  {Acosta}, {Allsman}, {Alonso}, {AlSayyad}, {Anderson}, {Andrew}, {Angel},
  {Angeli}, {Ansari}, {Antilogus}, {Araujo}, {Armstrong}, {Arndt}, {Astier},
  {Aubourg}, {Auza}, {Axelrod}, {Bard}, {Barr}, {Barrau}, {Bartlett}, {Bauer},
  {Bauman}, {Baumont}, {Bechtol}, {Bechtol}, {Becker}, {Becla}, {Beldica},
  {Bellavia}, {Bianco}, {Biswas}, {Blanc}, {Blazek}, {Blandford}, {Bloom},
  {Bogart}, {Bond}, {Booth}, {Borgland}, {Borne}, {Bosch}, {Boutigny},
  {Brackett}, {Bradshaw}, {Brandt}, {Brown}, {Bullock}, {Burchat}, {Burke},
  {Cagnoli}, {Calabrese}, {Callahan}, {Callen}, {Carlin}, {Carlson},
  {Chandrasekharan}, {Charles-Emerson}, {Chesley}, {Cheu}, {Chiang}, {Chiang},
  {Chirino}, {Chow}, {Ciardi}, {Claver}, {Cohen-Tanugi}, {Cockrum}, {Coles},
  {Connolly}, {Cook}, {Cooray}, {Covey}, {Cribbs}, {Cui}, {Cutri}, {Daly},
  {Daniel}, {Daruich}, {Daubard}, {Daues}, {Dawson}, {Delgado}, {Dellapenna},
  {de Peyster}, {de Val-Borro}, {Digel}, {Doherty}, {Dubois},
  {Dubois-Felsmann}, {Durech}, {Economou}, {Eifler}, {Eracleous}, {Emmons},
  {Fausti Neto}, {Ferguson}, {Figueroa}, {Fisher-Levine}, {Focke}, {Foss},
  {Frank}, {Freemon}, {Gangler}, {Gawiser}, {Geary}, {Gee}, {Geha}, {Gessner},
  {Gibson}, {Gilmore}, {Glanzman}, {Glick}, {Goldina}, {Goldstein}, {Goodenow},
  {Graham}, {Gressler}, {Gris}, {Guy}, {Guyonnet}, {Haller}, {Harris},
  {Hascall}, {Haupt}, {Hernandez}, {Herrmann}, {Hileman}, {Hoblitt}, {Hodgson},
  {Hogan}, {Howard}, {Huang}, {Huffer}, {Ingraham}, {Innes}, {Jacoby}, {Jain},
  {Jammes}, {Jee}, {Jenness}, {Jernigan}, {Jevremovi{\'c}}, {Johns}, {Johnson},
  {Johnson}, {Jones}, {Juramy-Gilles}, {Juri{\'c}}, {Kalirai}, {Kallivayalil},
  {Kalmbach}, {Kantor}, {Karst}, {Kasliwal}, {Kelly}, {Kessler}, {Kinnison},
  {Kirkby}, {Knox}, {Kotov}, {Krabbendam}, {Krughoff}, {Kub{\'a}nek},
  {Kuczewski}, {Kulkarni}, {Ku}, {Kurita}, {Lage}, {Lambert}, {Lange},
  {Langton}, {Le Guillou}, {Levine}, {Liang}, {Lim}, {Lintott}, {Long},
  {Lopez}, {Lotz}, {Lupton}, {Lust}, {MacArthur}, {Mahabal}, {Mandelbaum},
  {Markiewicz}, {Marsh}, {Marshall}, {Marshall}, {May}, {McKercher}, {McQueen},
  {Meyers}, {Migliore}, {Miller}, {Mills}, {Miraval}, {Moeyens}, {Moolekamp},
  {Monet}, {Moniez}, {Monkewitz}, {Montgomery}, {Morrison}, {Mueller},
  {Muller}, {Mu{\~n}oz Arancibia}, {Neill}, {Newbry}, {Nief}, {Nomerotski},
  {Nordby}, {O'Connor}, {Oliver}, {Olivier}, {Olsen}, {O'Mullane}, {Ortiz},
  {Osier}, {Owen}, {Pain}, {Palecek}, {Parejko}, {Parsons}, {Pease},
  {Peterson}, {Peterson}, {Petravick}, {Libby Petrick}, {Petry},
  {Pierfederici}, {Pietrowicz}, {Pike}, {Pinto}, {Plante}, {Plate}, {Plutchak},
  {Price}, {Prouza}, {Radeka}, {Rajagopal}, {Rasmussen}, {Regnault}, {Reil},
  {Reiss}, {Reuter}, {Ridgway}, {Riot}, {Ritz}, {Robinson}, {Roby}, {Roodman},
  {Rosing}, {Roucelle}, {Rumore}, {Russo}, {Saha}, {Sassolas}, {Schalk},
  {Schellart}, {Schindler}, {Schmidt}, {Schneider}, {Schneider}, {Schoening},
  {Schumacher}, {Schwamb}, {Sebag}, {Selvy}, {Sembroski}, {Seppala}, {Serio},
  {Serrano}, {Shaw}, {Shipsey}, {Sick}, {Silvestri}, {Slater}, {Smith},
  {Smith}, {Sobhani}, {Soldahl}, {Storrie-Lombardi}, {Stover}, {Strauss},
  {Street}, {Stubbs}, {Sullivan}, {Sweeney}, {Swinbank}, {Szalay}, {Takacs},
  {Tether}, {Thaler}, {Thayer}, {Thomas}, {Thornton}, {Thukral}, {Tice},
  {Trilling}, {Turri}, {Van Berg}, {Vanden Berk}, {Vetter}, {Virieux},
  {Vucina}, {Wahl}, {Walkowicz}, {Walsh}, {Walter}, {Wang}, {Wang}, {Warner},
  {Wiecha}, {Willman}, {Winters}, {Wittman}, {Wolff}, {Wood-Vasey}, {Wu},
  {Xin}, {Yoachim}, \& {Zhan}}]{Ivezic2019ApJ873111I}
{Ivezi{\'c}}, {\v{Z}}., {Kahn}, S.~M., {Tyson}, J.~A., {et~al.} 2019, \apj,
  873, 111

\bibitem[{{Jarvis} {et~al.}(2016){Jarvis}, {Sheldon}, {Zuntz}, {Kacprzak},
  {Bridle}, {Amara}, {Armstrong}, {Becker}, {Bernstein}, {Bonnett}, {Chang},
  {Das}, {Dietrich}, {Drlica-Wagner}, {Eifler}, {Gangkofner}, {Gruen},
  {Hirsch}, {Huff}, {Jain}, {Kent}, {Kirk}, {MacCrann}, {Melchior}, {Plazas},
  {Refregier}, {Rowe}, {Rykoff}, {Samuroff}, {S{\'a}nchez}, {Suchyta},
  {Troxel}, {Vikram}, {Abbott}, {Abdalla}, {Allam}, {Annis}, {Benoit-L{\'e}vy},
  {Bertin}, {Brooks}, {Buckley-Geer}, {Burke}, {Capozzi}, {Carnero Rosell},
  {Carrasco Kind}, {Carretero}, {Castander}, {Clampitt}, {Crocce}, {Cunha},
  {D'Andrea}, {da Costa}, {DePoy}, {Desai}, {Diehl}, {Doel}, {Fausti Neto},
  {Flaugher}, {Fosalba}, {Frieman}, {Gaztanaga}, {Gerdes}, {Gruendl},
  {Gutierrez}, {Honscheid}, {James}, {Kuehn}, {Kuropatkin}, {Lahav}, {Li},
  {Lima}, {March}, {Martini}, {Miquel}, {Mohr}, {Neilsen}, {Nord}, {Ogando},
  {Reil}, {Romer}, {Roodman}, {Sako}, {Sanchez}, {Scarpine}, {Schubnell},
  {Sevilla-Noarbe}, {Smith}, {Soares-Santos}, {Sobreira}, {Swanson}, {Tarle},
  {Thaler}, {Thomas}, {Walker}, \& {Wechsler}}]{Jarvis2016MNRAS4602245J}
{Jarvis}, M., {Sheldon}, E., {Zuntz}, J., {et~al.} 2016, \mnras, 460, 2245

\bibitem[{Joe(2014)}]{joe2014dependence}
Joe, H. 2014, Dependence modeling with copulas (CRC press)

\bibitem[{{Kaiser}(1992)}]{Kaiser1992ApJ...388..272K}
{Kaiser}, N. 1992, \apj, 388, 272

\bibitem[{{Kaiser}(2000)}]{Kaiser2000ApJ537555K}
{Kaiser}, N. 2000, \apj, 537, 555

\bibitem[{{Kannawadi} {et~al.}(2019){Kannawadi}, {Hoekstra}, {Miller}, {Viola},
  {Fenech Conti}, {Herbonnet}, {Erben}, {Heymans}, {Hildebrandt}, {Kuijken},
  {Vakili}, \& {Wright}}]{Kannawadi2019AA}
{Kannawadi}, A., {Hoekstra}, H., {Miller}, L., {et~al.} 2019, \aap, 624, A92

\bibitem[{{Kilbinger}(2015)}]{Kilbinger2015RPPh78h6901K}
{Kilbinger}, M. 2015, Reports on Progress in Physics, 78, 086901

\bibitem[{{Kitching} {et~al.}(2012){Kitching}, {Balan}, {Bridle}, {Cantale},
  {Courbin}, {Eifler}, {Gentile}, {Gill}, {Harmeling}, {Heymans}, {Hirsch},
  {Honscheid}, {Kacprzak}, {Kirkby}, {Margala}, {Massey}, {Melchior},
  {Nurbaeva}, {Patton}, {Rhodes}, {Rowe}, {Taylor}, {Tewes}, {Viola},
  {Witherick}, {Voigt}, {Young}, \& {Zuntz}}]{Kitching2012MNRAS4233163K}
{Kitching}, T.~D., {Balan}, S.~T., {Bridle}, S., {et~al.} 2012, \mnras, 423,
  3163

\bibitem[{{Kitching} {et~al.}(2008){Kitching}, {Miller}, {Heymans}, {van
  Waerbeke}, \& {Heavens}}]{Kitching2008MNRAS390149K}
{Kitching}, T.~D., {Miller}, L., {Heymans}, C.~E., {van Waerbeke}, L., \&
  {Heavens}, A.~F. 2008, \mnras, 390, 149

\bibitem[{{Komatsu} {et~al.}(2009){Komatsu}, {Dunkley}, {Nolta}, {Bennett},
  {Gold}, {Hinshaw}, {Jarosik}, {Larson}, {Limon}, {Page}, {Spergel},
  {Halpern}, {Hill}, {Kogut}, {Meyer}, {Tucker}, {Weiland}, {Wollack}, \&
  {Wright}}]{Komatsu2009ApJS..180..330K}
{Komatsu}, E., {Dunkley}, J., {Nolta}, M.~R., {et~al.} 2009, \apjs, 180, 330

\bibitem[{{Kuijken}(2011)}]{Kuijken2011Msngr1468K}
{Kuijken}, K. 2011, The Messenger, 146, 8

\bibitem[{{Kuijken} {et~al.}(2019){Kuijken}, {Heymans}, {Dvornik},
  {Hildebrandt}, {de Jong}, {Wright}, {Erben}, {Bilicki}, {Giblin}, {Shan},
  {Getman}, {Grado}, {Hoekstra}, {Miller}, {Napolitano}, {Paolilo}, {Radovich},
  {Schneider}, {Sutherland}, {Tewes}, {Tortora}, {Valentijn}, \& {Verdoes
  Kleijn}}]{Kuijken2019AA625A2K}
{Kuijken}, K., {Heymans}, C., {Dvornik}, A., {et~al.} 2019, \aap, 625, A2

\bibitem[{{Kuijken} {et~al.}(2015){Kuijken}, {Heymans}, {Hildebrandt},
  {Nakajima}, {Erben}, {de Jong}, {Viola}, {Choi}, {Hoekstra}, {Miller}, {van
  Uitert}, {Amon}, {Blake}, {Brouwer}, {Buddendiek}, {Conti}, {Eriksen},
  {Grado}, {Harnois-D{\'e}raps}, {Helmich}, {Herbonnet}, {Irisarri},
  {Kitching}, {Klaes}, {La Barbera}, {Napolitano}, {Radovich}, {Schneider},
  {Sif{\'o}n}, {Sikkema}, {Simon}, {Tudorica}, {Valentijn}, {Verdoes Kleijn},
  \& {van Waerbeke}}]{Kuijken2015MNRAS4543500K}
{Kuijken}, K., {Heymans}, C., {Hildebrandt}, H., {et~al.} 2015, \mnras, 454,
  3500

\bibitem[{{Lagos} {et~al.}(2019){Lagos}, {Robotham}, {Trayford}, {Tobar},
  {Bravo}, {Bellstedt}, {Davies}, {Driver}, {Elahi}, {Obreschkow}, \&
  {Power}}]{Lagos2019MNRAS}
{Lagos}, C. d.~P., {Robotham}, A. S.~G., {Trayford}, J.~W., {et~al.} 2019,
  \mnras, 489, 4196

\bibitem[{{Lagos} {et~al.}(2018){Lagos}, {Tobar}, {Robotham}, {Obreschkow},
  {Mitchell}, {Power}, \& {Elahi}}]{Lagos2018MNRAS}
{Lagos}, C. d.~P., {Tobar}, R.~J., {Robotham}, A. S.~G., {et~al.} 2018, \mnras,
  481, 3573

\bibitem[{{Laigle} {et~al.}(2016){Laigle}, {McCracken}, {Ilbert}, {Hsieh},
  {Davidzon}, {Capak}, {Hasinger}, {Silverman}, {Pichon}, {Coupon}, {Aussel},
  {Le Borgne}, {Caputi}, {Cassata}, {Chang}, {Civano}, {Dunlop}, {Fynbo},
  {Kartaltepe}, {Koekemoer}, {Le F{\`e}vre}, {Le Floc'h}, {Leauthaud}, {Lilly},
  {Lin}, {Marchesi}, {Milvang-Jensen}, {Salvato}, {Sanders}, {Scoville},
  {Smolcic}, {Stockmann}, {Taniguchi}, {Tasca}, {Toft}, {Vaccari}, \&
  {Zabl}}]{Laigle2016ApJS}
{Laigle}, C., {McCracken}, H.~J., {Ilbert}, O., {et~al.} 2016, \apjs, 224, 24

\bibitem[{{Lange} {et~al.}(2016){Lange}, {Moffett}, {Driver}, {Robotham},
  {Lagos}, {Kelvin}, {Conselice}, {Margalef-Bentabol}, {Alpaslan}, {Baldry},
  {Bland-Hawthorn}, {Bremer}, {Brough}, {Cluver}, {Colless}, {Davies},
  {H{\"a}u{\ss}ler}, {Holwerda}, {Hopkins}, {Kafle}, {Kennedy}, {Liske},
  {Phillipps}, {Popescu}, {Taylor}, {Tuffs}, {van Kampen}, \&
  {Wright}}]{Lange2016MNRAS4621470L}
{Lange}, R., {Moffett}, A.~J., {Driver}, S.~P., {et~al.} 2016, \mnras, 462,
  1470

\bibitem[{{Laureijs} {et~al.}(2011){Laureijs}, {Amiaux}, {Arduini},
  {Augu{\`e}res}, {Brinchmann}, {Cole}, {Cropper}, {Dabin}, {Duvet}, {Ealet},
  {Garilli}, {Gondoin}, {Guzzo}, {Hoar}, {Hoekstra}, {Holmes}, {Kitching},
  {Maciaszek}, {Mellier}, {Pasian}, {Percival}, {Rhodes}, {Saavedra Criado},
  {Sauvage}, {Scaramella}, {Valenziano}, {Warren}, {Bender}, {Castander},
  {Cimatti}, {Le F{\`e}vre}, {Kurki-Suonio}, {Levi}, {Lilje}, {Meylan},
  {Nichol}, {Pedersen}, {Popa}, {Rebolo Lopez}, {Rix}, {Rottgering},
  {Zeilinger}, {Grupp}, {Hudelot}, {Massey}, {Meneghetti}, {Miller}, {Paltani},
  {Paulin-Henriksson}, {Pires}, {Saxton}, {Schrabback}, {Seidel}, {Walsh},
  {Aghanim}, {Amendola}, {Bartlett}, {Baccigalupi}, {Beaulieu}, {Benabed},
  {Cuby}, {Elbaz}, {Fosalba}, {Gavazzi}, {Helmi}, {Hook}, {Irwin}, {Kneib},
  {Kunz}, {Mannucci}, {Moscardini}, {Tao}, {Teyssier}, {Weller}, {Zamorani},
  {Zapatero Osorio}, {Boulade}, {Foumond}, {Di Giorgio}, {Guttridge}, {James},
  {Kemp}, {Martignac}, {Spencer}, {Walton}, {Bl{\"u}mchen}, {Bonoli},
  {Bortoletto}, {Cerna}, {Corcione}, {Fabron}, {Jahnke}, {Ligori}, {Madrid},
  {Martin}, {Morgante}, {Pamplona}, {Prieto}, {Riva}, {Toledo}, {Trifoglio},
  {Zerbi}, {Abdalla}, {Douspis}, {Grenet}, {Borgani}, {Bouwens}, {Courbin},
  {Delouis}, {Dubath}, {Fontana}, {Frailis}, {Grazian}, {Koppenh{\"o}fer},
  {Mansutti}, {Melchior}, {Mignoli}, {Mohr}, {Neissner}, {Noddle}, {Poncet},
  {Scodeggio}, {Serrano}, {Shane}, {Starck}, {Surace}, {Taylor},
  {Verdoes-Kleijn}, {Vuerli}, {Williams}, {Zacchei}, {Altieri}, {Escudero
  Sanz}, {Kohley}, {Oosterbroek}, {Astier}, {Bacon}, {Bardelli}, {Baugh},
  {Bellagamba}, {Benoist}, {Bianchi}, {Biviano}, {Branchini}, {Carbone},
  {Cardone}, {Clements}, {Colombi}, {Conselice}, {Cresci}, {Deacon}, {Dunlop},
  {Fedeli}, {Fontanot}, {Franzetti}, {Giocoli}, {Garcia-Bellido}, {Gow},
  {Heavens}, {Hewett}, {Heymans}, {Holland}, {Huang}, {Ilbert}, {Joachimi},
  {Jennins}, {Kerins}, {Kiessling}, {Kirk}, {Kotak}, {Krause}, {Lahav}, {van
  Leeuwen}, {Lesgourgues}, {Lombardi}, {Magliocchetti}, {Maguire}, {Majerotto},
  {Maoli}, {Marulli}, {Maurogordato}, {McCracken}, {McLure}, {Melchiorri},
  {Merson}, {Moresco}, {Nonino}, {Norberg}, {Peacock}, {Pello}, {Penny},
  {Pettorino}, {Di Porto}, {Pozzetti}, {Quercellini}, {Radovich}, {Rassat},
  {Roche}, {Ronayette}, {Rossetti}, {Sartoris}, {Schneider}, {Semboloni},
  {Serjeant}, {Simpson}, {Skordis}, {Smadja}, {Smartt}, {Spano}, {Spiro},
  {Sullivan}, {Tilquin}, {Trotta}, {Verde}, {Wang}, {Williger}, {Zhao},
  {Zoubian}, \& {Zucca}}]{Laureijs2011arXiv11103193L}
{Laureijs}, R., {Amiaux}, J., {Arduini}, S., {et~al.} 2011, arXiv e-prints,
  arXiv:1110.3193

\bibitem[{{Le F{\`e}vre} {et~al.}(2013){Le F{\`e}vre}, {Cassata}, {Cucciati},
  {Garilli}, {Ilbert}, {Le Brun}, {Maccagni}, {Moreau}, {Scodeggio}, {Tresse},
  {Zamorani}, {Adami}, {Arnouts}, {Bardelli}, {Bolzonella}, {Bondi},
  {Bongiorno}, {Bottini}, {Cappi}, {Charlot}, {Ciliegi}, {Contini}, {de la
  Torre}, {Foucaud}, {Franzetti}, {Gavignaud}, {Guzzo}, {Iovino}, {Lemaux},
  {L{\'o}pez-Sanjuan}, {McCracken}, {Marano}, {Marinoni}, {Mazure}, {Mellier},
  {Merighi}, {Merluzzi}, {Paltani}, {Pell{\`o}}, {Pollo}, {Pozzetti},
  {Scaramella}, {Tasca}, {Vergani}, {Vettolani}, {Zanichelli}, \&
  {Zucca}}]{LeFevre2013AA}
{Le F{\`e}vre}, O., {Cassata}, P., {Cucciati}, O., {et~al.} 2013, \aap, 559,
  A14

\bibitem[{{Le F{\`e}vre} {et~al.}(2015){Le F{\`e}vre}, {Tasca}, {Cassata},
  {Garilli}, {Le Brun}, {Maccagni}, {Pentericci}, {Thomas}, {Vanzella},
  {Zamorani}, {Zucca}, {Amorin}, {Bardelli}, {Capak}, {Cassar{\`a}},
  {Castellano}, {Cimatti}, {Cuby}, {Cucciati}, {de la Torre}, {Durkalec},
  {Fontana}, {Giavalisco}, {Grazian}, {Hathi}, {Ilbert}, {Lemaux}, {Moreau},
  {Paltani}, {Ribeiro}, {Salvato}, {Schaerer}, {Scodeggio}, {Sommariva},
  {Talia}, {Taniguchi}, {Tresse}, {Vergani}, {Wang}, {Charlot}, {Contini},
  {Fotopoulou}, {L{\'o}pez-Sanjuan}, {Mellier}, \& {Scoville}}]{lefevre2015AA}
{Le F{\`e}vre}, O., {Tasca}, L.~A.~M., {Cassata}, P., {et~al.} 2015, \aap, 576,
  A79

\bibitem[{{MacCrann} {et~al.}(2022){MacCrann}, {Becker}, {McCullough}, {Amon},
  {Gruen}, {Jarvis}, {Choi}, {Troxel}, {Sheldon}, {Yanny}, {Herner},
  {Dodelson}, {Zuntz}, {Eckert}, {Rollins}, {Varga}, {Bernstein}, {Gruendl},
  {Harrison}, {Hartley}, {Sevilla-Noarbe}, {Pieres}, {Bridle}, {Myles},
  {Alarcon}, {Everett}, {S{\'a}nchez}, {Huff}, {Tarsitano}, {Gatti}, {Secco},
  {Abbott}, {Aguena}, {Allam}, {Annis}, {Bacon}, {Bertin}, {Brooks}, {Burke},
  {Carnero Rosell}, {Carrasco Kind}, {Carretero}, {Costanzi}, {Crocce},
  {Pereira}, {De Vicente}, {Desai}, {Diehl}, {Dietrich}, {Doel}, {Eifler},
  {Ferrero}, {Fert{\'e}}, {Flaugher}, {Fosalba}, {Frieman},
  {Garc{\'\i}a-Bellido}, {Gaztanaga}, {Gerdes}, {Giannantonio}, {Gschwend},
  {Gutierrez}, {Hinton}, {Hollowood}, {Honscheid}, {James}, {Lahav}, {Lima},
  {Maia}, {March}, {Marshall}, {Martini}, {Melchior}, {Menanteau}, {Miquel},
  {Mohr}, {Morgan}, {Muir}, {Ogando}, {Palmese}, {Paz-Chinch{\'o}n}, {Plazas},
  {Rodriguez-Monroy}, {Roodman}, {Samuroff}, {Sanchez}, {Scarpine}, {Serrano},
  {Smith}, {Soares-Santos}, {Suchyta}, {Swanson}, {Tarle}, {Thomas}, {To},
  {Wilkinson}, {Wilkinson}, \& {DES Collaboration}}]{MacCrann2022MNRAS5093371M}
{MacCrann}, N., {Becker}, M.~R., {McCullough}, J., {et~al.} 2022, \mnras, 509,
  3371

\bibitem[{{Mandelbaum}(2018)}]{Mandelbaum2018ARAA56393M}
{Mandelbaum}, R. 2018, \araa, 56, 393

\bibitem[{{Mandelbaum} {et~al.}(2018){Mandelbaum}, {Lanusse}, {Leauthaud},
  {Armstrong}, {Simet}, {Miyatake}, {Meyers}, {Bosch}, {Murata}, {Miyazaki}, \&
  {Tanaka}}]{Mandelbaum2018MNRAS4813170M}
{Mandelbaum}, R., {Lanusse}, F., {Leauthaud}, A., {et~al.} 2018, \mnras, 481,
  3170

\bibitem[{{Mandelbaum} {et~al.}(2015){Mandelbaum}, {Rowe}, {Armstrong}, {Bard},
  {Bertin}, {Bosch}, {Boutigny}, {Courbin}, {Dawson}, {Donnarumma}, {Fenech
  Conti}, {Gavazzi}, {Gentile}, {Gill}, {Hogg}, {Huff}, {Jee}, {Kacprzak},
  {Kilbinger}, {Kuntzer}, {Lang}, {Luo}, {March}, {Marshall}, {Meyers},
  {Miller}, {Miyatake}, {Nakajima}, {Ngol{\'e} Mboula}, {Nurbaeva}, {Okura},
  {Paulin-Henriksson}, {Rhodes}, {Schneider}, {Shan}, {Sheldon}, {Simet},
  {Starck}, {Sureau}, {Tewes}, {Zarb Adami}, {Zhang}, \&
  {Zuntz}}]{Mandelbaum2015MNRAS4502963M}
{Mandelbaum}, R., {Rowe}, B., {Armstrong}, R., {et~al.} 2015, \mnras, 450, 2963

\bibitem[{{Massey} {et~al.}(2007){Massey}, {Heymans}, {Berg{\'e}}, {Bernstein},
  {Bridle}, {Clowe}, {Dahle}, {Ellis}, {Erben}, {Hetterscheidt}, {High},
  {Hirata}, {Hoekstra}, {Hudelot}, {Jarvis}, {Johnston}, {Kuijken},
  {Margoniner}, {Mandelbaum}, {Mellier}, {Nakajima}, {Paulin-Henriksson},
  {Peeples}, {Roat}, {Refregier}, {Rhodes}, {Schrabback}, {Schirmer}, {Seljak},
  {Semboloni}, \& {van Waerbeke}}]{Massey2007MNRAS37613M}
{Massey}, R., {Heymans}, C., {Berg{\'e}}, J., {et~al.} 2007, \mnras, 376, 13

\bibitem[{{Massey} {et~al.}(2013){Massey}, {Hoekstra}, {Kitching}, {Rhodes},
  {Cropper}, {Amiaux}, {Harvey}, {Mellier}, {Meneghetti}, {Miller},
  {Paulin-Henriksson}, {Pires}, {Scaramella}, \&
  {Schrabback}}]{Massey2013MNRAS429661M}
{Massey}, R., {Hoekstra}, H., {Kitching}, T., {et~al.} 2013, \mnras, 429, 661

\bibitem[{{Masters} {et~al.}(2017){Masters}, {Stern}, {Cohen}, {Capak},
  {Rhodes}, {Castander}, \& {Paltani}}]{Masters2017ApJ}
{Masters}, D.~C., {Stern}, D.~K., {Cohen}, J.~G., {et~al.} 2017, \apj, 841, 111

\bibitem[{{Masters} {et~al.}(2019){Masters}, {Stern}, {Cohen}, {Capak},
  {Stanford}, {Hernitschek}, {Galametz}, {Davidzon}, {Rhodes}, {Sanders},
  {Mobasher}, {Castander}, {Pruett}, \& {Fotopoulou}}]{Masters2019ApJ}
{Masters}, D.~C., {Stern}, D.~K., {Cohen}, J.~G., {et~al.} 2019, \apj, 877, 81

\bibitem[{{McConnell} \& {Ma}(2013)}]{McConnell2013ApJ764184M}
{McConnell}, N.~J. \& {Ma}, C.-P. 2013, \apj, 764, 184

\bibitem[{{McCracken} {et~al.}(2012){McCracken}, {Milvang-Jensen}, {Dunlop},
  {Franx}, {Fynbo}, {Le F{\`e}vre}, {Holt}, {Caputi}, {Goranova}, {Buitrago},
  {Emerson}, {Freudling}, {Hudelot}, {L{\'o}pez-Sanjuan}, {Magnard}, {Mellier},
  {M{\o}ller}, {Nilsson}, {Sutherland}, {Tasca}, \&
  {Zabl}}]{McCracken2012AA544A156M}
{McCracken}, H.~J., {Milvang-Jensen}, B., {Dunlop}, J., {et~al.} 2012, \aap,
  544, A156

\bibitem[{{McFarland} {et~al.}(2013){McFarland}, {Verdoes-Kleijn}, {Sikkema},
  {Helmich}, {Boxhoorn}, \& {Valentijn}}]{McFarland2013ExA3545M}
{McFarland}, J.~P., {Verdoes-Kleijn}, G., {Sikkema}, G., {et~al.} 2013,
  Experimental Astronomy, 35, 45

\bibitem[{{Melchior} \& {Viola}(2012)}]{Melchior2012MNRAS4242757M}
{Melchior}, P. \& {Viola}, M. 2012, \mnras, 424, 2757

\bibitem[{{Miller} {et~al.}(2013){Miller}, {Heymans}, {Kitching}, {van
  Waerbeke}, {Erben}, {Hildebrandt}, {Hoekstra}, {Mellier}, {Rowe}, {Coupon},
  {Dietrich}, {Fu}, {Harnois-D{\'e}raps}, {Hudson}, {Kilbinger}, {Kuijken},
  {Schrabback}, {Semboloni}, {Vafaei}, \& {Velander}}]{Miller2013MNRAS4292858M}
{Miller}, L., {Heymans}, C., {Kitching}, T.~D., {et~al.} 2013, \mnras, 429,
  2858

\bibitem[{{Miller} {et~al.}(2007){Miller}, {Kitching}, {Heymans}, {Heavens}, \&
  {van Waerbeke}}]{Miller2007MNRAS382315M}
{Miller}, L., {Kitching}, T.~D., {Heymans}, C., {Heavens}, A.~F., \& {van
  Waerbeke}, L. 2007, \mnras, 382, 315

\bibitem[{{Navarro} {et~al.}(1995){Navarro}, {Frenk}, \&
  {White}}]{Navarro1995MNRAS275720N}
{Navarro}, J.~F., {Frenk}, C.~S., \& {White}, S. D.~M. 1995, \mnras, 275, 720

\bibitem[{{Navarro} {et~al.}(1996){Navarro}, {Frenk}, \&
  {White}}]{Navarro1996ApJ462563N}
{Navarro}, J.~F., {Frenk}, C.~S., \& {White}, S. D.~M. 1996, \apj, 462, 563

\bibitem[{{Navarro} {et~al.}(1997){Navarro}, {Frenk}, \&
  {White}}]{Navarro1997ApJ490493N}
{Navarro}, J.~F., {Frenk}, C.~S., \& {White}, S. D.~M. 1997, \apj, 490, 493

\bibitem[{{Obreschkow} {et~al.}(2009){Obreschkow}, {Kl{\"o}ckner}, {Heywood},
  {Levrier}, \& {Rawlings}}]{Obreschkow2009ApJ...703.1890O}
{Obreschkow}, D., {Kl{\"o}ckner}, H.~R., {Heywood}, I., {Levrier}, F., \&
  {Rawlings}, S. 2009, \apj, 703, 1890

\bibitem[{{Paulin-Henriksson} {et~al.}(2008){Paulin-Henriksson}, {Amara},
  {Voigt}, {Refregier}, \& {Bridle}}]{Paulin2008AA48467P}
{Paulin-Henriksson}, S., {Amara}, A., {Voigt}, L., {Refregier}, A., \&
  {Bridle}, S.~L. 2008, \aap, 484, 67

\bibitem[{{Planck Collaboration}(2016)}]{Planck2016AA}
{Planck Collaboration}. 2016, \aap, 594, A13

\bibitem[{{Poulton} {et~al.}(2018){Poulton}, {Robotham}, {Power}, \&
  {Elahi}}]{Poulton2018PASA3542P}
{Poulton}, R. J.~J., {Robotham}, A. S.~G., {Power}, C., \& {Elahi}, P.~J. 2018,
  \pasa, 35, e042

\bibitem[{{Raichoor} {et~al.}(2014){Raichoor}, {Mei}, {Erben}, {Hildebrandt},
  {Huertas-Company}, {Ilbert}, {Licitra}, {Ball}, {Boissier}, {Boselli},
  {Chen}, {C{\^o}t{\'e}}, {Cuillandre}, {Duc}, {Durrell}, {Ferrarese},
  {Guhathakurta}, {Gwyn}, {Kavelaars}, {Lan{\c{c}}on}, {Liu}, {MacArthur},
  {Muller}, {Mu{\~n}oz}, {Peng}, {Puzia}, {Sawicki}, {Toloba}, {Van Waerbeke},
  {Woods}, \& {Zhang}}]{Raichoor2014ApJ}
{Raichoor}, A., {Mei}, S., {Erben}, T., {et~al.} 2014, \apj, 797, 102

\bibitem[{{Refregier}(2003)}]{Refregier2003ARAA41645R}
{Refregier}, A. 2003, \araa, 41, 645

\bibitem[{{Refregier} {et~al.}(2012){Refregier}, {Kacprzak}, {Amara}, {Bridle},
  \& {Rowe}}]{Refregier2012MNRAS4251951R}
{Refregier}, A., {Kacprzak}, T., {Amara}, A., {Bridle}, S., \& {Rowe}, B. 2012,
  \mnras, 425, 1951

\bibitem[{{Robin} {et~al.}(2003){Robin}, {Reyl{\'e}}, {Derri{\`e}re}, \&
  {Picaud}}]{Robin2003AA409523R}
{Robin}, A.~C., {Reyl{\'e}}, C., {Derri{\`e}re}, S., \& {Picaud}, S. 2003,
  \aap, 409, 523

\bibitem[{{Robotham} {et~al.}(2020){Robotham}, {Bellstedt}, {Lagos}, {Thorne},
  {Davies}, {Driver}, \& {Bravo}}]{Robotham2020MNRAS495905R}
{Robotham}, A.~S.~G., {Bellstedt}, S., {Lagos}, C. d.~P., {et~al.} 2020,
  \mnras, 495, 905

\bibitem[{{Rowe} {et~al.}(2015){Rowe}, {Jarvis}, {Mandelbaum}, {Bernstein},
  {Bosch}, {Simet}, {Meyers}, {Kacprzak}, {Nakajima}, {Zuntz}, {Miyatake},
  {Dietrich}, {Armstrong}, {Melchior}, \& {Gill}}]{galsim_2015AC10121R}
{Rowe}, B.~T.~P., {Jarvis}, M., {Mandelbaum}, R., {et~al.} 2015, Astronomy and
  Computing, 10, 121

\bibitem[{{Samuroff} {et~al.}(2018){Samuroff}, {Bridle}, {Zuntz}, {Troxel},
  {Gruen}, {Rollins}, {Bernstein}, {Eifler}, {Huff}, {Kacprzak}, {Krause},
  {MacCrann}, {Abdalla}, {Allam}, {Annis}, {Bechtol}, {Benoit-L{\'e}vy},
  {Bertin}, {Brooks}, {Buckley-Geer}, {Carnero Rosell}, {Carrasco Kind},
  {Carretero}, {Crocce}, {D'Andrea}, {da Costa}, {Davis}, {Desai}, {Doel},
  {Fausti Neto}, {Flaugher}, {Fosalba}, {Frieman}, {Garc{\'\i}a-Bellido},
  {Gerdes}, {Gruendl}, {Gschwend}, {Gutierrez}, {Honscheid}, {James}, {Jarvis},
  {Jeltema}, {Kirk}, {Kuehn}, {Kuhlmann}, {Li}, {Lima}, {Maia}, {March},
  {Marshall}, {Martini}, {Melchior}, {Menanteau}, {Miquel}, {Nord}, {Ogando},
  {Plazas}, {Roodman}, {Sanchez}, {Scarpine}, {Schindler}, {Schubnell},
  {Sevilla-Noarbe}, {Sheldon}, {Smith}, {Soares-Santos}, {Sobreira}, {Suchyta},
  {Tarle}, {Thomas}, {Tucker}, \& {DES
  Collaboration}}]{Samuroff2018MNRAS4754524S}
{Samuroff}, S., {Bridle}, S.~L., {Zuntz}, J., {et~al.} 2018, \mnras, 475, 4524

\bibitem[{{Schirmer}(2013)}]{Schirmer2013ApJS20921S}
{Schirmer}, M. 2013, \apjs, 209, 21

\bibitem[{{S{\'e}rsic}(1963)}]{sersic1963BAAA641S}
{S{\'e}rsic}, J.~L. 1963, Boletin de la Asociacion Argentina de Astronomia La
  Plata Argentina, 6, 41

\bibitem[{{Sheldon} \& {Huff}(2017)}]{Sheldon2017ApJ84124S}
{Sheldon}, E.~S. \& {Huff}, E.~M. 2017, \apj, 841, 24

\bibitem[{{Silverman} {et~al.}(2015){Silverman}, {Kashino}, {Sanders},
  {Kartaltepe}, {Arimoto}, {Renzini}, {Rodighiero}, {Daddi}, {Zahid}, {Nagao},
  {Kewley}, {Lilly}, {Sugiyama}, {Baronchelli}, {Capak}, {Carollo}, {Chu},
  {Hasinger}, {Ilbert}, {Juneau}, {Kajisawa}, {Koekemoer}, {Kovac}, {Le
  F{\`e}vre}, {Masters}, {McCracken}, {Onodera}, {Schulze}, {Scoville},
  {Strazzullo}, \& {Taniguchi}}]{Silverman2015ApJS}
{Silverman}, J.~D., {Kashino}, D., {Sanders}, D., {et~al.} 2015, \apjs, 220, 12

\bibitem[{Sklar(1959)}]{sklar1959functions}
Sklar, M. 1959, Publ. inst. statist. univ. Paris, 8, 229

\bibitem[{{Spergel} {et~al.}(2015){Spergel}, {Gehrels}, {Baltay}, {Bennett},
  {Breckinridge}, {Donahue}, {Dressler}, {Gaudi}, {Greene}, {Guyon}, {Hirata},
  {Kalirai}, {Kasdin}, {Macintosh}, {Moos}, {Perlmutter}, {Postman},
  {Rauscher}, {Rhodes}, {Wang}, {Weinberg}, {Benford}, {Hudson}, {Jeong},
  {Mellier}, {Traub}, {Yamada}, {Capak}, {Colbert}, {Masters}, {Penny},
  {Savransky}, {Stern}, {Zimmerman}, {Barry}, {Bartusek}, {Carpenter}, {Cheng},
  {Content}, {Dekens}, {Demers}, {Grady}, {Jackson}, {Kuan}, {Kruk}, {Melton},
  {Nemati}, {Parvin}, {Poberezhskiy}, {Peddie}, {Ruffa}, {Wallace}, {Whipple},
  {Wollack}, \& {Zhao}}]{Spergel2015arXiv150303757S}
{Spergel}, D., {Gehrels}, N., {Baltay}, C., {et~al.} 2015, arXiv e-prints,
  arXiv:1503.03757

\bibitem[{{Stanford} {et~al.}(2021){Stanford}, {Masters}, {Darvish}, {Stern},
  {Cohen}, {Capak}, {Hernitschek}, {Davidzon}, {Rhodes}, {Sanders}, {Mobasher},
  {Castander}, {Paltani}, {Aghanim}, {Amara}, {Auricchio}, {Balestra},
  {Bender}, {Bodendorf}, {Bonino}, {Branchini}, {Brinchmann}, {Capobianco},
  {Carbone}, {Carretero}, {Casas}, {Castellano}, {Cavuoti}, {Cimatti},
  {Cledassou}, {Conselice}, {Corcione}, {Costille}, {Cropper}, {Degaudenzi},
  {Douspis}, {Dubath}, {Dusini}, {Fosalba}, {Frailis}, {Franceschi},
  {Franzetti}, {Fumana}, {Garilli}, {Giocoli}, {Grupp}, {Haugan}, {Hoekstra},
  {Holmes}, {Hormuth}, {Hudelot}, {Jahnke}, {Kiessling}, {Kilbinger},
  {Kitching}, {Kubik}, {K{\"u}mmel}, {Kunz}, {Kurki-Suonio}, {Laureijs},
  {Ligori}, {Lilje}, {Lloro}, {Maiorano}, {Marggraf}, {Markovic}, {Massey},
  {Meneghetti}, {Meylan}, {Moscardini}, {Niemi}, {Padilla}, {Pasian},
  {Pedersen}, {Pettorino}, {Pires}, {Poncet}, {Popa}, {Pozzetti}, {Raison},
  {Roncarelli}, {Rossetti}, {Saglia}, {Scaramella}, {Schneider}, {Secroun},
  {Seidel}, {Serrano}, {Sirignano}, {Sirri}, {Taylor}, {Teplitz}, {Tereno},
  {Toledo-Moreo}, {Valentijn}, {Valenziano}, {Verdoes Kleijn}, {Wang},
  {Zamorani}, {Zoubian}, {Brescia}, {Congedo}, {Conversi}, {Copin}, {Kermiche},
  {Kohley}, {Medinaceli}, {Mei}, {Moresco}, {Morin}, {Munari}, {Polenta},
  {Sureau}, {Tallada Cresp{\'\i}}, {Vassallo}, {Zacchei}, {Andreon}, {Aussel},
  {Baccigalupi}, {Balaguera-Antol{\'\i}nez}, {Baldi}, {Bardelli}, {Biviano},
  {Borsato}, {Bozzo}, {Burigana}, {Cabanac}, {Camera}, {Cappi}, {Carvalho},
  {Casas}, {Castignani}, {Colodro-Conde}, {Coupon}, {Courtois}, {Cuby}, {Da
  Silva}, {de la Torre}, {Di Ferdinando}, {Duncan}, {Dupac}, {Fabricius},
  {Farina}, {Farrens}, {Ferreira}, {Finelli}, {Flose-Reimberg}, {Fotopoulou},
  {Galeotta}, {Ganga}, {Gillard}, {Gozaliasl}, {Graci{\'a}-Carpio}, {Keihanen},
  {Kirkpatrick}, {Lindholm}, {Mainetti}, {Maino}, {Martinet}, {Marulli},
  {Maturi}, {Maurogordato}, {Metcalf}, {Nakajima}, {Neissner}, {Nightingale},
  {Nucita}, {Patrizii}, {Potter}, {Renzi}, {Riccio}, {Romelli}, {S{\'a}nchez},
  {Sapone}, {Schirmer}, {Schultheis}, {Scottez}, {Stanco}, {Tenti}, {Teyssier},
  {Torradeflot}, {Valiviita}, {Viel}, {Whittaker}, \&
  {Zucca}}]{Stanford2021ApJS}
{Stanford}, S.~A., {Masters}, D., {Darvish}, B., {et~al.} 2021, \apjs, 256, 9

\bibitem[{{Tallada} {et~al.}(2020){Tallada}, {Carretero}, {Casals},
  {Acosta-Silva}, {Serrano}, {Caubet}, {Castander}, {C{\'e}sar}, {Crocce},
  {Delfino}, {Eriksen}, {Fosalba}, {Gazta{\~n}aga}, {Merino}, {Neissner}, \&
  {Tonello}}]{Tallada2020AC3200391T}
{Tallada}, P., {Carretero}, J., {Casals}, J., {et~al.} 2020, Astronomy and
  Computing, 32, 100391

\bibitem[{{van den Busch} {et~al.}(2020){van den Busch}, {Hildebrandt},
  {Wright}, {Morrison}, {Blake}, {Joachimi}, {Erben}, {Heymans}, {Kuijken}, \&
  {Taylor}}]{Busch2020AA642A200V}
{van den Busch}, J.~L., {Hildebrandt}, H., {Wright}, A.~H., {et~al.} 2020,
  \aap, 642, A200

\bibitem[{{van den Busch} {et~al.}(2022){van den Busch}, {Wright},
  {Hildebrandt}, {Bilicki}, {Asgari}, {Joudaki}, {Blake}, {Heymans},
  {Kannawadi}, {Shan}, \& {Tr{\"o}ster}}]{Busch2022AA...664A.170V}
{van den Busch}, J.~L., {Wright}, A.~H., {Hildebrandt}, H., {et~al.} 2022,
  \aap, 664, A170

\bibitem[{{van der Wel} {et~al.}(2016){van der Wel}, {Noeske}, {Bezanson},
  {Pacifici}, {Gallazzi}, {Franx}, {Mu{\~n}oz-Mateos}, {Bell}, {Brammer},
  {Charlot}, {Chauk{\'e}}, {Labb{\'e}}, {Maseda}, {Muzzin}, {Rix}, {Sobral},
  {van de Sande}, {van Dokkum}, {Wild}, \& {Wolf}}]{vanderwel2016ApJS}
{van der Wel}, A., {Noeske}, K., {Bezanson}, R., {et~al.} 2016, \apjs, 223, 29

\bibitem[{{Viola} {et~al.}(2015){Viola}, {Cacciato}, {Brouwer}, {Kuijken},
  {Hoekstra}, {Norberg}, {Robotham}, {van Uitert}, {Alpaslan}, {Baldry},
  {Choi}, {de Jong}, {Driver}, {Erben}, {Grado}, {Graham}, {Heymans},
  {Hildebrandt}, {Hopkins}, {Irisarri}, {Joachimi}, {Loveday}, {Miller},
  {Nakajima}, {Schneider}, {Sif{\'o}n}, \& {Verdoes
  Kleijn}}]{Viola2015MNRAS.452.3529V}
{Viola}, M., {Cacciato}, M., {Brouwer}, M., {et~al.} 2015, \mnras, 452, 3529

\bibitem[{{Wright} {et~al.}(2018){Wright}, {Driver}, \&
  {Robotham}}]{Wright2018MNRAS4803491W}
{Wright}, A.~H., {Driver}, S.~P., \& {Robotham}, A.~S.~G. 2018, \mnras, 480,
  3491

\bibitem[{{Wright} {et~al.}(2019){Wright}, {Hildebrandt}, {Kuijken}, {Erben},
  {Blake}, {Buddelmeijer}, {Choi}, {Cross}, {de Jong}, {Edge},
  {Gonzalez-Fernandez}, {Gonz{\'a}lez Solares}, {Grado}, {Heymans}, {Irwin},
  {Kupcu Yoldas}, {Lewis}, {Mann}, {Napolitano}, {Radovich}, {Schneider},
  {Sif{\'o}n}, {Sutherland}, {Sutorius}, \& {Verdoes
  Kleijn}}]{Wright2019AA632A34W}
{Wright}, A.~H., {Hildebrandt}, H., {Kuijken}, K., {et~al.} 2019, \aap, 632,
  A34

\bibitem[{{Wright} {et~al.}(2020){Wright}, {Hildebrandt}, {van den Busch}, \&
  {Heymans}}]{Wright2020AA637A100W}
{Wright}, A.~H., {Hildebrandt}, H., {van den Busch}, J.~L., \& {Heymans}, C.
  2020, \aap, 637, A100

\bibitem[{{Wright} \& {Brainerd}(2000)}]{Wright2000ApJ53434W}
{Wright}, C.~O. \& {Brainerd}, T.~G. 2000, \apj, 534, 34

\end{thebibliography}
